# Technical Design Study for the

# $\overline{\text{P}}$ANDA
# Time Projection Chamber

GEM-TPC Collaboration

June 29, 2012

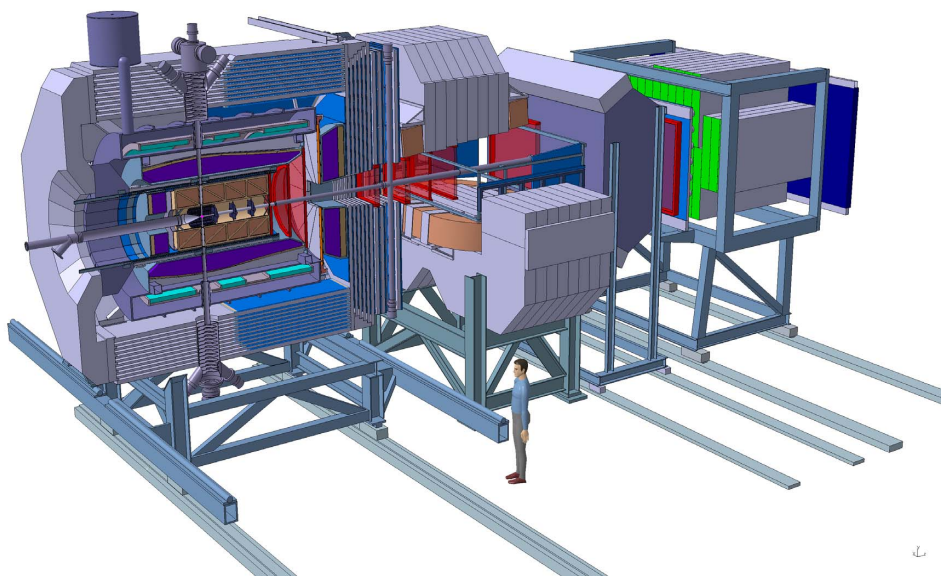

# The GEM-TPC Collaboration

Technische Universität **München**, Germany

M. Ball, F. V. Böhmer, S. Dørheim, C. Höppner, B. Ketzer, I. Konorov, S. Neubert, S. Paul, J. Rauch, S. Uhl, M. Vandenbroucke

Excellence Cluster "Universe", TU **München**, Germany

M. Berger, J.-C. Berger-Chen, F. Cusanno, L. Fabbietti, R. Münzer

GSI Helmholtzzentrum für Schwerionenforschung GmbH, **Darmstadt**, Germany

R. Arora, J. Frühauf, M. Kiš, Y. Leifels, V. Kleipa, J. Hehner, J. Kunkel, N. Kurz, K. Peters, H. Risch, C. J. Schmidt, L. Schmitt, S. Schwab, D. Soyk, B. Voss, J. Weinert

Helmholtz-Institut für Strahlen- und Kernphysik, **Bonn**, Germany

R. Beck, D. Kaiser, M. Lang, R. Schmitz, D. Walther

Stefan Meyer Institut für Subatomare Physik, **Vienna**, Austria

P. Bühler, P. Müllner, J. Zmeskal

Universität **Heidelberg**, Germany

N. Hermann

# Preface

This document illustrates the technical layout and the expected performance of the Time Projection Chamber as the central tracking system of the $\overline{\mathsf{P}}$ANDA experiment.

# Contents

# 1 Introduction

A Time Projection Chamber (TPC) [1] with its low material budget constitutes an ideal device for tracking charged particles in 3-dimensional space, fulfilling all the requirements on $\overline{\text{P}}$ANDA tracking. TPCs have been and currently are successfully employed in many experiments such as PEP-4 [2], ALEPH [3], DELPHI [4], NA49 [5, 6], STAR [7, 8], ALICE [9]. In its standard form, a TPC consists of a large gas-filled cylindrical volume inside a solenoid magnetic field, surrounding the interaction point, and covering the full $4\pi$ solid angle [10]. Fig. 1.1 shows a schematic view of a TPC.

An electric field along the cylinder axis separates positive gas ions from electrons created by ionizing particles traversing the gas volume. The primary electrons then drift towards the readout anode located at end cap of the cylinder, the transverse diffusion being reduced by the strong magnetic field parallel to the drift direction. At the end cap avalanche amplification occurs typically in Multiwire Proportional Chambers (MWPCs), the induced signals being detected by an arrangement of pad electrodes measuring the projection of the track onto the end plane. The third coordinate of the track comes from a measurement of the drift time of each primary electron cluster, requiring a precise knowledge of the electric drift field in the chamber. Distortions of the electric field in the drift volume due to the accumulation of space charge from primary ions or avalanche ions drifting back into the drift volume deteriorate the resolution and have to be kept at a minimum. To this end all TPCs up to now have been operated in a pulsed mode, where an electrostatic gate to the readout region is opened only when an interaction in the target has occurred, and is closed immediately thereafter, preventing avalanche ions from penetrating the drift volume.

Owing to the beam properties at the High Energy Storage Ring (HESR) of FAIR, with its high luminosity of $2 \cdot 10^{32}\,\text{cm}^{-2}\,\text{s}^{-1}$ corresponding to $2 \cdot 10^7$ $\overline{\text{p}}$ annihilations per second, the TPC in $\overline{\text{P}}$ANDA has to operate continuously, i. e. the technique of gating cannot be applied. Ions created in the multiplication region have to be prevented from drifting back into the drift volume by other means.

This will be achieved by using Gas Electron Multiplier (GEM) [11] foils as charge amplifier instead of conventional MWPCs. The GEM consists of a $50\,\mu\text{m}$ thin insulating Polyimide foil with Cu-coated surfaces, typically $2-5\,\mu\text{m}$ thick. The foil is perforated by photo-lithographic processing, forming a dense, regular pattern of (double-conical) holes. Usually the holes have an inner diameter of $\sim 50\,\mu\text{m}$. Figure 1.2 shows an electron microscope photograph of a GEM foil, also indicating the typical dimensions.

The small dimensions of the amplification structures lead to very large field strengths $\mathcal{O}(50\,\text{kV/cm})$ inside the holes of the GEM foil when a moderate voltage difference of typically $300-400\,\text{V}$ is applied between the metal layers, sufficient for avalanche creation inside the GEM holes. The avalanche electrons are extracted from the bottom side of the foil, and can be collected or transferred to the next amplification stage. Typically three GEM foils are combined in a stack, leading to effective gains (see Eq. 11.1) of $\sim 10^4$ and at the same time guaran-teeing a stable operation without the occurrence of discharges [12].

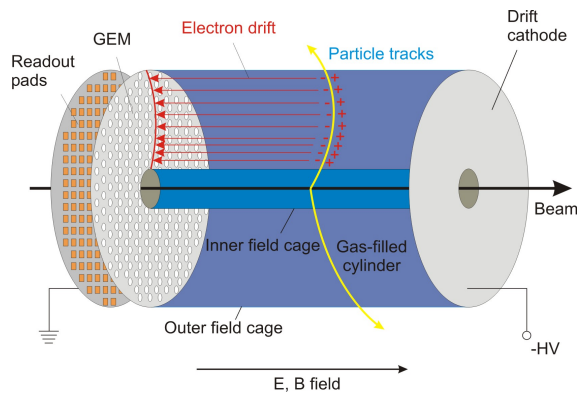

**Figure 1.1:** Schematic view of a GEM-based TPC.

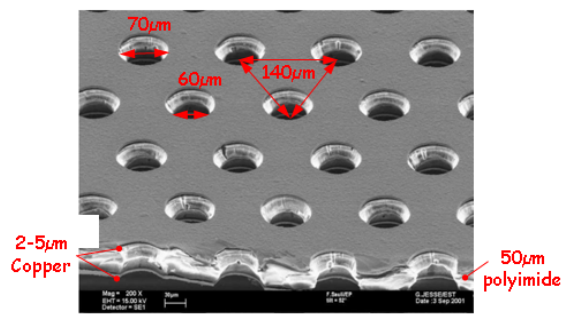

**Figure 1.2:** Electron microscope photograph of a GEM foil with typical dimensions.

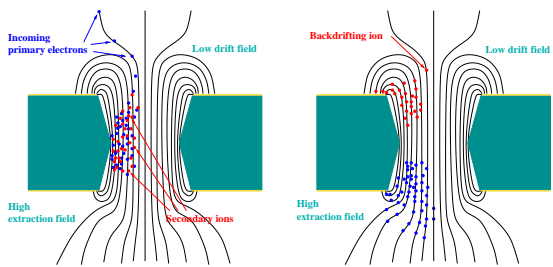

**Figure 1.3:** Working principle of a GEM: (left) electrons are guided into the holes by the low drift field, where avalanches of electron-ion pairs are generated. (right) The asymmetric field configuration of low drift field and higher extraction field together with the small ion mobility lead to efficient back flow suppression.

The back flow of ions from the amplification region is suppressed in a configuration with asymmetric electric fields above and below the GEM foil, as shown in Fig. 1.3. Electrons are guided into the GEM holes by the drift field, where they produce an avalanche. The ions created in the avalanche are mainly collected on the top side of the GEM foil, if the extraction field is much higher than the drift field.

Detectors based on GEM amplification have been pioneered by the COMPASS experiment at CERN [13, 14, 15, 16], and are now routinely used in several particle physics experiments like LHCb [17], PHENIX [18], and TOTEM [19]. New applications include the use of GEM-based detectors in KLOE-2 [20] and CMS [21]. Its main features, rendering possible a continuously operating TPC, are:

- suppression of ion back flow in an asymmetric field configuration,

- high granularity,

- high rate capability,

- no preferred direction (as for wires), therefore isotropic $\boldsymbol{E} \times \boldsymbol{B}$ effects.

The GEM readout scheme is also envisaged for the TPC of the International linear Collider ILC [22], and combined efforts are currently concentrating on further developing this technique. The increased granularity of such a detector necessarily leads to larger event sizes and requires substantial data reduction already at the level of front-end electronics.

The ungated, continuous operation mode of the TPC at the envisaged event rates at $\overline{\mathrm{P}}$ANDA gives rise to about 3000 tracks which are superimposed in the drift volume at any given time. The association of these tracks to distinct physics events ("event deconvolution") requires fast online tracking capabilities of the data acquisition system.

In conclusion, a TPC read out by the Gas Electron Multiplier (GEM) will fulfill all the requirements to the Central Tracker of the $\overline{\mathrm{P}}$ANDA experiment. Its stand-alone momentum resolution is sufficiently high such that also the momenta of tracks not traversing the Micro-Vertex Detector (MVD), like from $\Lambda$ decays, can be determined with the required precision. Its low material budget will minimize multiple scattering of charged particles and photon conversion, thus optimizing the resolution of the spectrometer both for charged and neutral particles. The large number of 3-D space points ($\sim 50 - 100$) measured for each track greatly simplifies pattern recognition in a complex and dense environment. This is especially important for low-momentum particles which do not leave the Central Tracker, but spiral with a small radius of the order of a few cm along the $z$ direction, and for the detection of neutral particle decays or kinks. Monte-Carlo simulations have shown that about 40% of charmonium decays have tracks going in forward direction, i.e. passing the Central Tracker through its forward end cap. In this region of phase space, the bending power of the solenoid magnet is small and the number of points measured on a given track is small. Therefore it is vital that the Central Tracker is able to contribute with a substantial number of hits, also providing $z$ information for these hits. Finally, the TPC will contribute to the identification of charged particles by measuring the specific energy loss, $\mathrm{d}E/\mathrm{d}x$ of each particle track, especially at momenta below $1\,\mathrm{GeV}/c$, which is essential for background suppression.

# 2 General Detector Layout

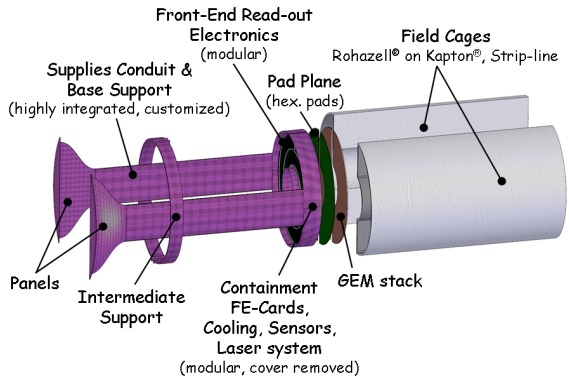

**Figure 2.1:** Conceptual Design of the $\overline{\text{P}}$ANDA GEM-TPC; only the basic modules of the detector are shown.

## 2.1 Design rules

Throughout the layout the following goals where formulated and respective general design rules were deduced from them.

Minimization of risks and failures will be done by maximizing the modularity of functional units and keeping diversity in the design wherever it is possible and feasible. We try to minimize any bias on physical quantities by reducing the material budget of all structural elements maximizing at the same time the geometrical yield, acceptance and active volume. In all phases of the life cycle of the detector system such as: design, part production, mounting, adjustment, calibration, operation and maintenance we try to minimize the required effort, work load and costs. This immediately calls for flexibility in the overall design and a maximum compactness of all modules. We try to take advantage of symmetries especially with respect to the required support structures, and make use of pre-fabrication and pre-mounting as much as reasonably possible.

## 2.2 Detector structure

The envisaged sections of the $\overline{\text{P}}$ANDA GEM-based TPC are visualized schematically in Fig. 2.1. They are composed of:

- A detection volume where traversing particles ionize the gaseous detection medium producing positive and negative charge carriers in an electrical drift field. This field, which is oriented parallel to the beam-axis, is spawned between a cathode end cap downstream of the target position and the anode layer upstream of it. A field-defining system, the so called 'field cage', is required to keep the gradient of the field constant throughout the active volume. All electrode structures, made by metallic material patterned on a carrier material, are optimized in terms of achievable field homogeneity and material budget (see Sec. 3).

- An amplification stage that serves to multiply the ionization electrons after they have traversed the drift volume. In case of a GEM-TPC this is done by a GEM stack which is composed of a plane-parallel set of GEM foils oriented perpendicular to the beam axis (see Sec. 4 for details).

- A patterned electrode structure on the anode layer, the so called pad plane (see Sec. 4.4), where the signals corresponding to the collected charges are detected.

- The electronic readout system (see Sec. 8).

- The support and alignment structures and its fixation points.

- The supplies like gaseous and fluid media lines, low and high voltage and the respective cables, pipes and terminals (panels).

- A system of various detector-near sensors monitoring all relevant operation parameters like temperature, pressure and flow of gas and cooling media, high and low voltages (see Sec. 7).

## 2.3 Overall configuration

The configuration of the $\overline{\text{P}}$ANDA GEM-TPC is shown in Fig. 2.1 and Fig. 2.2. The whole structure will be self-supporting; the overall center of gravity of the set up will be close to the backward end-cap. As a consequence, the external support structures may be minimized, mainly requiring only a few fixation points: one carrying the main load of approximately $150\,\text{kg}$ (field cage: $2 \times 22\,\text{kg}$, rest: $106\,\text{kg}$) including cabling and terminals, one on the intermediate support structure giving the orientation, and one installed in the vicinity of the cathode

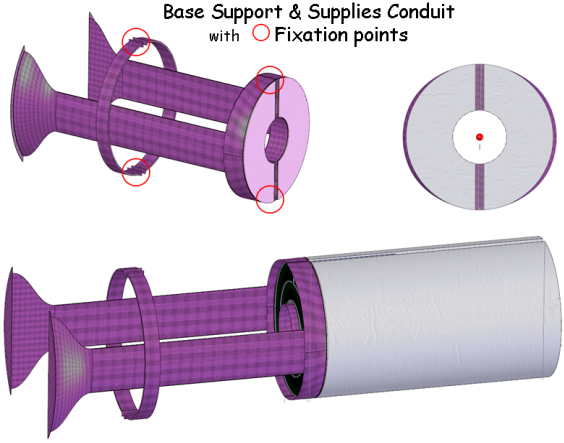

**Figure 2.2:** Conceptual Design of the $\overline{P}$ANDA GEM-TPC. The supplies-conduit and base support is made from one piece of carbon reinforced plastics (CRP) to assure stability and proper adjustment. The upper right part shows the front view visualizing the shape of the two independent drift volumes mounted on the same support. The lower part of the figure shows the overall assembly of the TPC.

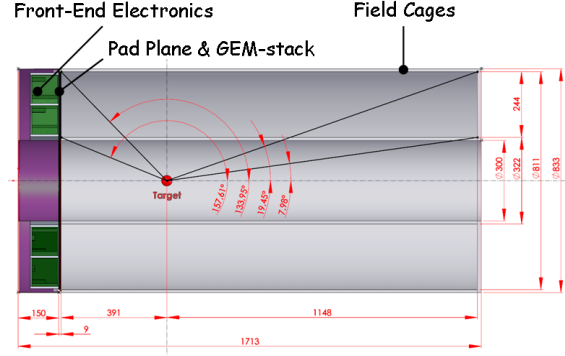

**Figure 2.3:** Schematic side view of the $\overline{P}$ANDA GEM-TPC describing the positioning inside the Target Spectrometer and the resulting acceptance in the (y,z)-plane.

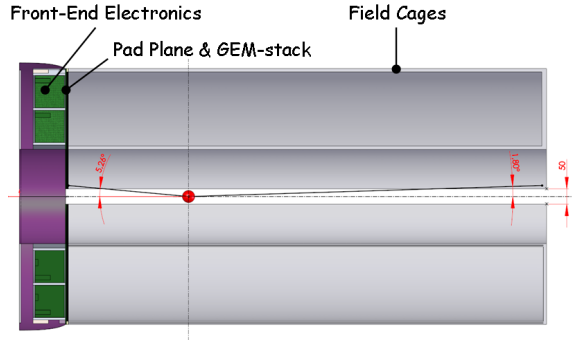

**Figure 2.4:** Schematic top view of the $\overline{P}$ANDA GEM-TPC describing the positioning inside the Target Spectrometer and the resulting acceptance in the (x,z)-plane.

end-cap. The fixation points are represented by the red circles in Fig. 2.2.

Due to the target tube traversing the TPC vertically, the cylindrical field cage of the GEM-TPC is split into two identical and independent volumes of half-circular cross-section. They are both flanged to a common support structure as shown in Fig. 2.2. The whole apparatus therefore falls into three mechanically separable parts:

- Two half cylinder-like individual vessels containing the detection gas, with an inner bore of 300 mm diameter, an outer diameter of 833 mm, and a length of 1554 mm. With a structural thickness of the barrel walls of 11 mm, and of the end-cap of 15 mm, the active gas volume extends from an inner diameter of 322 mm to an outer diameter of 811 mm, with a length of 1539 mm. A volume-to-surface ratio of the active drift volume of less than 0.1 m is obtained on the base of the geometry and the thicknesses of the walls of the detection volume (see Sec. 3.1).

- The base support comprising the structural elements like e.g. support, supplies, readout, and amplification stage; it has an overall length of about 1.9 m in beam direction including the readout electronics as well as the supplies and patch panels.

The overall length of the TPC will be 1704 mm without the support and conduit structures and 3454 mm including them.

We have foreseen a security shell of additional 4 mm free space enclosing all the elements shown in Fig. 2.2 which is not shown there. Furthermore, we are aiming for dimensional accuracies of about 0.1 mm in all parts to be manufactured.

Figs. 2.3 and 2.4 show a schematic side and top view of the active volume of the $\overline{P}$ANDA GEM-TPC, respectively. The angular acceptance of the detector with respect to the nominal position of the target point will range from 8° to 158° in polar angle, and from 5° to 175° and from 185° to 355° in azimuthal angle with a cone-shaped cut-out in the forward and backward direction of 1.8° and 5.3°, respectively. These numbers are also compiled in Tab. 2.1.

The rear view of the GEM-TPC in beam direction is shown in Fig. 2.5. The readout structure will be built out of one piece. This way, the routing necessary for the number of channels envisaged is best accommodated (see Sec. 4.4); the cooling structures and the overall mounting and adjustment can be simplified saving valuable space and material inside the magnetic field. The stiff containment will be made from carbon reinforced plastics (CRP) with a shape conformal to the volume available for the detector system, taking into account the space requirements of e.g. the backward Electromagnetic Calorimeter (EMC).

**Table 2.1:** Compilation of basic figures of the $\overline{\text{P}}$ANDA GEM-TPC. The numbers in brackets give effective values taking into account structural and functional thicknesses, e.g. of walls.

| Feature | Value |
|---|---|
| | (mm, kg, °) |
| Inner diameter | 300 (322) |
| Outer diameter | 833 (811) |
| Length of drift | 1554 (1539) |
| Length ($z$) of readout electronics | 150 |
| Length ($z$) of detector | 1704 |
| Length ($z$) overall | 3454 |
| Angular acceptance $\theta$ ($y, z$) | 8 - 158 |
| Angular acceptance $\phi$ ($x, y$) | 5 - 175 |
| Weight | 150 |

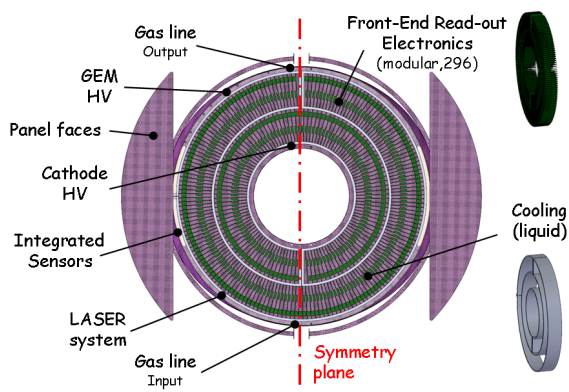

**Figure 2.5:** Conceptual design of the $\overline{\text{P}}$ANDA GEM-TPC. The cross section at the longitudinal position of the read-out electronics is shown in the left part of the picture. The upper right part shows some details of the radial circular arrangement of the front-end electronic cards (FEC). The lower right part of the figure shows a sketch of the cooling surfaces and guidance structures for the FECs.

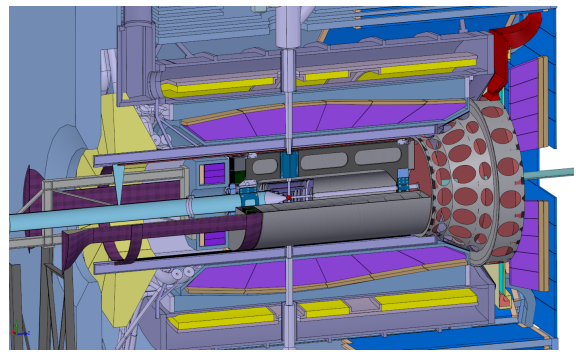

**Figure 2.6:** $\overline{\text{P}}$ANDA GEM-TPC mounted inside the Target Spectrometer. The upper-right part of the TPC detector has been cut away to allow view to the support and holding structure of the central tracker, the target-pipe crossing, the beam line tubes as well as the Micro-Vertex Detector (MVD).

Fig. 2.6 shows the $\overline{\text{P}}$ANDA GEM-TPC mounted inside the Target Spectrometer. The base support already comprises all the supply lines condensed into the stiff supplies conduit which ends in two identical terminals outside the magnet structures. This facilitates mounting and saves space inside the magnetic volume.

# 3 Field Cage

The field cage (FC) is one of the main components of the TPC and combines multiple functionality. Its main purpose is to define a homogeneous electrostatic field in the contained gas volume in order to ensure constant transport properties of the generated charges. The material composition and layer sequences are optimized with respect to the radiation length requirements. Great care has been taken in the choice of the insulator with respect to the surface resistivity to avoid charge build-up and breakdowns. The walls are gas tight and mechanically stable against changes in gas pressure and temperature. Several supply lines are incorporated in this structure, like the high voltage trace to the cathode end-cap, a gas distribution system and light guides for the laser infrastructure. These lines have been placed preferably at the edges of the vessels and are evenly distributed inside the inner sandwich layer. Since they are glued to the surrounding matrix, they also serve as mechanical stiffeners. The tunnels for the gas distribution system coincide with the SMD resistors soldered to the outer surface of the FC foils, providing a homogeneous gas distribution and cooling of the resistors at the same time. The vessel is fully electrically insulated from the high-voltage lines. It is shielded by an aluminum coating on its outside surface which is connected to detector-ground potential and which forms a Faraday cage to shield the detector from external electromagnetic signals.

## 3.1 Mechanical Structure

The layer stacking envisaged for the walls of the field cage is shown in Fig. 3.1. This baseline version may still be subject to further optimization. In order to minimize the losses of active volume, a part of the infrastructure has to be incorporated into the walls, such as e.g. laser light-guides, high-voltage cables, gas contribution system etc. Tables 3.1 and 3.2 list the materials used for the structures of the walls and the cathode end-cap.

## 3.2 Radiation Length

With the baseline design, the contribution to the radiation length is less than 1% for a single sidewall in the direction perpendicular to the surface and about 1% for the layers forming the cathode end-cap. This is roughly twice the value achieved for the large prototype GEM-TPC (see Sec. 12.2), which is due to the fact that the 60 kV operation voltage of the final P̄ANDA TPC is higher than the one for the prototype detector (30 kV).

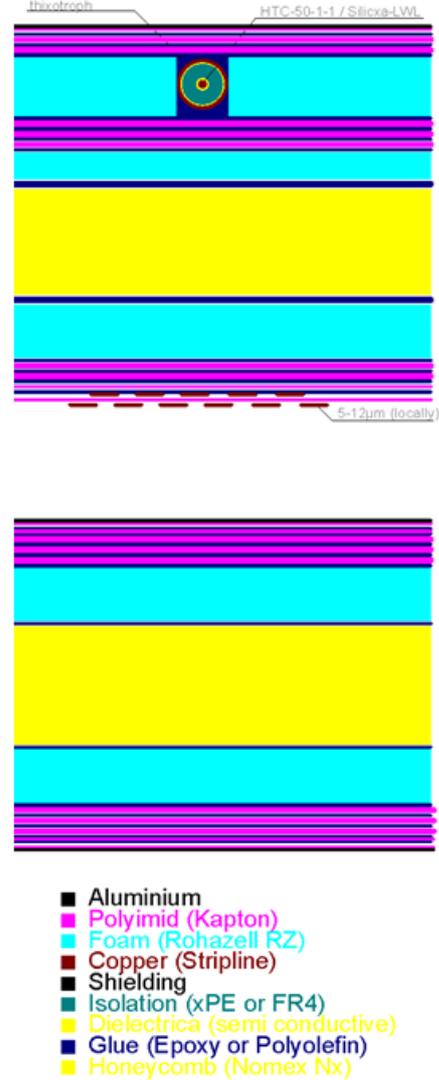

**Figure 3.1:** Cross section of the structural layout of the barrel walls (upper) and the cathode end-cap (lower). The layer stacking and the sequence of the materials used are shown not to scale.

The material budget has been evaluated within the PandaRoot analysis framework [23]. A detailed geometry of the GEM-TPC was created and the material traversed was evaluated along tracks with dif-

**Table 3.1:** Materials and corresponding thicknesses for the barrel walls. The items are ordered from the outside to the inside of the field cage.

| Material | Thickness | |
|---|---|---|
| | (µm) | (% $X_0$) |
| Aluminum | 0.2 | 0.00022 |
| Kapton | 275 | 0.09679 |
| Glue | 60 | 0.01209 |
| Rohacell | 2000 | 0.02587 |
| Glue | 80 | 0.01611 |
| Kapton | 375 | 0.13199 |
| Rohacell | 1000 | 0.01294 |
| Glue | 100 | 0.02014 |
| Nomex honeycomb | 5000 | 0.06762 |
| Glue | 50 | 0.01007 |
| Rohacell | 1000 | 0.02587 |
| Glue | 80 | 0.01611 |
| Kapton | 275 | 0.09679 |
| Copper | 12 | 0.08362 |
| Kapton | 50 | 0.01759 |
| Copper | 12 | 0.08362 |

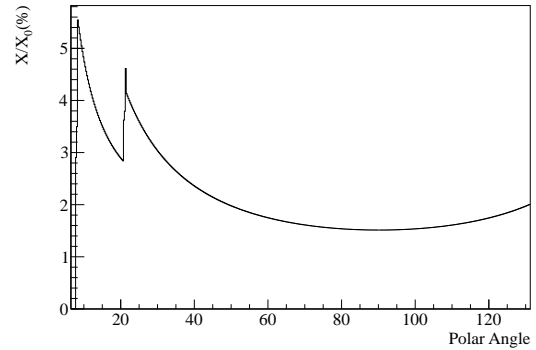

**Figure 3.2:** Material thickness in units of a radiation length of the GEM-TPC plotted versus the polar angle.

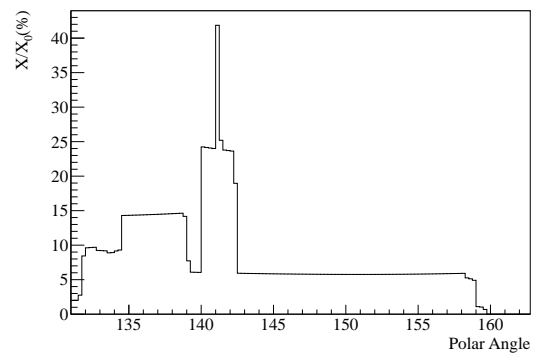

**Figure 3.3:** Material thickness in units of a radiation length of the GEM-TPC plotted versus the polar angle for the backward part of the chamber.

**Table 3.2:** Materials and corresponding thicknesses for the cathode end-cap. The items are ordered from the outside to the inside of the field cage.

| Material | Thickness | |
|---|---|---|
| | (µm) | (% $X_0$) |
| Aluminum | 0.2 | 0.00022 |
| Kapton | 400 | 0.14079 |
| Glue | 80 | 0.01611 |
| Rohacell | 2000 | 0.02587 |
| Glue | 50 | 0.01007 |
| Nomex honeycomb | 10000 | 0.13525 |
| Glue | 50 | 0.01007 |
| Rohacell | 2000 | 0.02587 |
| Glue | 100 | 0.02014 |
| Kapton | 425 | 0.14959 |
| Aluminum | 0.2 | 0.00022 |

## 3.3 Strip Foil and Voltage Divider

The innermost layer of the field cage vessel is composed of a polyimide (Kapton) foil coated with metal on both sides. The metal surfaces are patterned in two sets of parallel strips aligned transversely to the beam axis, and step-wise degrading the potential from the cathode voltage to the one close to the first GEM. In order to improve the homogeneity of the field close to the field cage walls, the two sets of strips on both sides of the foil are shifted by half a pitch between strips, a design that has proved to be superior to a single set of metallic strips with insulator in between. The thickness of the electrodes in the baseline design is 12 µm. We are investigating the possibility to adopt a minimum thickness of the copper layers of 2 µm, which is technically feasible, or switch to aluminum-based electrode structures; both measures would decrease the material budget with respect to the ferent emission angles. In this way, one obtains the polar angle dependency of the material budget. Figure 3.2 shows this dependency for angles up to about 130°, where the material at the backward end-cap (GEM foils and pad plane) is not traversed by the particles. The material budget for particles going through this part of the TPC is shown in Fig. 3.3.

values quoted in Table 3.1.

Due to limitations in the availability of the base material and imposed by the production machinery which limits the width of the foils to about 600 mm, the field cage has to be assembled from stitched foils. At least three sections of equal width alongside the beam axis will be realized. This technique was already successfully applied for the large prototype (cf. Sec. 12.1.1).

Calculations of the electrostatic field using finite element methods (FEM) have been employed to evaluate the field inhomogeneity for different strip-line layouts. The obtained results lead to the choice of 1 mm wide electrodes separated by a pitch of 1.5 mm. Figure 3.4 shows the resulting electric field. The outreach of noticeable field inhomogeneities into the drift field region is of the order of 10% of the structural sizes which translates into a distance of 1 mm from the inner surface. This holds true alongside the field cage surface except for the corner regions. There, a 10% variation of the field reaches approximately a distance of 10 mm into the active detector volume. These static deformations of the field bend the drift path of the ionization electrons and thus spoil the resolution. These effects can be calibrated out by recording cosmic tracks or by a laser calibration system (see e.g. Sec. 6.1).

Figure 3.5 shows an enlarged view of the strip-line foil implemented in the large prototype TPC which will also be employed for the $\overline{\text{P}}$ANDA TPC. The base-material is Kapton with a thickness of 50 μm, the electrodes are made of copper forming layers of 2 × 25 μm thickness including the gold-plating and the cover-lay.

In order to achieve a sufficient uniformity of the field, the variations of the inter-strip resistivity should be of the order of 0.1 %. Unfortunately, such high-precision resistors are not available in the MΩ regime and for the small required dimensions (SMD 0805 or even SMD 0603). A lower-precision specimen was chosen (typically 1 %) and the resistors were grouped in parallel networks composing split chains with paired tolerances.

Figure 3.7 shows the sketch of the resistor network to be realized for the $\overline{\text{P}}$ANDA GEM-TPC.

We foresee to make use of the same resistor types (SMD 0805) and individual values (4.2 MΩ) that have been already employed for the large prototype TPC. This results in a total resistance of the strip-line field cage of ≈ 2.15 GΩ. Given a drift field of 400 V/cm and thus applying a drift voltage of 62 kV to the cathode, a current of approximately 30 μA flows through the resistor network.

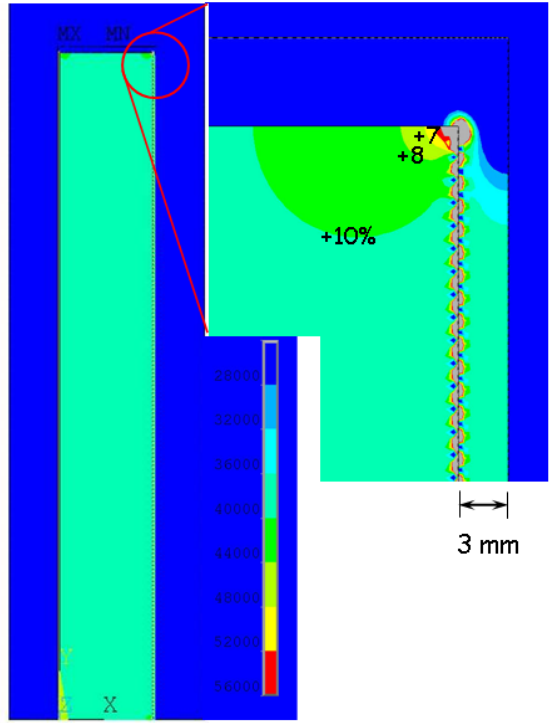

**Figure 3.4:** Results from a 2-D simulation of the electric field inside the drift volume and at its borders. The color code displays the field inhomogeneities. The width of the strips staggered at the inner and outer surface is 1 mm, the pitch is 1.5 mm. Most of the inhomogeneities occur within 1 mm from the innermost surfaces except for the corners where a 10% variation is reached at a distance of 10 mm from the surface.

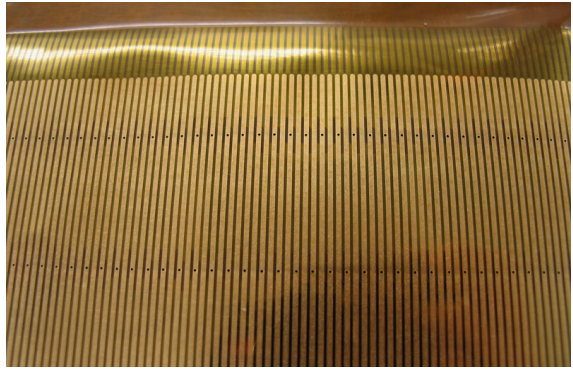

**Figure 3.5:** Photograph of the strip-line field-defining electrodes. The surface facing the drift volume is shown. The strips of copper metal of 2 μm thickness have a width of 1 mm and a gap of 0.5 mm. The holes show the location of the bores used for the gas-distribution system integrated into the walls of the gas-tight vessel of the field cages. On the upper part the picture one clearly recognized the staggered structure of the strip-line electrodes between the front and backside of the foil.

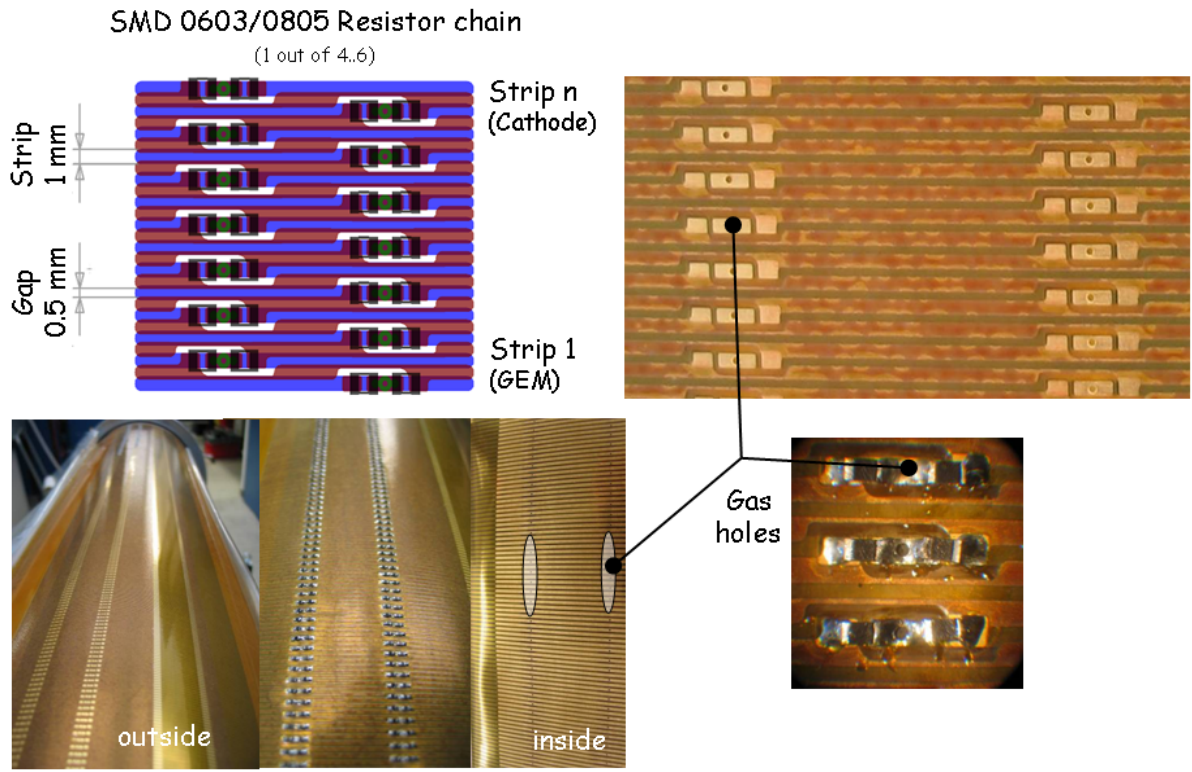

**Figure 3.6:** Photograph of the strip-line field-defining electrodes of the large prototype GEM-TPC with SMD 0805 resistors mounted on its surface. The layer oriented towards the outer circumference of the field cage vessels shown except for the part labeled 'inside' which shows the lines of holes of the gas distribution system integrated into the vessel-wall structure. The upper left part gives the sketch on details of the principle layout, the upper right part the realization. The lower right shows a photograph of an enlarged view on the SMD 0805 resistors soldered to their respective pads.

## 3.4 Temperature Stabilization

The heat dissipation produced by the persistent currents flowing through the resistor chains should be minimized. Direct cooling of the resistors is not favorable because it would require a metallic or ceramic contact introducing a quite large amount of material. Moreover, the electrical insulation would be quite challenging. To cope with this problem we will install multiple rows of resistors of the same value in parallel connection on the surface of the strip-line foil spreading the heat generated.

In order to monitor the temperature gradient inside the field-cage vessel, SMD-mounted 1-wire sensors will be integrated inside the walls. This allows for respective measurements with a precision of $\approx 0.2\,°C$. This concept has been proven helpful and reliable in the framework of the tests done with the large prototype GEM-TPC detector (see Fig. 12.9).

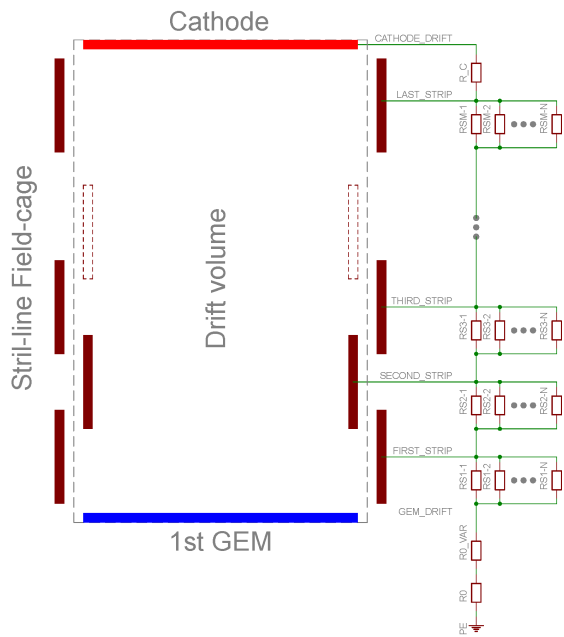

**Figure 3.7:** Sketch of the resistor network to be realized for the $\overline{\text{P}}$ANDA GEM-TPC. The field-defining electrodes on the strip-line foils as well as the cathode and the surface of the GEM-stack facing the Drift volume are visualized to the left. The right part shows the scheme of the various resistors to be installed. Please note that the GEM-stack itself is supplied by a separate circuit (see Fig. 4.5).

# 4 GEM Stage

## 4.1 Mechanical Layout

The mechanical layout of the GEM stage follows the general design rules as presented in Sec. 2.1. Especially with respect to the acceptance of the detector, we maximize the active volume and avoid the introduction of zones with reduced detection efficiency. The GEM foils will be pre-stretched and glued to frames made of glass-fiber reinforced plastics (GRP). The width of the frames is chosen to be 11 mm to match the thickness of the drift volume walls. The total thickness of the frames for one GEM layer is constrained by the distance between the layers in the stack of 1 mm. The so-dimensioned frames can sustain the in-plane stretching forces of 18 N/cm.

Mechanical FEM simulations have been carried out to determine the deformation of the GEM frames. Figure 4.1 shows the results obtained assuming a 1 mm thick frame made from glass-fiber reinforced plastics (GRP) with isotropic E-modulus. A huge deformation of more than 10 mm is expected if the the original half-circular 'kidney' shaped design is assumed, as proposed in [24]. The GEM foil stretched with circumferential forces of 1 N/mm is assumed to be sandwiched between two frames of equal thickness of 0.5 mm in a way that the resulting forces in the direction perpendicular to the foils surface cancel out.

The situation changes drastically by adopting a design symmetric in the plane perpendicular to the beam axis as shown in Fig. 4.2. The respective simulations exhibit a reduction of the deformations to values around 1 mm horizontally and even below 1 mm vertically. These deformations are acceptable and can be treated assuming a proper suspension of the single framed GEM-foils.

This design will allow a modular GEM section which is favorable for maintenance purposes and in terms of the risk assessment.

The size of the TPC would allow for the installation of GEM foils with a maximum diameter of 900 mm, which, however, is above the limits of today's technologies. The obvious solution is to follow the symmetries of the detector and cut the foils in half along the vertical direction ending up with $900 \times 450 \ \text{mm}^2$ foil sizes. This is well in accordance with what has already been achieved at the CERN production site in recent developments for other ex-periments [20, 21].

| Maximum deformation (mm) | | | |
|---|---|---|---|
| $U_x$ | $U_y$ | $U_z$ | $U_{res}$ |
| -700 | ± 220 | ± 8 | 733 |

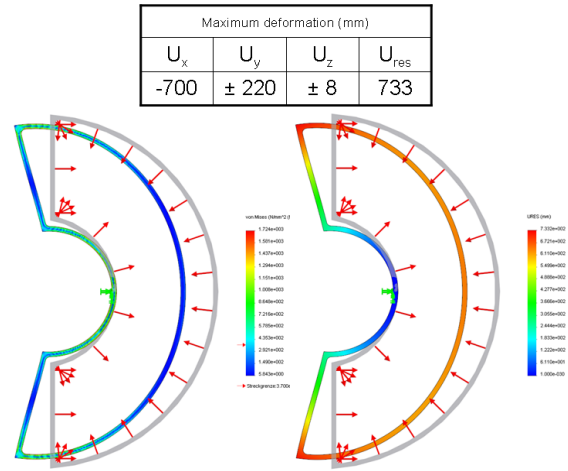

**Figure 4.1:** FEM simulation of the shape deformation of the GEM carrier frames made of glass-fiber reinforced plastic with a half-circular shape. The maximum bending exceeds several centimeters. The off-plane deformation emerging from the asymmetric boundary conditions applied is irrelevant because it will level out using double framing keeping the overall thickness of the set.

| Maximum deformation (mm) | | | |
|---|---|---|---|
| $U_x$ | $U_y$ | $U_z$ | $U_{res}$ |
| ±1,0 | ± 0,16 | ± 0,05 | 1,3 |

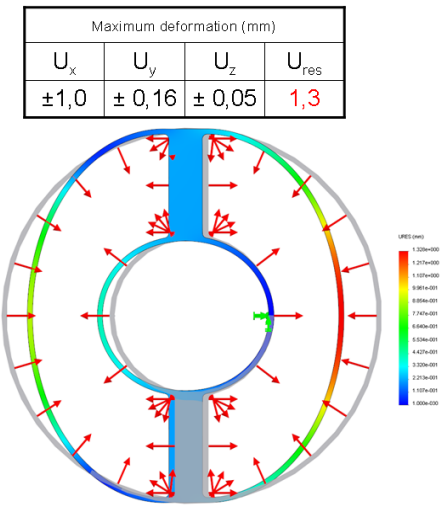

**Figure 4.2:** FEM simulation of the shape deformation of the GEM carrier frames composed of glass-fiber reinforced plastic in case of symmetrical layout. The maximum bending is kept at a bearable level of 1.5 mm in $x$-direction.

The technology applied for the production of these large-size foils is different from the standard GEM production technique used for most of the exist-ing GEM detectors, including the large prototype GEM-TPC described in Sec. 12.1. The conventional method of GEM foil patterning requires photolithographic processes based on two masks with identical hole patterns, placed on the two sides of the copper-coated base foil, and aligned with a precision of 1 μm with respect to each other in order for the holes to be perpendicular to the surface. The maximum size of the mask sets one of the limitations to the process due to the extreme requirements on their relative positioning of 1 μm over ∼ 1 m. Another important constraint is the size of the industrially available base material and the machinery required for the processing, both being basically limited to a width of 600 mm.

These limitations can be bypassed by employing a single-mask technique [25]. Indeed, this technique has been shown to deliver similarly good results with respect to homogeneity and gain performance of the GEM-foils Areas of $350 \times 700 \, \text{mm}^2$ for the single mask GEM have been achieved in the framework of R&D for the KLOE-2 by the RD51 collaboration [26]. A small decrease in gain by 25 % has been observed in comparison with a standard GEM at the same operation conditions, which can easily be compensated for by a slight increase of the operation voltage by 20 V. No discharges were observed up to a gain of $4 \cdot 10^4$. Similar results were obtained in the framework of developments for the CMS muon system with foils of an active area of $990 \times (220\text{-}455) \text{mm}^2$ [21]. This already corresponds to the size needed to cover the active area of each of the two half-cylinders of the $\overline{\text{P}}$ANDA TPC. The gap of 50 mm between the two halve cylinders leaves sufficient space to stitch together two single foils without introducing any inefficient regions.

## 4.2 Electrical layout

Extensive studies have shown that reducing the capacitance between the two metal surfaces of a GEM foil significantly reduces the probability of discharges in a GEM detector [12]. To this end, one side of the GEM foil is segmented into individually powered sectors with a surface of ∼ 100 cm². This restriction limits the amount of charge which is involved in case of occasional sparks. The inter-sector distances are reduced down to 100 μm. As a result, approximately 36 sectors are required to cover the active cross-section of the GEM-TPC drift volume.

The optimization of the structuring shows that sectioning the surface into 1/4 - circular rings or vertical strips of equal area as shown in Fig. 4.3, re-spectively, allows to make electrical contact in the middle bar. In this way, also the contact problem for stacked foils is minimized.

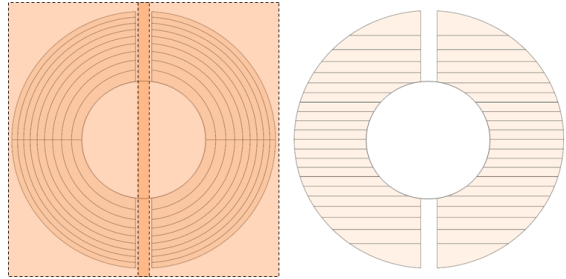

**Figure 4.3:** Sectioning of the GEM-foils. According to the design recommendations each sector has a surface of ≈ 100 cm². Two different schemes of adjacent foils in a GEM stack are proposed which avoid the generation of potentially less efficient zones due to an overlap of the section borders. The backside electrode remains unstructured in both cases. We will either use one single foil large enough to cover the whole surface or stitch two foils together (see left) at the vertical symmetry line.

In contrast to the design chosen for the large prototype, where the space limitations were not as stringent as for the final TPC and contact to the high voltage (HV) supplies is done via flaps to a common distribution structure embracing the GEM frame, we choose to contact via the middle section of the GEM frame here.

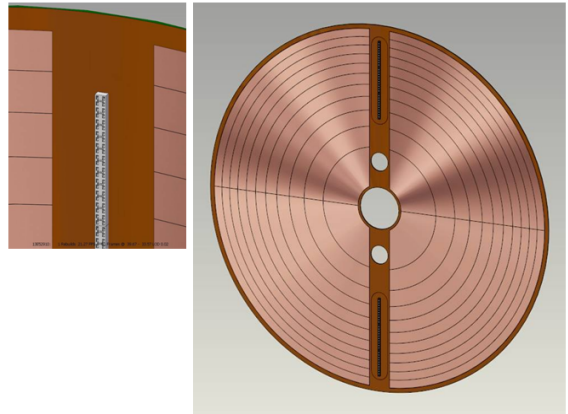

**Figure 4.4:** Framing and contact scheme of the GEM sectors shown for the circular design.

The potentials on the GEM electrodes are defined by individual external HV supplies. Figure 4.5 shows the schematics of the HV supply scheme.

The potential on the sectored side of the GEM foil, which is facing the drift electrode, is defined through large 100 MΩ loading resistors. The potential on the unsectorized side, facing the readout plane, is supplied directly, and has a high-ohmic connection to ground. This scheme has the following features:

- In case of a discharge between the two sides of a GEM foil, a voltage drop occurs only on the top side, whereas the bottom side remains at its nominal potential. This prevents the propagation of the discharge to the next GEM foil or to the readout circuit.

- In case of a temporary or permanent short across a GEM foil in one or several sectors, the remaining sectors still remain fully operational. The resistor to ground for the bottom side of the GEM foil is necessary in order to avoid reverse currents into the HV supply.

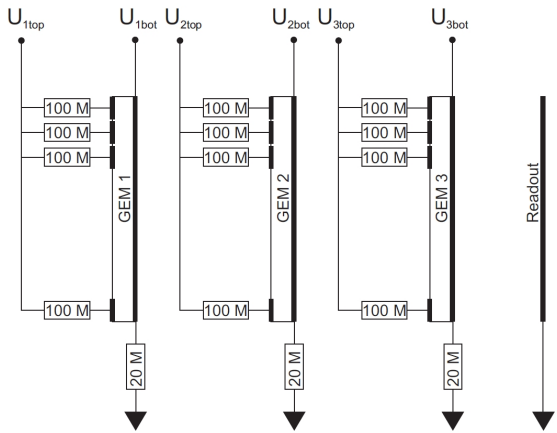

**Figure 4.5:** Schematics of the HV distribution scheme for the GEM detectors.

| Drift Field | 0.4 kV/cm |
|---|---|
| $\Delta U_{GEM1}$ | 244 V |
| Transfer Field 1 | 4.5 kV/cm |
| $\Delta U_{GEM2}$ | 278 V |
| Transfer Field 2 | 0.16 kV/cm |
| $\Delta U_{GEM3}$ | 333 V |
| Collection/Induction Field | 5.0 kV/cm |

**Table 4.1:** High voltage settings for a minimal ion backflow at an effective gain of about $2 \cdot 10^3$.

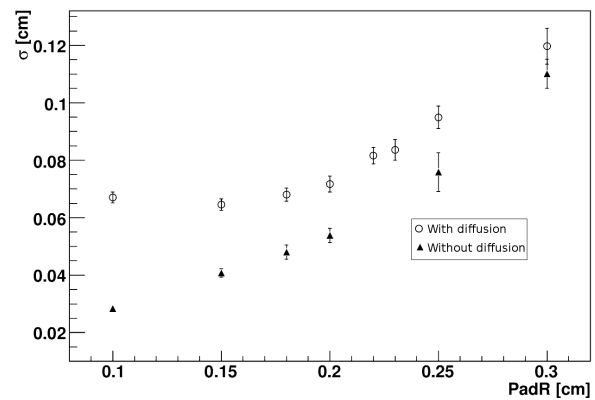

**Figure 4.6:** Cluster residual width along $x$ versus pad outer radius, with diffusion in a 2 T solenoid field (open circles) and without diffusion (full triangles).

## 4.3 Ion Backflow

In order to minimize the backflow of ions produced in the amplification region, the electric field configuration of the GEM stack as well as the sharing of the gain among the three amplification stages has been optimized. Measurements on small test detectors, reported in Sec. 11.2, indicate a minimum ion backflow at the voltage settings shown in table 4.1, which correspond to an effective gain of about 2000 in the GEM stack for the envisaged $Ne/CO_2$ (90/10) mixture. With an ion backflow of 0.25%, this corresponds to four back drifting ions per electron reaching the GEM stack.

A further reduction of the ion backflow to the level of $10^{-5}$ can be achieved using additional pattering on one side of a GEM foil, as will be discussed in

Sec. 13.5. This very promising strategy will be investigated using the large prototype. If successful, it essentially eliminates the problem of ion backflow for the TPC.

## 4.4 Pad Plane

The electrons emerging from the last GEM stage induce a fast signal on the readout pads. In order to achieve a more uniform distribution of the signal on neighboring pads, a hexagonal symmetry is foreseen. In this way, the distance to all neighboring pads is constant, in contrast to non-uniform distances for a rectangular pattern.

The optimum pad size was determined by Monte Carlo simulations of pions with $0.5\,\mathrm{GeV}/c$ momentum traversing the $\overline{\mathrm{P}}$ANDA TPC at a polar angle of $40°$ in a magnetic field of 2 T. Figure 4.6 shows the r.m.s. width of the residual distribution of clusters along $x$, i.e. perpendicular to the drift direction, as a function of the circumscribed circle radius of the hexagonal pads, both with and without diffusion in the drift region. Below a pad outer radius of 1.5 mm, the diffusion starts to dominate the resolu-tion, while the pad size dominates for larger values. A pad outer radius of 1.5 mm was chosen accordingly. It should be noted here that these simulations have been performed using slightly different parameters for the gas mixture[1]. Therefore, the absolute value of the r.m.s. width is to be taken with a grain of salt, but the conclusions towards the saturation remain valid. The total number of pads in the active area of the two halves of the TPC is about 80000. As for the large prototype (cf. Sec. 12), it is foreseen to measure the temperature at various positions on the pad plane inside the gas volume.

Since the pad plane also encloses the gas volume of the detector at the backward end-cap, it is designed as a multi-layer PCB, thus avoiding through-going vias for the signal lines. Signals from different pads are routed to the connectors in way to minimize cross talk. The capacitance of the pads and the signal lines corresponds to 10 pF. The signals are read out by 320 front-end cards, mounted perpendicularly to the readout plane to minimize heat transfer to the readout plane, and each reading 256 electronic channels.

---

1. for the large prototype TPC at a field of 2.5 T

# 5 Gas System

## 5.1 Key Issues

For the design and operation of the gas system the following points are important and will be further discussed in the following sections.

- As chamber gas a $Ne/CO_2$ mixture (with a mole ratio 90:10) will be used.

- The TPC volume amounts to $0.7\,m^3$.

- A gas exchange rate to replace the TPC volume once every two hours is foreseen, the average gas flow amounts to $0.15\,m^3/h$.

- A closed cycle gas system with a purification unit is needed.

- $O_2$ and $H_2O$ contamination will be controlled and monitored to be kept below a level of 5 ppm.

- The chamber pressure will be controlled to a value of $5\pm1$ mbar above atmospheric pressure, to avoid a distortion of the chamber wall.

## 5.2 Choice of Gas Mixture

The choice of detector gas is crucial to the design of the TPC [27]. It not only influences the performance of the detector, but has impact on the design and electrical properties of the field cage, the amplification region, and the readout electronics. The requirements to the gas mixture for a TPC are manifold, and partly contradictory:

- high electron drift velocity: fast clearing of the gas volume,

- low electron diffusion: spatial resolution,

- constant gain: variations of $T$, $p$, material,

- low electron attachment,

- high ion mobility: accumulation of space charge controllable,

- specific ionization: $dE/dx$ vs. space charge,

- low density, high radiation length: multiple scattering,

- no aging.

**Table 5.1:** Properties of gases commonly used in TPCs at normal temperature and pressure (20°C, 1013.25 mbar). Density $\rho$, radiation length $X_0$, total number of electron-ion pairs for MIPs $N_T$.

| Gas | $\rho$ [g/l] | $X_0$ [g/cm²] | $X_0$ [m] | $N_T$ [1/cm] |
|---|---|---|---|---|
| He | 0.1785 | 94.32 | 5280 | 8 |
| Ne | 0.89990 | 28.94 | 322 | 40 |
| Ar | 1.784 | 19.55 | 110 | 97 |
| $CH_4$ | 0.717 | 46.22 | 645 | 54 |
| $CO_2$ | 1.977 | 36.2 | 183 | 100 |
| $C_2H_6$ | 1.356 | 45.47 | 335 | 112 |
| $CF_4$ | 3.93 | 36 | 90 | 120 |

Only the noble gases Argon and Neon qualify as the main component. Helium is difficult to contain and hence has a high leak rate, Krypton and Xenon are rare and hence not affordable in larger quantities. In addition, their high density leads to significant multiple scattering. Due to its lower abundance, Neon is about a factor of 8 more expensive than Argon, making a closed gas system indispensable. Table 5.1 summarizes the important parameters of some noble gases and common admixtures.

The radiation length is about a factor of three larger for Ne than for Ar, leading to reduced photon conversion and multiple scattering in Ne. The primary ionization rate is about a factor of two smaller for Ne ($n_{mp} = 16$) than for Ar ($n_{mp} = 38$), requiring a higher gain in the case of Ne. On the other hand, space charge effects will be significantly lower for Ne than for Ar, supplemented by mobility of Ne ions which is a factor of 2-3 larger than that for Ar ($\mu_{Ne} \sim 4\,cm^2/Vs$ for $E < 10^4\,V/cm$). Following these arguments, the preferred main gas component for the PANDA TPC will be Ne.

As quenchers, either organic gases such as Methane ($CH_4$), Ethane ($C_2H_6$), Isobutane ($iC_4H_{10}$), freons such as $CF_4$, or $CO_2$ are most widely used. $CH_4$, used in e.g. P10 ($Ar/CH_4$ 90/10) is a very convenient quencher, since it has a maximum drift velocity of $5\,cm/\mu s$ at an electric field of only $125\,V/cm$, greatly simplifying the design of the field cage. A strong longitudinal magnetic field is necessary, however, in order to reduce transverse diffusion. The drawback of organic mixtures is that they may cause aging by forming polymeric deposits on the electrodes of the chamber when the accumulated charge is large. CF$_4$ is in principle interesting as quencher, since it provides a high drift velocity at low electric fields in mixtures with Ar or Ne. In addition, it is sometimes used in attempts to etch polymer deposits from electrodes. Its high reactivity, however, makes its use very dangerous to other components of the gas system, like glass joints, or Aluminum parts. In addition, fluorine is electronegative and attaches electrons. CO$_2$ remains an uncritical quencher, with the advantage of low diffusion, and the disadvantage of lower drift velocities. Hence, CO$_2$ will be used as quencher for the $\overline{\text{P}}$ANDA TPC.

**Table 5.2:** Gas parameters for the $\overline{\text{P}}$ANDA TPC at a magnetic solenoid field of 2 T.

| | |
|---|---|
| Gas mixture | Ne/CO$_2$ (90/10) |
| Drift field | 400 V/cm |
| Electron drift velocity | 2.731 cm/µs |
| Transverse diffusion | 130 µm/$\sqrt{\text{cm}}$ |
| Longitudinal diffusion | 230 µm/$\sqrt{\text{cm}}$ |
| Ne$^+$ ion drift velocity in Ne | 1.767 cm/ms |

The mixture ratio is Ne/CO$_2$ 90/10 (by weight). This mixture, which is also used in the NA49 and ALICE TPCs, features attractive charge transport properties at reasonable drift fields, and is non-flammable. Figure 5.1 shows the drift velocity, Fig. 5.2 the transverse and longitudinal diffusion constant of Ne/CO$_2$ (90/10) in a 2 T longitudinal magnetic field as a function of the electric drift field, calculated with MAGBOLTZ [28, 29].

Table 5.2 summarizes important parameters of this mixture for the $\overline{\text{P}}$ANDA TPC.

## 5.3 Operational Requirements

One of the main requirements to the gas system, including the detector vessel itself, is the limitation of oxygen and water vapor concentration. In the case of Ne/CO$_2$ mixtures for example the electron attachment is of the order of 1% per ppm oxygen concentration and per meter drift length. This has severe implications on the particle identification power, because it degrades the d$E$/d$x$-measurement. The contamination mainly originates from bad sealing joints and/or permeation of water and oxygen through the TPC wall material. The volume-to-surface ratio of the $\overline{\text{P}}$ANDA TPC is only about 0.1 m, compared to existing large TPCs, such the ALICE TPC (0.70 m), where an oxygen contamination lower than 5 ppm was achieved. This value is therefore regarded as the upper limit for the $\overline{\text{P}}$ANDA TPC.

In addition to the gas purity, the accuracy and stability of the Ne/CO$_2$ gas mixture is crucial, as it determines important parameters like the drift velocity or diffusion within the gas.

## 5.4 Design and Layout

The large detector volume of about 0.7 m$^3$ as well as the usage of a high-cost gas mixture leads to a closed circulation gas system as the natural choice. Figure 5.3 presents the basic layout of such a system. It is built up of the gas supply unit, the purification system and the gas analyzing devices (oxygen and water vapor monitors and a residual gas analyzer), which will be located outside of the experimental area in the gas supply building. For simplicity the TPC is drawn as one gas volume instead of two. While full lines mark the gas pipelines, the dashed lines correspond to control connections. For normal running the gas will be recycled through a purifier and go back to the main loop with a moderate 'fresh' gas injection of only a few percent. Therefore the return gas from the detector has to be compressed and pumped back to the gas supply building. The detector is planned to be operated constantly at 5 mbar over atmospheric pressure. This is achieved by an independent control loop consisting of a pressure transducer, an adjustable electro-magnetic valve and a stand-alone control unit, which is installed at the output line of the TPC. The transducer measures the pressure close to the TPC volume, submits the value to the control unit which then regulates the valve. In addition, to check the resulting pressure regulation, pressure measurements are done permanently at three different positions by transducers. In addition, the total gas flow is controlled by a flow meter and adjusted by an electro-magnetic valve. The main TPC gas monitor system collects all this information and also checks the gas quality at different points within the circulation system. Like the stand-alone pressure control system, it should be easily accessible inside a general control area close to the experiment.

### 5.4.1 Purification Unit

Closed loop gas circulation systems require gas purification in order to achieve high regeneration rates of the "old" gas and low requirement of "fresh" gas. The main impurities of concern, which ac-cumulate in the system, are oxygen and water vapor. These pollutants infiltrate the system through the detector sealing and due to the under-pressure at the compressor entrance. A set of two purifier cartridges (see Fig.5.4) are filled with two cleaning agents: molecular sieve (3 Å) to remove the water vapor and activated copper to remove the oxygen content in the gas stream. The advantage of having two parallel cylinders in each cleaning stage is to run the gas mixture through one, while the other one is regenerated. Both agents in the same cylinder can be regenerated at the same time by heating the columns to 220 °C while purged with a Ne - H$_2$ gas mixture with a mixing ratio of 9:1. The operating cycle to clean the purifiers will be controlled by the slow control unit, which will follow exactly the same protocol each time.

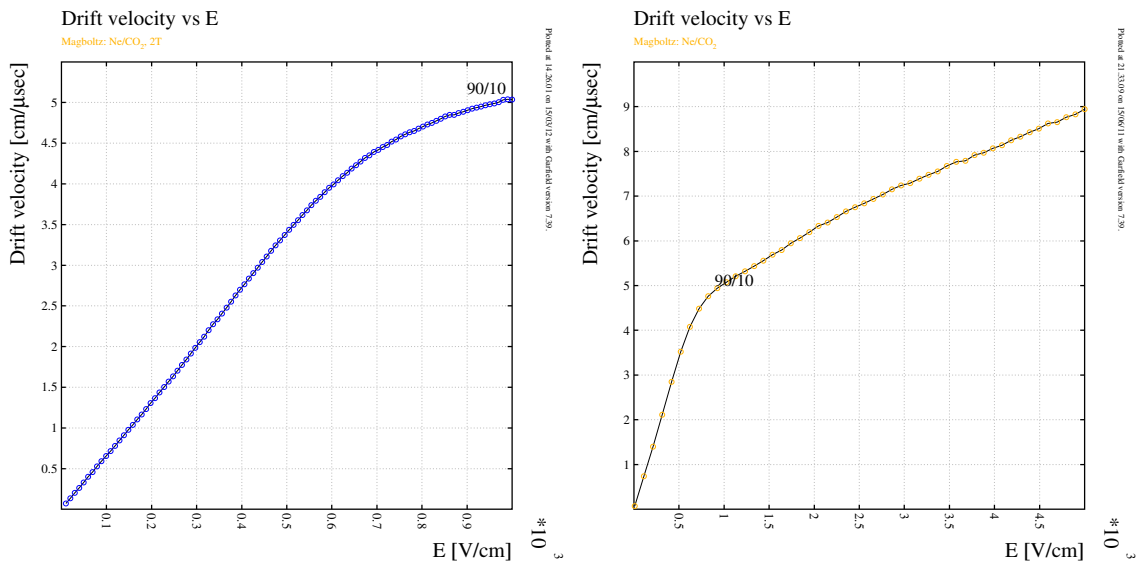

**Figure 5.1:** Drift velocity as a function of the electric drift field for a Ne/CO$_2$ (90/10) gas mixture. Low-field drift region (left) and region extended to fields in the GEM stack (right).

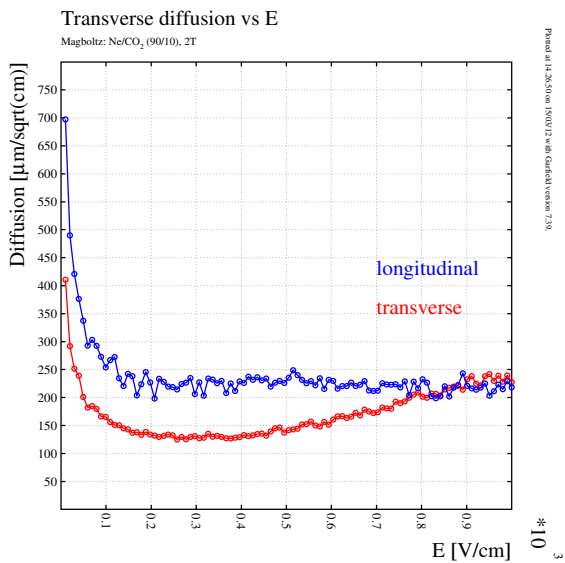

**Figure 5.2:** Transverse (red) and longitudinal (blue) diffusion as a function of the electric drift field for a Ne/CO$_2$ (90/10) mixture.

### 5.4.2 Gas Mixing Unit

The design of the gas mixing unit is shown in Fig.5.5. It will mix the gas components in the desired proportions with the help of mass flow controllers. The device will have the possibility to run in two different modes: the "Fill"-mode with higher gas flow rates up to 350 l/h and the continuous "Run" mode with a lower rate of about 10 l/h, corresponding to the injection rate of fresh gas during normal operation. Additionally a separated purging line using N$_2$ will be installed. The contents of the gas mixture components are continuously monitored with a mass spectrometer (RGA) and the process parameters are recorded by the slow control system.

### 5.4.3 Emergency Devices

Very close to the TPC an emergency system has to be installed (at the input line of the system to avoid under pressure and at the output line and hence over pressure in the TPC). If for any reason the main system stops working (e.g. because of a power cut), this backup system keeps the pressure inside the TPC constant.

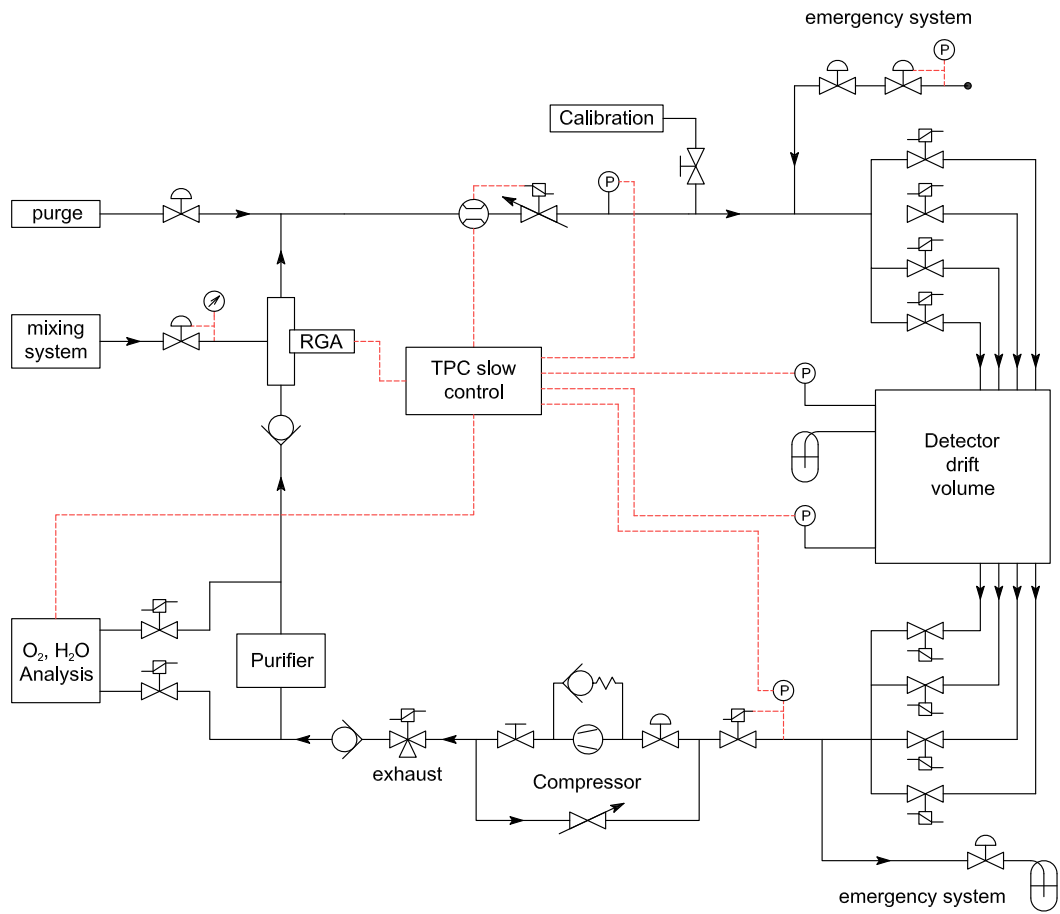

**Figure 5.3:** Basic layout of the cycle TPC gas system.

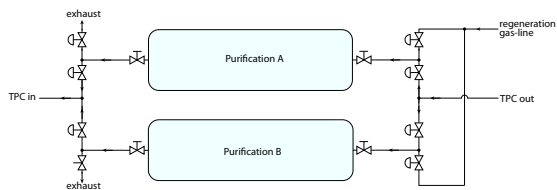

**Figure 5.4:** Gas purification twin-unit for the removal of oxygen and water vapor contamination in the TPC gas mixture.

In case of over pressure gas is relieved through a security bubbler which is adjusted to 5 mbar over atmospheric pressure. In case of under pressure pre-mixed Ne/CO$_2$ is injected from a bottle.

## 5.4.4 Filling, Purging and Calibration

Filling of the detector with operating gas will be done without recirculation and with a high inlet gas flow, compared to normal running conditions, from the mixing unit (see Fig. 5.3). For this a three way valve, positioned before the purifier input, can be set to "exhaust". At the same time the second set of flow meters ("fill" condition, see Fig. 5.5) within the mixing unit is used to increase the flow. By this a gas exchange of at least one whole TPC volume should be possible every two hours.

For purging the TPC, a separate input gas line is connected. From there for example N$_2$ is injected into the gas system, flushed through the detector and exhausted through a bubbler. In case of a longer shutdown one can fill the two TPC vessels with N$_2$ and then use the purge system to keep them at a constant pressure.

For gain calibrations of the TPC and its readout channels a radioactive noble gas isotope could be admixed to the operating gas. A bypass arrangement will serve to implement this feature.

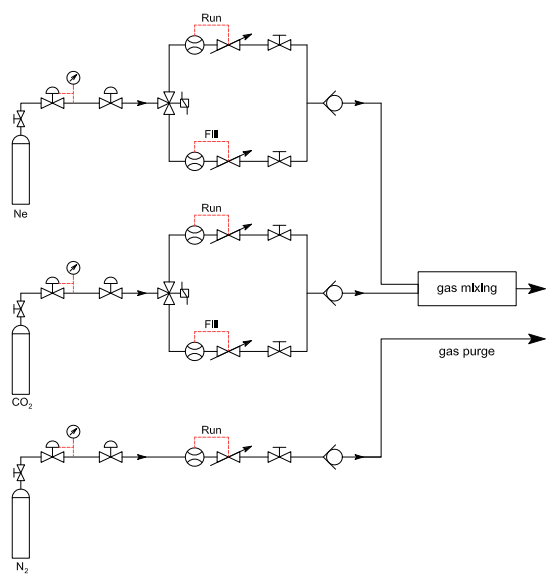

**Figure 5.5:** The gas mixing system with the possibility to switch between filling and continuous running application.

# 6 Calibration

## 6.1 Laser System

The drift velocity of charge carriers is a key parameter in the operation of a gaseous TPC. The knowledge of its absolute value within a relative precision of $5 \cdot 10^{-4}$ is crucial for a trigger-less system. Variations of the drift velocity can be introduced by imperfections and non-uniformities in the gas, temperature and the electric and magnetic fields.

Another source of deviations from the ideal homogeneous drift field are space charge effects arising from the ion backflow to the cathode. This can cause distortions in the drift path of the electrons (as described in Section 11.2) traveling to the anode. The effect is severe as its influence may vary with the beam intensity and time. It is also subject to changes in the operation conditions of the detector, especially the settings of the GEM-voltages. There is also a non-negligible static contribution originating from mechanical and electrical imperfections of the field cage itself (see Sec. 3).

All these effects need to be corrected in order to achieve good spatial and momentum resolution. The goal is to obtain a uniformity within a relative error of about $10^{-4}$ corresponding to a point resolution of $150\,\mu m$ and a z-resolution of $200\,\mu m$ for a total drift length of $1.5\,m$. A calibration system has to fulfill all of these requirements. It has to be usable for calibration of intrinsic non-uniformities as well as time dependent non-uniformities. Due to the continuous operation of the $\overline{P}$ANDA experiment, those calibrations will be performed in special calibration runs without beam and if possible during the beam operation.

There are various methods possible to achieve information on both quantities, drift velocity and field distortions, with different grades of feasibility and date of application.

- Reconstruction of cosmic tracks. Any deviation from a straight line of the reconstructed path may be attributed to local fields. Obviously this method is applicable only without the presence of a magnetic field limiting it to times without beam on target. The particles may probe the whole active volume basically in vertical direction. This method is very common and applied to all experiments of today.

- Generation of artificial tracks perpendicular to the electrical field lines in the drift volume probing the active volume in full 3-D. One method to obtain such a calibration is by a number of narrow ultraviolet rays at well known predefined positions.

- Generation of point- and/or strip-like 2-D patterns at the cathode end-cap surface. This method allows probing the active volume in an integral way.

For the last two methods a laser system [30] will be the best choice. Laser systems have been used in many TPCs as an essential tool for calibration. The main tasks of a laser calibration in a TPC are:

- determination of the drift velocity;

- alignment of different modules and sectors to each other;

- correction of spatial distortions.

Ideally, several of the above-mentioned calibration methods should be used in parallel for reasons of redundancy and systematic error minimization. For instance the drift velocity can be determined with reconstructed cosmic muons and the laser system. The combination of these two methods will help to reduce the systematic uncertainties and therefore allow a much higher precision on this critical parameter. As the drift velocity will not be in a plateau for the $\overline{P}$ANDA TPC, the redundancy in the calibration methods will be of highest importance.

In the following, two possible laser calibration systems will be discussed. The 3-D laser system, on the one hand, would be the preferable choice as it would offer the maximum amount of information needed to perfectly calibrate the TPC. It will, however, require a high level of complexity in the mechanical implementation. A system like employed for the T2K TPC, on the other hand, requires less complexity, but might provide less information than the 3-D system. Therefore it can be seen as a potential fallback solution.

### 6.1.1 3-D Laser Calibration

Numerous systems based on the reconstruction of laser-generated tracks have been implemented and successfully operated in the past years, e.g. for the STAR TPC [31, 32], and for the ALICE TPC [9]. It is the aim of such a setup to measure several hundred simultaneously generated laser tracks throughout the TPC drift volume.

While the molecules of the $Ne/CO_2$ gas mixture have ionization potentials above $10\,eV$, organic impurities have values in the range of $5–8\,eV$. A monochromatic laser beam with a wavelength of $266\,nm$ ($E = 4.66\,eV$) will ionize the gas impurities by a two-photon absorption process, which proceeds via a virtual state with a very short lifetime of $\tau \approx 10^{-16}\,s$. The number of electrons $n_e$ created per unit length of the laser track varies quadratically with the total number of photons $N_\gamma$: given by Eq. 6.1 [33]:

$$n_e = \frac{\sigma^2 n_0 N_\gamma^2 \tau}{s T} \qquad (6.1)$$

Here, $\sigma_1 = \sigma_2 = \sigma \approx 10^{-16} - 10^{-17}\,cm^2$ are the cross sections for the excitation of the virtual state and the ionization from the virtual state, respectively, $n_0$ is the number of impurity molecules per $cm^3$, $T$ the laser pulse width and $s$ its area. Energy densities of about $20\,\mu J/mm^2$ for a $5\,ns$ pulse are sufficient to obtain an ionization, which corresponds to several minimum ionizing particles [9].

The implementation of such a system for the $\overline{P}ANDA$ TPC is at the cutting-edge of today's technology. The major constraints are given by the available installation space. Whereas the above-mentioned laser systems for TPCs are implemented on cm/m scales, one has to deal with a mm window in case of the $\overline{P}ANDA$ TPC. For this purpose the laser light is generated outside the TPC and distributed through light-guide fibers, which are incorporated in the wall of the FC vessels. Afterward it enters the gas volume under $90°$ to the beam axis at different $z$ positions and providing tracks of constant drift time. Figure 6.1 illustrates the conceptual design of such a laser system.

It is important to minimize the extraction of electrons within the chamber by the photoelectric effect which takes place on metals that release electrons at energies below $4.66\,eV$. A considerable amount of low-energy electrons has to be expected from diffusely scattered UV light hitting metallic surfaces. In the current setup these are mainly formed by the GEM layers (Cu: $4.3–4.5\,eV$) and the aluminum coating (Al: $3.0–4.2\,eV$) of the cathode end-cap and the copper (aluminum) strips of the electrode forming the field defining system of the field cage. Since the strip-lines currently are coated by a thin layer of Ni/Au with bigger values (Ni: $5.0\,eV$, Au: $4.8–5.4\,eV$), the emission from the cathode remains the main contribution.

### 6.1.2   2-D Laser Calibration

The other potential laser system, which is an option for the $\overline{P}ANDA$ TPC is based on the laser calibration systems developed for the HARP TPC [35] and further optimized for the T2K [34] TPC. The benefit of this concept is that it is comparatively simple and easy to implement. The basic concept consists of a fiber system coupled in at the anode side of the TPC. The fiber system will follow basically the same implementation as shown in Fig. 6.1. The fibers however will not enter the drift volume, but instead produce a homogeneous light distribution at the beginning of the drift volume. The UV light distribution will illuminate a pattern of aluminum strips and dots imprinted on the copper cathode using photo-lithographic methods. The aluminum has a low extraction energy of $3.0–4.2\,eV$, so that charges comparable to MIPs would be produced. The strip and dot structure forms a specific pattern, that after reconstruction can be used for a precise determination of absolute spatial distortions along the full drift length of TPC. As the laser pulse length is short ($\sim 5\,ns$) compared to the total drift time variations of extracted electrons, it can be used to monitor the drift velocity.

Also for this system the fibers will have to penetrate through the GEM frame to have a direct connection the gas volume. For the T2K system, SILICA solar resistant fibers with diameters of 600 and $800\,\mu m$ have been tested. They have a damage threshold of 2.7 resp. $4.9\,mJ$ for $5\,ns$ pulses of $266\,nm$ wavelength. This damage threshold is far above the total energy density needed for the illumination of the total cathode plane. To have a similar ionization like a MIP, roughly $10\,e^-$ per mm must be produced. The strips should have a width of about $2\,mm$, so that the width of the charge cloud is dominated by diffusion and not by the strip width. The photon efficiency of aluminum is about $3 \cdot 10^{-8}\,e^-/\gamma$, and with an energy of $4.6\,eV$ for each $\gamma$ an energy density of $12.446\,nJ/cm^2$ is needed to produce a MIP-like charge deposition. With a total area of $789.75\,cm^2$ an integrated energy of about $30\,\mu J$ is needed to illuminate the whole cathode. There are several companies, that can provide Nd:YAG laser with an energy of several mJ per pulse.

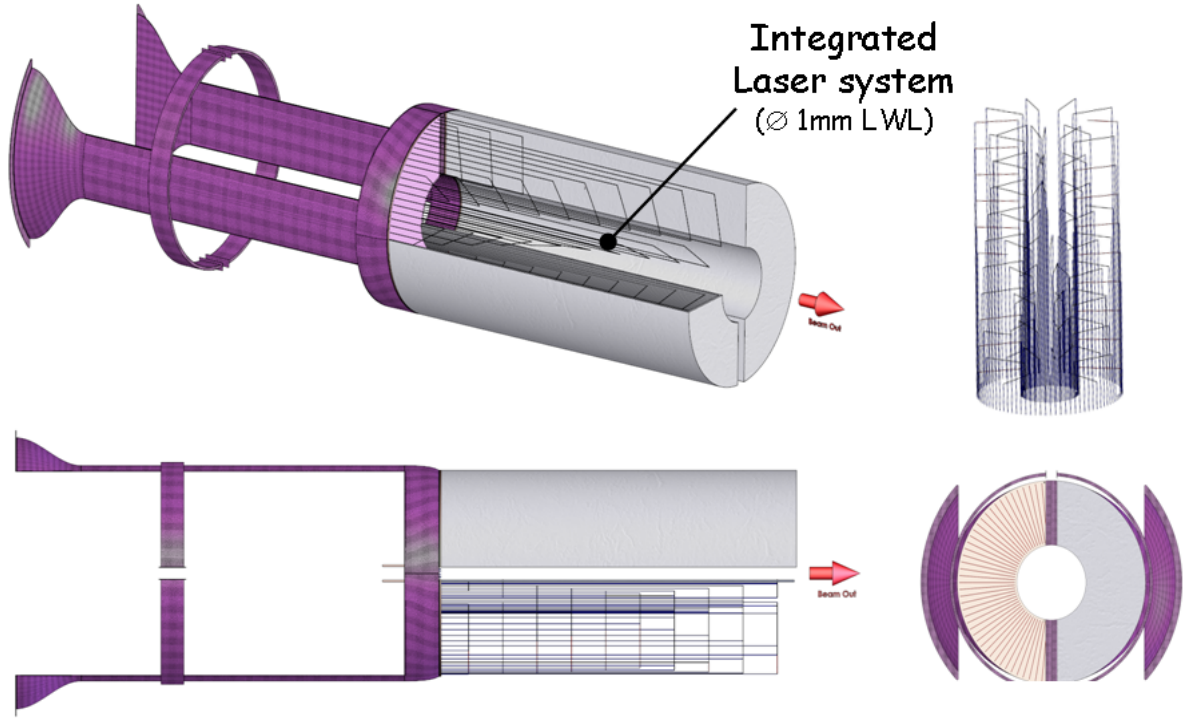

**Figure 6.1:** Conceptual design of the integration of the fiber light guide-based laser calibration system into the P̄ANDA GEM-TPC is shown. The fibers are integrated into the walls of the field cage vessels. The lines traversing the drift volume visualize the laser beams and are shown to guide the eye only. Currently we foresee a granularity in $z$ of 100–150 mm and 40° which allows for a total of 900 generated tracks, 45 per drift volume. The difference in angle between adjacent $z$ layers is chosen to be 4°.

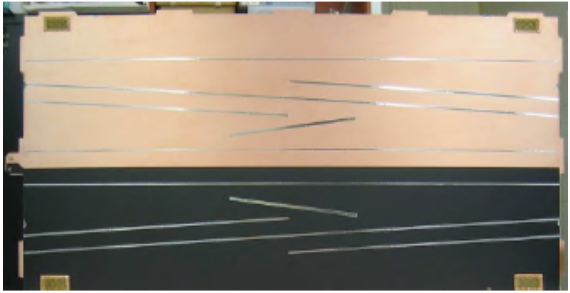

**Figure 6.2:** The picture shows the central cathode of the T2K TPC with aluminum strips [34].

## 6.2 Krypton Calibration

Absolute and relative gain calibration of a GEM-TPC is an important factor to improve the overall energy, $dE/dx$ and spatial resolution. Spatial fluctuations in the gain factors for each channel can arise e.g. from different circuit lengths, gain variations in the ADCs and front-end electronics or sector borders of the GEM foils. Furthermore systematic GEM profile studies have shown that the spatial gain distribution of double and triple GEM detectors can have high variations. Gain variations up to 20% due to stresses on the GEM frame and bending of the foils could be observed. The bending does not change the gain of the GEM itself. However it modifies the effective transfer and induction fields between the GEMs. This leads to different attachment and extraction coefficients thus modifying locally the effective gain [36]. These effects need to be corrected in order to reach an optimal energy resolution.

One method to perform such an energy calibration is by introducing radioactive $^{83m}Kr$ into the drift volume. This technique has already been used in various large TPCs (e.g. ALEPH [37], HARP, NA49, ALICE [38], STAR [39]).

A $^{83}$Rb source is mounted in a bypass of the TPC gas system. It decays with a half-life of 86.2 d into an isomeric 41.6 keV state of $^{83}$Kr, which decays into the stable ground state with a half-life of 1.83 h via a short-lived excited state at 9.4 keV. The decay spectrum has 4 main peaks between 9.4 keV and 41.6 keV, originating from conversion electrons, that can be used for gain equalization and calibration

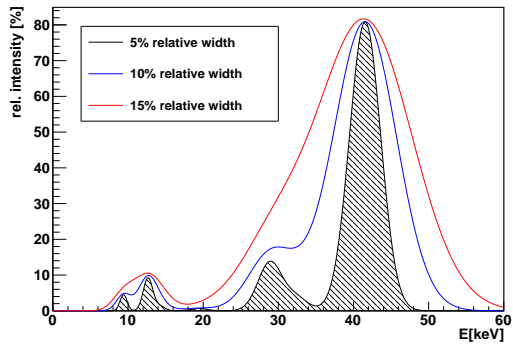

**Figure 6.3:** Charge spectrum from [83]Kr decays calculated combining the transition probabilities and fluorescence yields given in [40]. The photon conversion coefficients for the large TPC prototype with ArCO$_2$ 90/10 [41] were taken into account. The four main peaks between 9.4 keV and 41.6 keV are shown for different possible energy resolutions.

(see Fig. 6.3). A detailed discussion of the decay spectrum can be found in [40].

For calibration of the prototype TPC (first results see 12.2.3) a [83]Rb source with an activity of 2.5 MBq has been produced at the HISKP cyclotron[1] in Mar 2011 using an [81]Br$(\alpha, 2n)$[83]Rb reaction with $\sigma = 1300$ mb at an energy of 26 MeV. This production process has already been used before for a calibration source of the KATRIN experiment and is described in [42].

The container for the rubidium source can be seen in Fig. 6.4. The inner steel tube holding the radioactive material can be connected easily via two flanges to the gas system of the TPC. In case the system is not connected to the gas system it can be easily sealed with a dummy flange for transport. An outer shielding consisting of 13 cm of lead absorbs higher energetic decay photons that are emitted during the decay.

After connecting the container to the gas system the filling of the krypton-enriched gas is done by a three valve system as shown in Fig. 6.5. In the normal operation of the gas system the krypton container is bypassed from the system by two valves. In phases of krypton calibration runs the gas is circulated through the krypton container releasing [83]Kr in the TPC volume. After finishing the calibration runs the container is bypassed again while the remaining amount of krypton decays within a couple of tens of minutes.

Figure 6.6 shows the reactivation time needed to recover a 100 % [83]Kr activity after a complete ex-haustion of the krypton container.

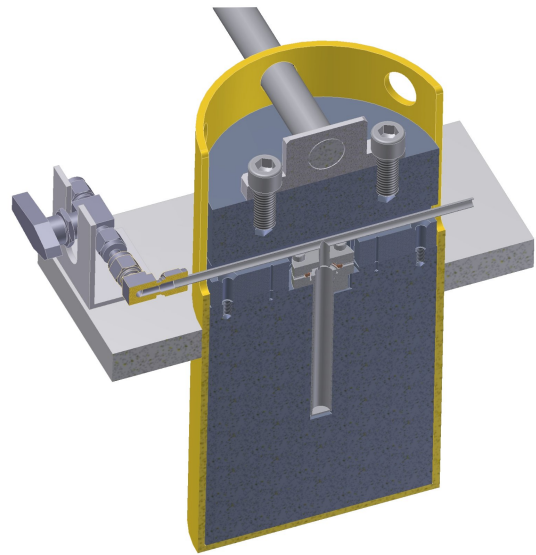

**Figure 6.4:** Lead-shielded container housing the [83]Rb source. The source is contained in a steel finger that can be attached to the gas inlet via a bypass system. The outer shielding absorbs higher energetic decay photons that are emitted during the decay.

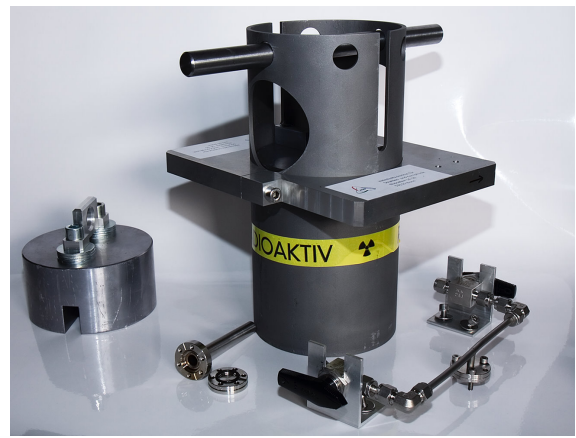

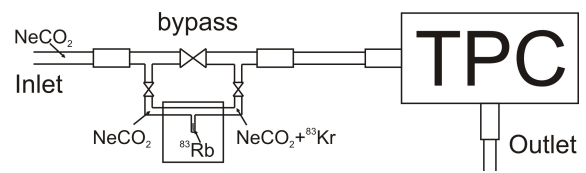

**Figure 6.5:** Sketch of the [83]Kr container connected to the TPC gas system.

---

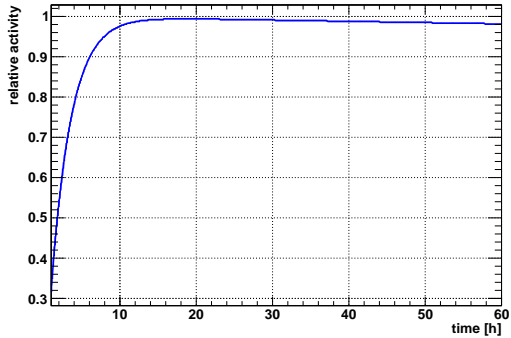

**Figure 6.6:** Reactivation time needed to recover a 100% $^{83}$Kr activity after a complete exhaustion of the krypton container.

## 6.3 Electronics Calibration

The readout plane of the final TPC detector is segmented into eighty thousand pads each one connected with the corresponding electronic channel. The readout system measures the charge induced on the readout pads and the absolute time when it occurs.

The charge value is very important for a precise determination of the $x$-$y$ cluster coordinate as well as for d$E$/d$x$ measurements. From our experience with the APV25 front-end chip the variations of charge sensitivity from channel to channel and from chip to chip are very small and do not require additional corrections. The $^{83}$Kr calibration takes into account small remaining channel-to-channel variations of the charge sensitivity. The absolute value of the sensitivity has to be determined during production of the front-end electronics and used for the full period of chamber operation.

The absolute arrival times of detector signals are measured with respect to a clock, common to all $\overline{\text{P}}$ANDA detectors and distributed by SODA (Time Distribution System of $\overline{\text{P}}$ANDA experiment) [43]. The distributed clock has a very small time jitter but due to the difference in cable lengths and propagation characteristics of electronic components there is an unknown relation between clock phases between different front-end cards. For the time calibration of the TPC front-end electronics one may use straight tracks of the laser calibration system or high energy cosmic tracks. In order to time in the TPC timing information with other detectors one has to use physics events or cosmics.

# 7  Slow Control

To ensure a safe and stable operation of the TPC, there are several parameters that have to be controlled, monitored and stored. Besides the temperature distribution in- and outside the detector and the parameters of the gas system, primarily the settings of the high voltages for the drift and the GEM amplification stack are essential. Also, the low voltages for the readout electronics have to be supervised.

## 7.1  High Voltage System

The operation of the TPC demands on the one hand a separate supervision of every high voltage that is applied to the GEM foils and the drift cathode and, on the other hand, the possibility to operate all channels simultaneously, especially during ramping up and down the chamber. To avoid any damages to the chamber, a fast emergency shutdown of the high voltage system is indispensable. Furthermore the system should allow the configuration of this emergency behavior.

The high voltage system for the GEM stack requires good voltage stability (ripple and noise < 50 mV), high precision current measurement (resolution 1 nA), adjustable ramp speeds, full remote controllability and output voltages up to 6 kV. For the drift cathode a high voltage system with a voltage of up to 70 kV and currents up to 1 mA is needed. Since both systems need to be connected for a synchronous operation and trip behavior an ISEG EHS 8060n HV module and an ISEG HPn700, controlled by a W-Ie-Ne-R MPOD Crate, would be a suitable choice. Both systems have a fast hardware current trip with a channel-wise adjustable current trip limit and can be controlled via SNMP commands over Ethernet or directly over CAN bus. For the large prototype a similar system is in use (30 kV module ISEG HPn300 instead of HPn700) and shows good functionality and reliability.

At these high voltages the security systems and the emergency behavior are important issues. The current architecture of an emergency high voltage shutdown system offers two solutions. A purely hardware based approach and a hardware plus software approach.

The hardware approach couples the HV system of the cathode plus field cage with the HV system of the GEM stack by an interlock cable. In case of a trip in one of the two systems the combined system is shutdown immediately and has to be ramped up from zero again. Since a trip in the GEM stack in this scenario also causes the drift voltage to trip, the ramping of which takes several hours, this introduces a significant dead time of the detectors, which is clearly disfavored during the time of operation.

The hardware plus software approach decouples the HV systems for the GEM stack and the drift cathode from each other. In case of a trip of one of the systems only the tripping system is shut down by hardware. To minimize the the potential danger of sustained discharges between the first GEM layer and the last field strip in case of a trip of the GEM stack, a software adjustment of the drift cathode voltage to a level of 60% has been implemented. This system has shown a good performance and has minimized the dead time significantly.

Besides the trip behavior, two security functions have been implemented purely on the software side: if the measured voltages rise above the set point voltage or if the current steps up unexpectedly the TPC should be ramped down within a short time to prevent severe damages. For the large prototype the voltage set point is compared with the moving average of the last measured values to detect these over-voltages. To detect any over-currents the average of the last ten measured values is compared with the average from ten seconds earlier.

Especially the currents are good indicators for a possible failure of the detector, because any short-cut between the GEM foils or at the field cage will result in a higher current. Therefore they have to be measured frequently and with high precision and be visible all the time.

To calculate the desired voltages from given values for the fields, distances and GEM foil potentials and to configure a ramp automation a GUI should be used (see Fig. 7.1). When ramping up the chamber, it is important to do this synchronously for all voltages (see Fig. 7.2) and with a slow ramp speed to avoid unwanted and dangerously high potentials between the GEM foils.

From the physical point of view it is necessary to monitor and save all voltages since the electron drift velocity arises from the applied drift field and the gain of GEM detectors is clearly linked to the adjusted settings. Furthermore it would be good to store these main parameters together with the taken data for the later analysis.

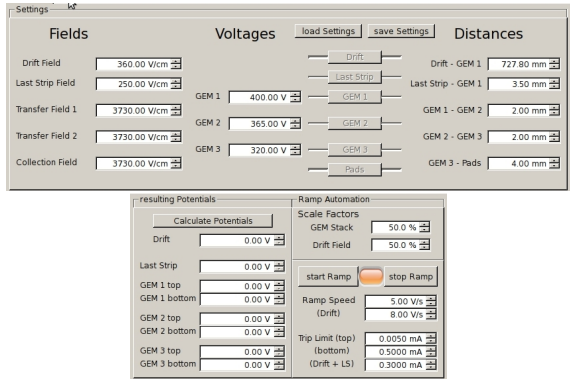

**Figure 7.1:** Graphical interface to set up the TPC high voltage for the large prototype. In the upper field the different settings can be adjusted. The lower box shows the calculated potentials and the settings for the ramp automation.

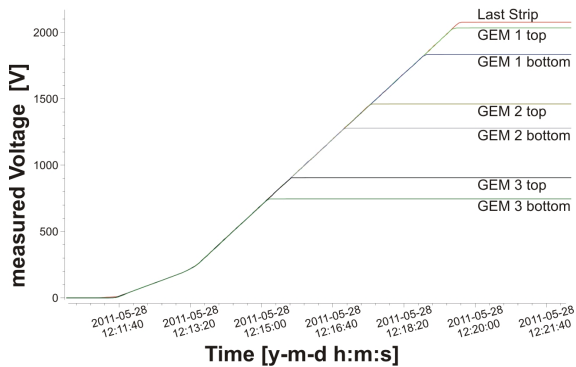

**Figure 7.2:** View of the voltages applied to the GEM foils when ramping up the chamber.

## 7.2 Low Voltage System

The low voltage for the Front-end electronics (FEE) will be provided by the $\overline{P}ANDA$ central low voltage supplies with a voltage of 48 V. Two or four such channels will be used, nevertheless each card can be switched on/off individually.

Three different powering schemes are being discussed at the moment.

- The first scheme is using air coils for DC-DC converters on the FEE. One has to investigate the influence of the high operating frequencies of the order of MHz on the noise of the FEE.

- The second solution is to put regular DC-DC converters just outside of the solenoid. The ca-

ble length from the DC-DC converters to the FEE will not exceed 2–3 m and the voltage drop over the cables will be sensed at destination.

- The last option is to power the FEE directly over long cables without voltage regulators at the FEE. This has been implemented successfully by the CMS tracker [44, 45].

For all options we foresee a maximum power consumption of 25 mW per channel, giving a maximal total power consumption of 2 kW. The power supplies will be floating with their ground being defined by the detector to avoid ground loops.

## 7.3 Front-end Electronics

Monitoring operation conditions and remote actions on all 320 front-end cards is important for stable data taking. It is foreseen to equip each front-end card with PT100 temperature sensors, voltage and current monitors. The slow control master will access these information with regular time intervals via optical interfaces. If one of the parameters exceeds its boundary conditions then the front-end card will be switched off by pulling down the voltage level of the safety wire.

Configuration of front-end electronics, reading status registers of front-end ASIC chips and injection of test pulses are also performed via the slow control interface.

## 7.4 Temperature Sensors

The temperature fluctuations of the detector gas have a major impact on the electron drift velocity. As the drift velocity is not on the plateau but on the slope region, monitoring of the gas temperature is crucial. Therefore several uniformly distributed temperature sensors are foreseen to get a precise distribution of the temperatures inside the TPC. Most of the sensors will be integrated into the walls of the field cage, since they should be as near to the active volume as possible but cannot be mounted inside the detector to preserve field homogeneity. More sensors will be installed on the pad plane to control the heat dissipation of the electronics. Also the temperature of the gas at the inlet and outlet will be monitored.

Due to the high number of needed sensors the use of PT100 or other analog temperature sensors would add many additional wires and material. Therefore digital readout sensors are advisable. For the proposed SMD mounted 1-wire sensors one can reduce the number to only two wires (ground and combined voltage/signal wire). Since the TPC is planned to be stabilized in the order of $\pm 0.2\,K$ the precision of the used sensors have be in the same order of magnitude.

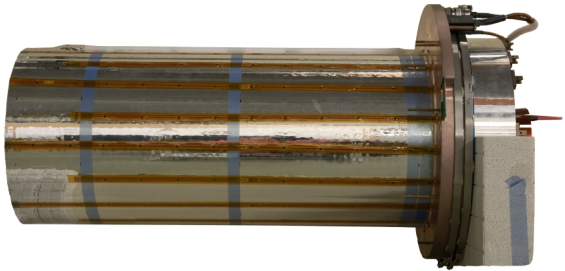

**Figure 7.3:** View of the TPC with the attached temperature sensors on the field cage. The sensors are placed every 5 cm on the yellow stripes.

At the large prototype 210 Dallas 1-Wire sensors (18B20U, see Fig. 7.3) with a USB readout are mounted on the outer field cage. Twelve analog PT100 sensors are attached to the pad plane. Unfortunately the 1-wire sensors induced too much noise during the prototype tests in the TPC FEE and the neighboring detectors due to the USB communication, which therefore had to be reduced to 9-bit and less frequent readout. For the $\overline{P}ANDA$ TPC the usage of thin shielded twisted wires will be investigated to overcome crosstalk between adjacent detectors.

## 7.5 Gas

The gas control will follow the general layout of the PANDA slow control. In particular, it will be similar to the internal target slow control system.

The main hardware of the gas control equipment consists of a Compact Reconfigurable I/O (cRIO) made by National Instruments as a front-end real-time embedded controller. This system consists of a real-time controller, a chassis with an integrated FPGA and several easily exchangeable I/O modules (see Fig.7.4). This allows a secure, standalone solution, which is independent from generally used computer networks. Signals from sensors and actuators (e.g. pressure sensors) will be read out by input modules and control signals (e.g. mass flow controllers) are sent to the correct devices via output modules.

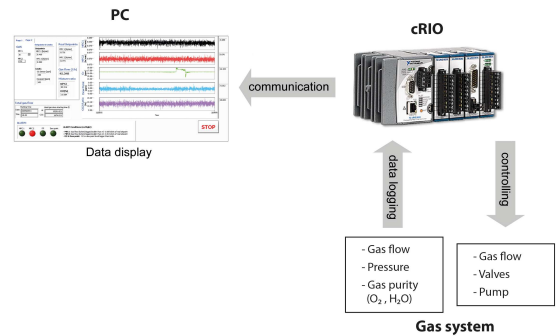

**Figure 7.4:** Block diagram of the main principle of the gas slow control.

On a higher level, users will be able to interact with the gas control system via a Graphical User Interface (using LabVIEW). This will ensure an easy integration of the gas control system into the general PANDA Control System.

## 7.6 Calibration

Any laser system as described in Sec. 6.1 requires remote control at least while it is active, which is basically realized by a simple trigger signal. In order to be able to deduce an intensity information, the stability of the laser has to be monitored either at the source or at the detectors place, e.g. by suitable photo diodes or something equivalent and the appropriate read-out electronics.

With respect to the calibration system based on Krypton gas as described in Sec. 6.2 simple remote controllable valves are sufficient to switch on and off this feature. In order to monitor the status of such a apparatus a few DC-level signal lines and a very simple electronic is needed.

# 8 Readout Electronics

## 8.1 Overview

The full TPC readout chain is shown in Fig. 8.1. It consists of the following components:

- 320 front-end cards, equipped with charge sensitive amplifiers, hit detection and digitization circuits,

- optical links,

- 40 data concentrator modules, performing feature extraction (time, amplitude) and cluster finding,

- high-speed serial links,

- data acquisition network,

- compute nodes for online reconstruction.

The front-end cards are mounted directly on the back side of the readout plane to minimize the input capacitance to the Charge Sensitive Amplifiers (CSAs). Since the electronic density is relatively low, the front-end cards are mounted perpendicularly to the readout plane (as it has been done on the large prototype chamber, cf. Sec. 12.1.3). Each front-end card reads 256 detector channels. According to the data-driven readout architecture of the $\overline{\text{P}}$ANDA experiment, hits have to be detected autonomously, i.e. without requiring an external trigger signal. The analog signals are digitized and marked by a time stamp. A single optical link connects each front-end card to a data concentrator module. The link is shared between three independent interfaces: timing distribution (SODA), slow control and data taking. Each data concentrator module reads data from 8 front-end cards. It extracts time and amplitude of a signal on a pad by pulse shape analysis (PSA) of the samples received from the front-end cards ("feature extraction"). These pad hits are then combined in time and space to clusters ("cluster finding"). From the data concentrator modules the data are sent through high-speed serial links to the $\overline{\text{P}}$ANDA data acquisition system, consisting of a large-bandwidth network fabric, which distributes the data from all sub-detectors to so-called compute nodes. These perform online processing of information combined in so-called super-bursts, corresponding to data from a time period of 0.5 ms, i.e. 10 full TPC drift frames with an overlap of 10% in order not to suffer from split tracks.

## 8.2 Front-end Electronics

The TPC measures three dimensional coordinates of charged particle trajectories. While the $x$ and $y$ coordinates are determined by processing charge information induced on a pad or group of pads of the readout plane, the $z$ coordinate is obtained from the electron drift time between the primary ionization position towards the gas amplification stage and eventually the readout plane. The signal induced on the pads by the electrons emerging from the last GEM foil has a fast rise time (less than 1 ns) and a duration of about 50 ns, which is given by the drift time of electrons between the last GEM and the readout plane.

The readout plane has a shape of a ring with inner and outer radius of 15.5 cm and 41 cm, respectively. It is segmented into 80000 hexagonal pads of 1.5 mm outer radius each. The pads are read out by a Charge Sensitive Amplifier (CSA) followed by a shaper. The parameters of these analogue circuits are optimized for low noise and fast shaping time. The noise performance of the electronics uniquely determines the minimum needed gas amplification of the GEM stack, which in turn defines the ion backflow and the performance of the TPC as tracking detector: the lower the noise the lower the ion backflow and consequently the lower the distortion of measured primary cluster coordinates. Thus it is very important to develop front-end electronics with a noise performance as low as possibly achievable. The minimum shaper peaking time for the TPC is about 120 ns, which is defined by the duration of the electron signal of $\sim 50$ ns and the average width in time of a charge cluster after 150 cm of drift due to longitudinal diffusion of $\sim 100$ ns. Shaping times longer than about 150 ns lead to excessive occupancies, especially for pads on the inner circumference of the chamber. The design of modern analogue circuits usually allows adjusting of the peaking time within this range and optimizing detector performance for particular operational conditions. A value of the ENC of 600–630 e$^-$ at 13–16 pF of input capacitance has been reached with the front-end electronics for the large prototype (cf. Sec. 12), including the digitization noise of $\sim 150$ e$^-$. Subtracting the digitization noise quadratically, the ENC of the front-end electronics is in the range 580–600 e$^-$. For the prototype, the main contribution to the input capacitance comes from the packaging of the ASIC itself (10 pF), and the traces on the PCB between the connector and the ASIC (3–5 pF). The pad contribution is only 0.2 pF and thus negligible. Optimizing the input capacitance to values between 3 and 5 pF, a noise of 400–450 e$^-$ can be reached. This value is thus defined as the maximum acceptable for the TPC detector.

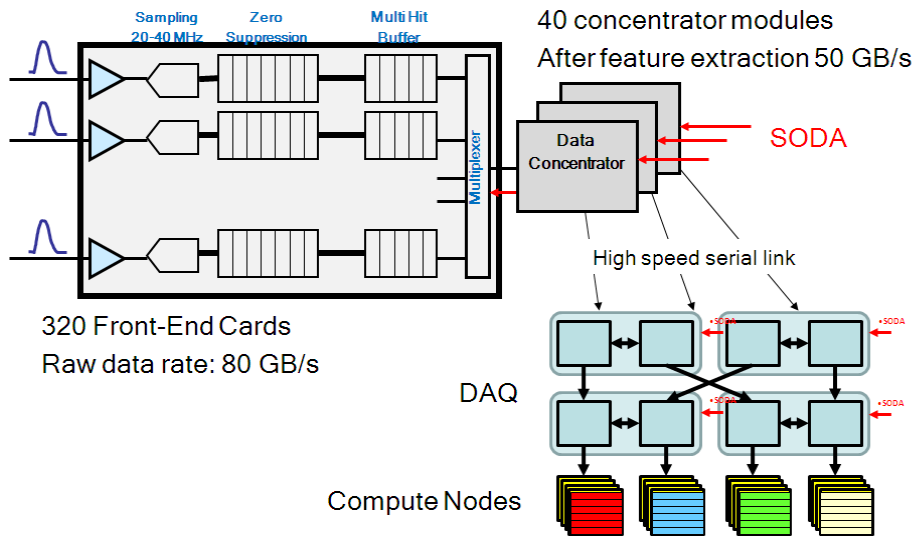

**Figure 8.1:** Schematic view of the TPC readout chain, including the P̄ANDA data acquisition system (lower right corner).

Another important requirement parameter is the average hit rate per pad. The P̄ANDA experiment will run at $2 \cdot 10^7$ s$^{-1}$ annihilations per second. The hit rate per pad for the inner pads is expected to be 200 kHz while for the outer pads it will be below 100 kHz. The PANDA experiment will operate in continuous, trigger-less mode requiring the development of a new type of data driven front-end electronics, which autonomously identifies detector signals and marks them with temporal information.

Table 8.1 lists the main requirements for the front-end electronics.

At the moment of writing there are three ASICs which are being developed for trigger-less experiments and which are thus considered for the P̄ANDA TPC :

- S-ALTRO,

- Transient Recorder,

- SPADIC.

| Parameter | Value |
|---|---|
| Number of channels | 80000 |
| Signal polarity | negative |
| SNR (for MIP) | 25 |
| Dynamic range | 10 bits |
| Noise (ENC) | 400 e$^-$ at 3 pF |
| Sensitivity | 400 e$^-$ per ADC bit |
| Shaping time | 120–150 ns |
| Sampling rate | 20–30 MHz |
| Power consumption | 15 mW/channel |

**Table 8.1:** Requirements for front-end electronics.

All those chips are mixed analogue-digital integrated circuits dedicated to run in continuous, data driven mode. The chip parameters are summarized in Table 8.2.

### 8.2.1 S-ALTRO

The S-ALTRO chip [46] is being developed by the CERN microelectronic group for the TPC at the International Linear Collider. Every channel of the chip includes an amplification stage with CSA and fourth order CR-RC[4] shaper, 40 MHz 10 bit pipelined ADC and advanced digital signal processing logic as shown in Fig. 8.2. The micro-electronic group recently received a first demonstrator chip with 16 channels and a total power consumption of

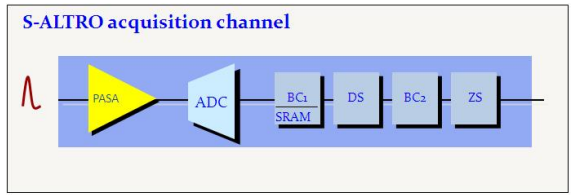

**Figure 8.2:** S-ALTRO ASIC architecture.

750 mW. 80% of the power consumption goes to ADCs. The main goal of this version of the chip is to demonstrate the feasibility and performance of the chip architecture.

The power consumption of the chip can be drastically reduced by using modern ADC designs which are able to achieve 4 mW/channel with similar resolution and sampling rate. If the development of the S-ALTRO chip is continued then the next generation of the chip will use low power consumption ADCs reducing the power to 12–15 mW/channel and increasing the integration level to 64 or even 128 channels per die.

### 8.2.2 Transient Recorder

The Transient Recorder ASIC is being developed by the GSI ASIC group. It is an alternative to the S-ALTRO approach, where zero suppression is done before digitization. After zero suppression the amount of information is significantly reduced allowing multiplexing of the detected samples from 32 channels into a single ADC. The architecture of the chip is shown in Figure 8.3; it consists of a Switched Capacitor Array, a zero suppression circuit and a pipelined ADC. The digitized information is transmitted to the next readout level via a serial link. The Transient Recorder requires a detector specific preamplifier-shaper chip for amplification and signal conditioning. It is foreseen to use the PCA16 ASIC, the front side of the S-ALTRO which includes a preamplifier and a shaper. The first prototype will be submitted for production in summer 2011.

### 8.2.3 SPADIC

The SPADIC ASIC is being developed at the University of Heidelberg for the Transition Radiation Detector of the CBM experiment at FAIR [47]. The chip has a similar architecture as the S-ALTRO, where every channel includes an amplification stage, a pipelined ADC and digital logic for signal processing. The chip is fabricated in UMC 0.18 μm technol-ogy and serves 32 detector channels. The dynamic range of the chip is 40 fC which is smaller than the charge created by low momentum particles on a pad by about a factor of two. Knowing the signal shape, however, it is possible to recover the amplitude information at a stage of feature extraction even with some channels in saturation. The advantage of the SPADIC chip is its low power consumption of 5 mW per ADC. The first prototype has been tested at the end 2010.

### 8.3 Feature Extraction

The data from a front-end card are transferred to so-called data concentrator modules. Here the following operations are performed:

- pulse shape analysis on the pad level to obtain the signal time and amplitude,

- sorting of data according to time slices,

- optionally cluster finding in order to further reduce the data rate at the output.

A first prototype of the data concentrator module is in development now at TU München. Figure 8.4 shows its functional diagram and a 3-D model. The module is based on Lattice FPGA featuring 16 high speed serial links. Up to 8 of such modules can be mounted on a single carrier card following the ATCA (Advanced Telecommunications Computing Architecture) standard. The optical transceivers are mounted on a separate module in the ATCA crate. The expected bandwidth of the module is 1.6 GB/s.

### 8.4 Data rates

The raw hit information is encoded into a 40 bit data word which includes 8 bits of channel number, 24 bits of three sample amplitudes and 7 bits of time tag. The raw data rate is about 80 GB/s at the output of the front-end cards. The first data reduction is done in a concentrator module by analyzing three amplitude values and extracting hit amplitude and time information. After feature extraction the data rate is reduced to about 50 GB/s. The next step in the data processing is clustering which is done after merging time correlated information into single data block. Clustering reduces the data rate down to 25 GB/s.

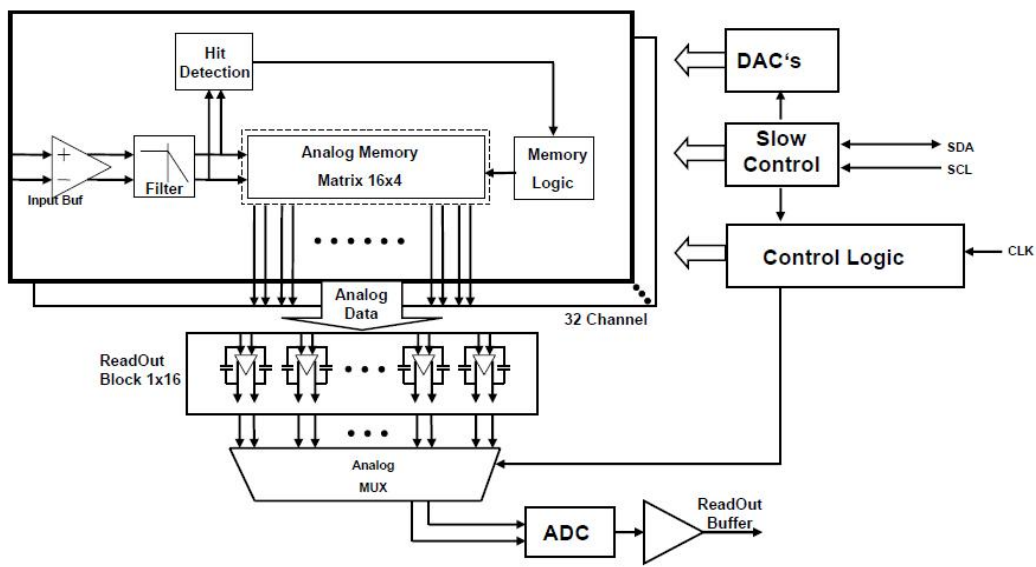

**Figure 8.3:** Transient Recorder ASIC functional diagram.

|  | S-ALTRO | PCA16 + TR | SPADIC |
|---|---|---|---|
| Process | IBM 0.13 μm | IBM 0.13 μm + UMC 0.18 μm | UMC 0.18 μm |
| Sensitivity | 12 mV/fC | 12 mV/fC | 0.08 fC/ADCch. |
| Noise(ENC) | 500 e⁻ at 5 pF | 500 e⁻ at 5 pF | 800 e⁻ at 30 pF |
| Dynamic range | 10 bits | 10 bits | 8(9) bits |
| Peaking time | 30–120 ns | 30–120 ns | 90 ns |
| Sampling rate | 20–40 MHz | 20–100 MHz | 25 MHz |
| Power consumption | 46 mW/ch | 10 mW/ch | 12 mW/ch |

**Table 8.2:** Parameters of ASICs considered for the $\overline{\text{P}}$ANDA TPC.

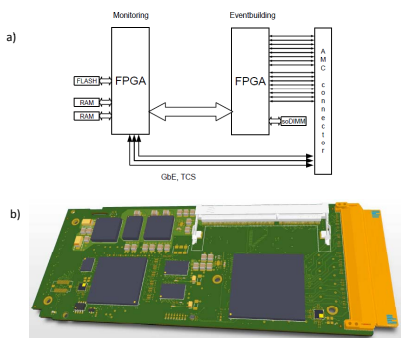

**Figure 8.4:** The TPC data concentrator module functional diagram and 3-D model.

## 8.5 Online Reconstruction

After digitization, feature extraction and cluster finding, the data of all $\overline{\text{P}}$ANDA detectors are sent via high-speed serial links to the data acquisition network (FPGA-based or Infiniband, about 500 GB/s total bandwidth), which distributes them in a time-sorted manner to the Compute Nodes. These are custom-made ATCA modules, which form the platform to run online reconstruction and event selection algorithms. For online track reconstruction, two options are currently being pursued inside $\overline{\text{P}}$ANDA:

- FPGA-based algorithms [48, 49, 50],
- GPU-based algorithms [51, 52].

Figure 8.5 illustrates the logic steps foreseen for the reconstruction of data from the TPC as well as the merging with the information from other detectors and the selection of interesting events.

The pattern recognition and piece-wise reconstruction of tracks (tracklets) first proceeds in parallel and completely independent in eight geographic sectors of the TPC. Here, local pattern recognition is performed by using advanced algorithms such as conformal mapping and Fast Hough Transform (cf. Sec 10.7.8 for details).

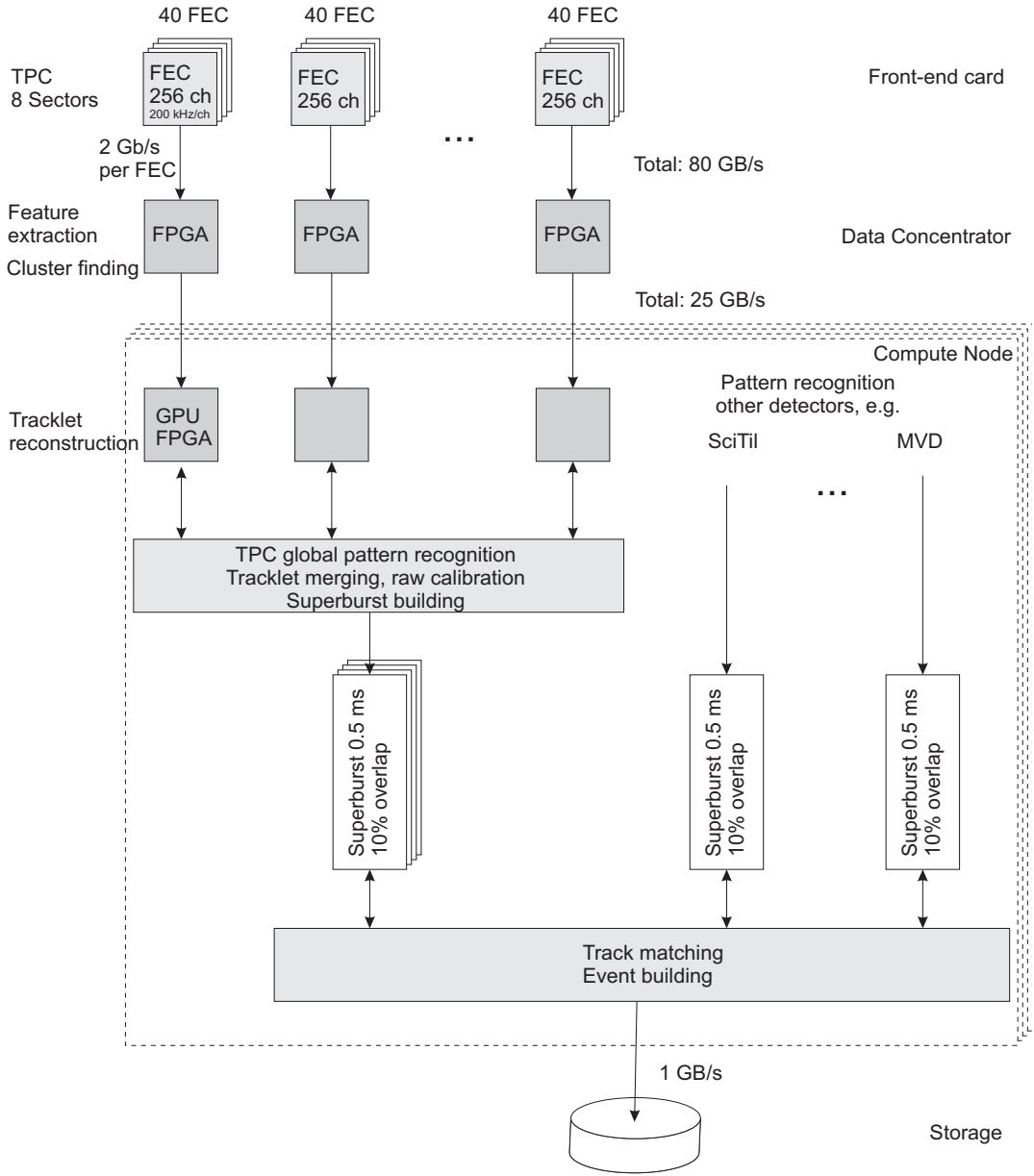

**Figure 8.5:** Flow diagram of the logic steps for online data reconstruction in the TPC, merging of data and event selection.

In the next stage, the tracklets from the individual sectors are merged making use of the preliminary helix parameters determined in the first step. At this step, first raw distortion corrections in the $r\phi$-plane will be applied to the tracks. A preliminary sorting of tracks is done based on the assumption that the tracks originate from the target ("target pointing"). At this stage, also more demanding algorithms selecting interesting event signatures in the TPC can be applied. These include

- decays of neutral particles, e.g. $\Lambda$s,
- kinks in charged particle tracks, e.g. from $\Xi$ decays,
- detection of $\pi \rightarrow \mu$ decays (important for background rejection in the muon range system of

$\overline{\text{P}}$ANDA).

In the next stage, the information from the TPC will be combined with that from other detectors, e.g. the Micro-Vertex Detector or the Barrel ToF Detector. For the MVD, this will be done following the same scheme which has already been applied on simulation level (cf. Sec. 10.13):

- selection of TPC tracks (after target pointing) within a given time window (e.g. $\pm 200\,\text{ns}$) around the time determined by the MVD (e.g. a displaced vertex),

- extrapolating the TPC track to the MVD and picking up MVD hits within a coarse roadwidth ($\mathcal{O}(1\,\text{mm})$).

In this way, events with tracks originating from the primary vertex were shown to be identified with an efficiency of $> 95\%$ in Monte Carlo simulations. In a similar way, information from the Barrel ToF with its fine granularity of $3 \times 3\,\text{cm}^2$ will also be used to assign an absolute time to tracks in the TPC.

Groups of tracks and signals which have been identified to bear a common physics signature are then stored or transferred for further analysis. In order to be able to apply full distortion corrections and advanced fitting algorithms, the preliminary track information determined in the global pattern recognition stage as well as the raw cluster data will be stored.

# 9 Cooling System

## 9.1 Requirements

The basic paradigm in operating the detector is to take away all heat introduced by the electronics into the detector and its surrounding in order to avoid that even a small percentage of it is transferred to the Pad-Plane and fed into the gas volume. This would cause a density gradient in the gas which is a source of inhomogeneities in drift velocity of the charge carriers.

The front-end electronics of the TPC is confined in a small volume where radiative or convection cooling is not effective enough nor applicable due to the high spatial density of the modules. Thus a closed loop cooling-system based on the liquid coolant HFE7100 has been set up. It fully exploits passive heat transfer processes from the electronics to a coolant circulating through a helix with small cross-sections optimized for low-material budget. The flow of the low-viscosity medium is strictly turbulent throughout the structure to support optimal heat-exchange. The design of the light-weight heat exchange structure is facilitating the easy exchange of the front-end cards due to its key and slot geometry. The walls are made from aluminum and are providing the shielding characteristics of a Faraday cage. No tubing is required from card to card and only two joints equipped with self-closing and dead-volume free valves are necessary on the detector, one at the input and one at the output. This is minimizing the risks taken by introducing liquid coolants in the vicinity of electronics and neighboring detector systems. The coolant itself is nontoxic, noncorrosive, electrically nonconducting and has a vapor pressure which is low enough that it is vaporizing residue-free at room temperature.

For simplicity a central chiller with cooling, heating and reservoir option which can digest a 10 kW heat load is working in a 1-step closed circuit without a separate heat-exchanger. The system is operating in overpressure mode in order to avoid problems which often occur in under-pressure systems arising from differences in height-levels within the experimental set up (as faced by other experiments).

The system is remote controllable within a wide range of operation conditions. The operation point is chosen in a way that at any position inside the PANDA magnet the local temperature at the surface of the cooling structures does not reach the dew point. This way only a small amount of thermal isolation is needed and no perspiration water is to be expected at surfaces.

## 9.2 Layout

A cooling system envisaged to be used with the PANDA-TPC has been set up for the Large Prototype TPC. It is designed to work with overpressure and can digest up to 5 kW of thermal energy. Flow is kept strictly turbulent inside the detector-near structures in order to optimize heat transfer from the cooled surfaces to the coolant. With a media input temperature of 12 °C the output temperature does not exceed 25 °C.

The coolant is of type HFE7100 and has the following advantages over a conventional water cooling medium:

- low viscosity (lower than water), best suited for low-profile channels

- low vapor pressure at the operation condition given, thus it would vanish without remains even if there is a leakage incident

- electrically isolating, non corrosive, best suited to be used in the vicinity of electronics

For the chiller we use a very compact module type Huber Unichiller UCHT operating with ambient air.

The main unit and several sensors measuring the temperatures and pressures in the system at various places is included into the detector control system to ease remote control and setting of operation conditions (see Sec. 7).

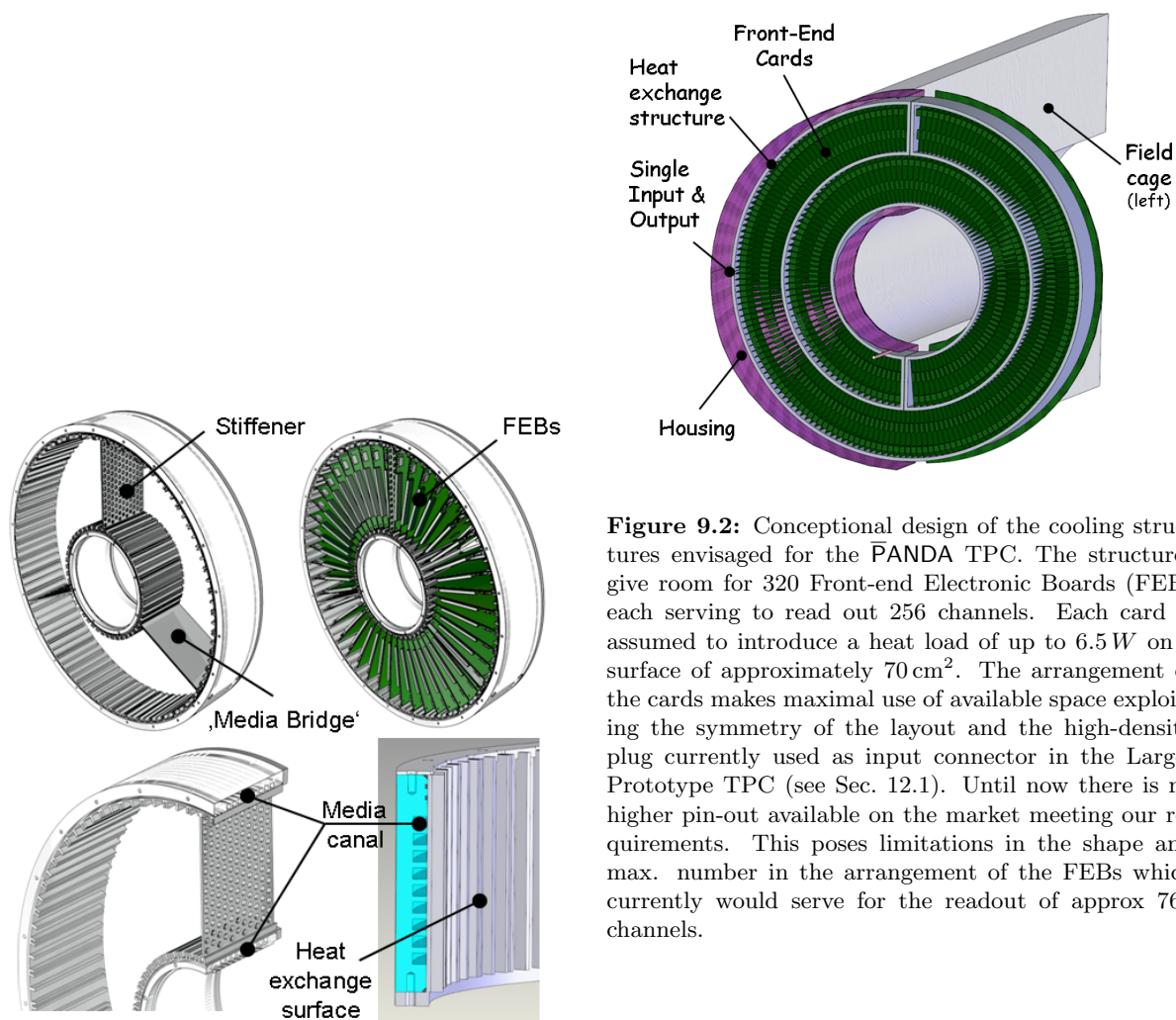

**Figure 9.2:** Conceptional design of the cooling structures envisaged for the $\overline{P}$ANDA TPC. The structures give room for 320 Front-end Electronic Boards (FEB) each serving to read out 256 channels. Each card is assumed to introduce a heat load of up to 6.5 W on a surface of approximately 70 cm$^2$. The arrangement of the cards makes maximal use of available space exploiting the symmetry of the layout and the high-density plug currently used as input connector in the Large-Prototype TPC (see Sec. 12.1). Until now there is no higher pin-out available on the market meeting our requirements. This poses limitations in the shape and max. number in the arrangement of the FEBs which currently would serve for the readout of approx 76k channels.

**Figure 9.1:** Design of the cooling structures without (upper left) and with (upper right) front-end boards (FEBs) installed. The lower part of the picture shows cutouts visualizing the helix of the media channels milled into the plastics part of the structure. The part facing the FEBs is made from chromatized Aluminum of 0.5 mm thickness of the heat-exchanger walls.

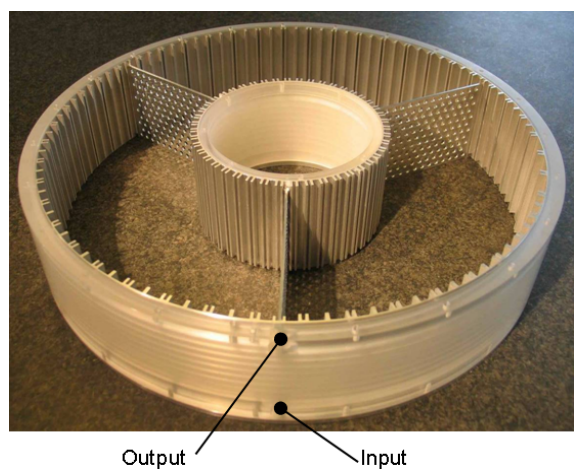

**Figure 9.3:** Photo of the realization of the system done for the Large Prototype TPC. Only two joints are required for the whole assembly which are labeled as 'Input' and 'Output', respectively.

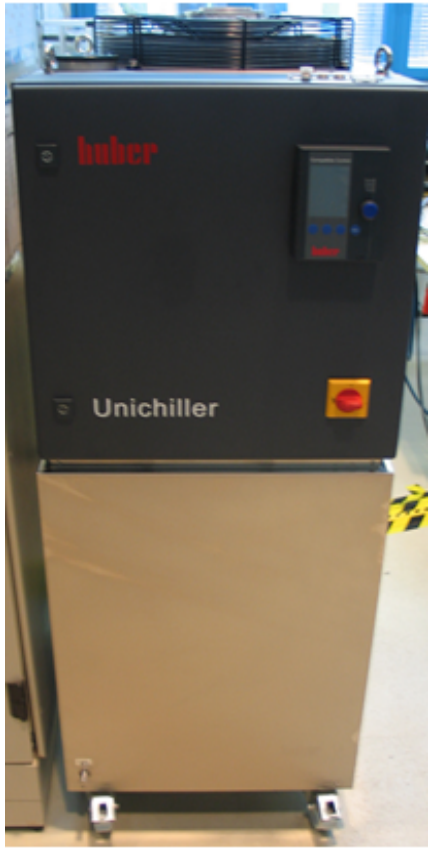

**Figure 9.4:** Photo of the heater/chiller module Unichiller UHCT with a cooling capability of $\approx 10\,\mathrm{kW}$.

# 10   Simulation of Detector Performance

An important tool to assess the general feasibility and estimate the actual performance of a complex detector system like the GEM-TPC for P̄ANDA are computer simulations. During R&D of the GEM-TPC detector, behavior has been modeled and studied in great detail using the tools provided by PandaRoot, the common `C++` computing framework developed within the P̄ANDA community.

In Secs. 10.1 to 10.4, we give an overview of the simulation chain as it is implemented in the PandaRoot framework. Various algorithms for the reconstruction of tracks have been developed and are described in Sec. 10.4 and following. Results from simulations are reported in Sec. 10.9.

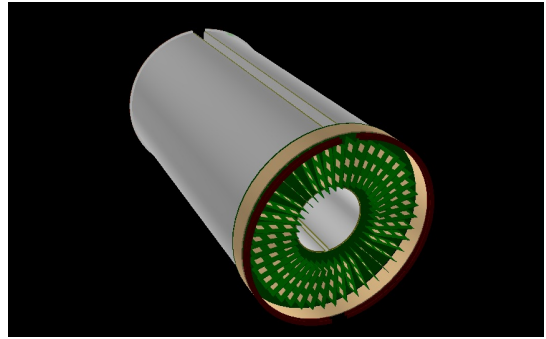

**Figure 10.1:** TPC geometry model as currently present in the Monte Carlo simulations. Placeholder volumes have been put where the final design is unclear (e.g. front-end electronics, cooling).

## 10.1   Simulation Chain Overview

We shall briefly describe the general structure of simulations, as implemented in the PandaRoot software framework [23]. Logically, the full simulation chain can be divided into three steps:

1. *"Monte Carlo"* simulations: particle generation and transportation,

2. *Digitization*: simulation of detector response,

3. *Reconstruction*: extraction of track parameters.

## 10.2   Monte Carlo Simulations

In the first step of the chain starting values for physical particles traveling inside the detector volumes are sampled from distributions from various models basically by "throwing the dice" (hence the term *"Monte Carlo simulations"*). These models range from simple "particle guns" simulating just one sort of particle at given momenta/angles to full-scale $\bar{p}p$-physics generators. The obtained generated particles (e.g. the starting values of their physical properties) are then extrapolated (*"transported"*) through the different layers of materials, based on a detailed geometrical model of the full P̄ANDA spectrometer. Figure 10.1 shows the geometrical model of the TPC as it is currently implemented in the simulation. The final result of this first step in the simulation chain is a set of space-points with attached information of associated momentum and energy loss, so-called *Monte Carlo (MC) Points.*

The piece of software responsible for this transportation is a widely-used community standard called *GEANT* [53]. GEANT has initially been developed at CERN as a package of FORTRAN routines and has been supported by CERN as a quasi-standard for many years up to its release version GEANT3. More recently, it has been ported to `C++` and extended by many features. The latest version available at CERN is GEANT 4 9.4.

GEANT4 has a number of advantages over its older FORTRAN-based predecessor. The one with the most impact for TPC simulations is a more realistic set of algorithms to treat energy loss due to ionization processes (Photo Absorption Ionization (PAI) model). Unfortunately, until very recently, there have been known problems involving these algorithms when working inside gaseous media, visible e.g. as un-physical energy loss distributions inside the TPC simulations. These problems have led us to work with GEANT in the previous version 3 for many years, however adapting an improved and more transparent method of calculating the energy loss inside the drift gas that was originally developed by the ALICE TPC collaboration [54], referred to as the *"ALICE model"* from here on. This model basically uses known gas properties to sample the step length between collisions (Poisson-statistics), using the Bethe-Bloch-Formula to construct the momentum-dependence. The energy loss for each collision is calculated using a modified Rutherford cross section: to mimic the atomic

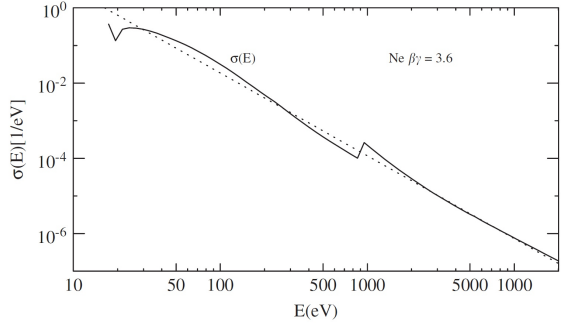

**Figure 10.2:** Comparison of inelastic collision cross-section $\sigma(E \,|\, \beta\gamma = 3.6)$ in Neon gas as obtained from the PAI method (solid line) and modified Rutherford cross-section of the ALICE MC model (Eq. 10.1). The ALICE coefficients have been chosen to match $\lambda$ of the PAI model at $\beta\gamma = 3.6$ [56].

binding of the electrons, the energy dependence of the cross section has been set to

$$\left(\frac{\mathrm{d}\sigma(E)}{\mathrm{d}E}\right)_{\text{ALICE}} \propto E^{-2.2}. \qquad (10.1)$$

[54, 55]. This choice has been found to reproduce the collision cross-section obtained with a PAI simulation reasonably well (c.f. Fig. 10.2). However, the resulting energy loss distributions ( *"straggling functions"* ) tend to peak at slightly lower values for the ALICE model when compared to PAI results [56], possibly decreasing energy deposition and hence the obtained energy loss resolution of the detector in the simulation.

More detailed comparisons between GEANT3 and GEANT4 will have to be made in the near future. The very recent update of GEANT4 in the simulation framework has unfortunately made it impossible to do full simulations with the superior ionization models already in the course of this work.

## 10.3 Digitization

The second step ( *"digitization"* ) comprises the modeling of all physical and conceptual processes taking place in the sub-detectors of $\overline{\text{P}}$ANDA, from the initial deposition of charge up to the signal creation in the readout electronics. The responsible simulation code from here on is organized in a modular fashion in so-called *tasks* to allow easy enabling/disabling or replacing of single steps or algorithms. In the case of the GEM-TPC, the procedure can be summarized by only a few essential building blocks as shown in 10.3:

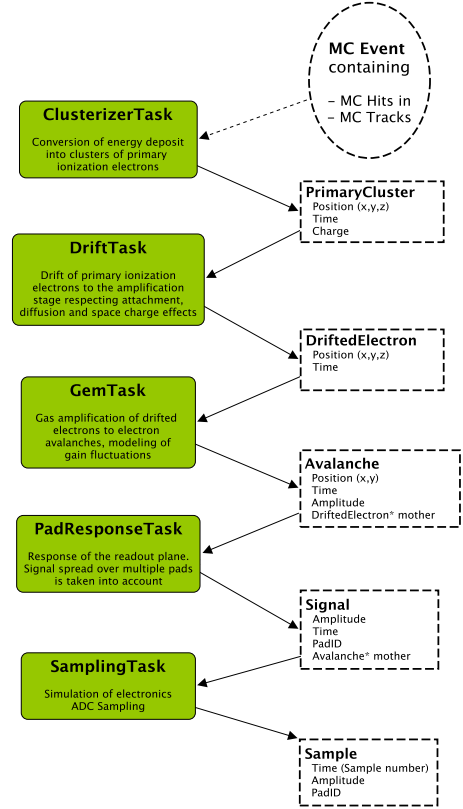

**Figure 10.3:** Standard work flow scheme of the $\overline{\text{P}}$ANDA TPC digitization. The output objects of each task are listed on the right. The simulation expert can chose individually, if they should be persistent, e.g. written to file.

- Conversion of deposited energy in the detector gas into primary electron clusters

- Drifting of the primary clusters towards the readout, taking into account attachment, diffusion and field effects.

- Charge amplification in the GEM-stack

- Response of the pickup pads to the arriving electron avalanches (pad-response-function)

- Signal creation

- ADC sampling

## 10.4 Reconstruction

After the digitization stage of the simulation chain the obtained data resembles the digitized front-end electronics output of a fully equipped $\overline{\text{P}}$ANDA GEM-TPC, i.e. amplitude information for pad-time bins ("samples"). Ideally, all code developed for the

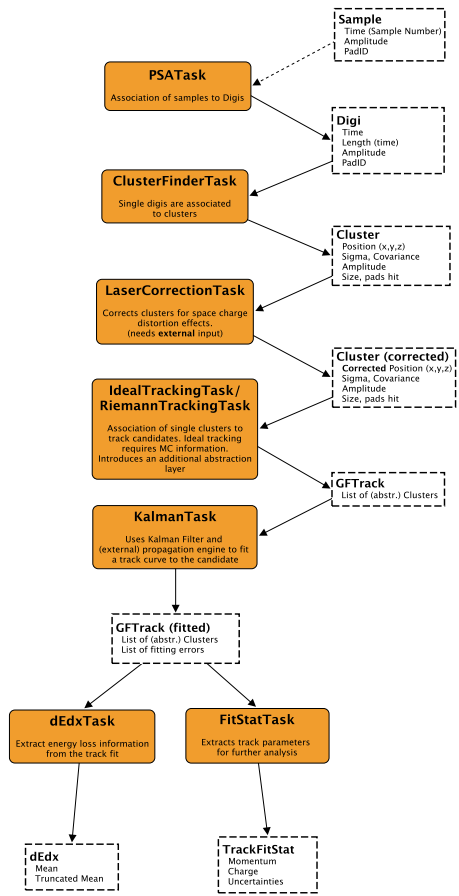

**Figure 10.4:** Standard work flow scheme of the $\overline{P}$ANDA TPC reconstruction. The output objects of each task are listed on the right. The simulation expert can chose individually, if they should be persistent, e.g. written to file.

*reconstruction* stage will be able to directly run on the data of the actual detector, once it has been built. As we will demonstrate in Sec. 12, this goal has already been achieved for the data recently taken with the large GEM-TPC prototype. The reconstruction stage comprises the following steps:

- Pulse Shape Analysis (PSA): adjacent samples in time on each pad are combined to pulse signals ("digits"), assigning to them a time and an amplitude.

- Cluster Finding: these digits are then grouped in space and time by the Cluster Finding task. The clusters, 3-D space points with a given amplitude, are the input to the

- Pattern Recognition algorithm and the

- Track Fitting algorithm, which finally extracts the physical properties of the track.

Figure 10.4 sketches the most important tasks of the reconstruction stage. All of these algorithms will have to run in close-to real-time during data taking with $\overline{P}$ANDA, as their results are needed as input for the online trigger system. In the following sections, the individual steps will be described in detail.

## 10.5 Pulse Shape Analysis

The first step is thus to combine single "samples" from the ADC to signals of a given time and amplitude. For the simulations and the analysis of the TPC Prototype data a simple PSA algorithm has been implemented. It starts a pulse if two consecutive samples are above a given threshold. Further samples are added to the pulse until a local minimum or a sample below threshold is found. The maximum amplitude is then assigned to be the amplitude of the signal (as opposed to a integration over the full signal width). Due to the symmetrical shape of the pulse, the time of the signal is calculated as the mean of the peak time and the center of the pulse. An example of this PSA algorithm used on real data taken with the TPC prototype can be seen in Fig. 10.5.

## 10.6 Cluster Finding

An important tool for data reduction in the TPC readout scheme is clustering, e.g. combining the digis origination from some spatially well-defined accumulation of charge in the chamber to one single object - a so-called *cluster*.

A cluster in this sense is fully defined by

- the position (charge center of gravity) with error, and

- the total amplitude.

The standard implementation of the cluster finding algorithm starts by presorting all available digis by decreasing amplitude. It then loops over all digis, checking if a digi is

- an immediate neighbor (pad-wise) to an already existing cluster,

- and it is close enough in $z$-direction, e.g. it lies within a certain *time slice* around an existing cluster.

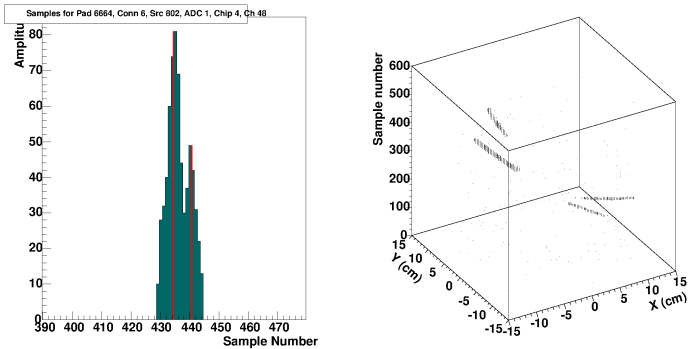

**Figure 10.5:** Example of the PSA on real data taken taken with the TPC-Prototype. On the left side one can see the time evolution of one pad with a signal from the steepest going track in the left figure. The red strips are the recognized pulses where the height is representing the amplitude. The data was taken with a peaking time of 116 ns, sampling at 16 MHz.

Consequently, the algorithm will start clustering around local maxima of deposited charge. If no matching cluster is found, a new cluster is created. If more than one cluster matches the digi, it is split, e.g. its amplitude is divided by the number of matching clusters and the digi is assigned to each of them. A threshold on cluster amplitude prevents isolated low-amplitude hits to form individual clusters, serving as an effective way of suppressing electronic noise.

As every digi needs to be checked against every cluster, the algorithm gets quite costly for large numbers of digis. To overcome this limitation, the digis are processed independently for each sector of the pad-plane and, additionally, a sectorization in $z$ is performed. This sectorization is subject to further optimization.

Figure 10.6 shows a 3D view of the clustering algorithm working on a track piece measured with the large prototype detector at FOPI, GSI, Germany (cf. Sec 12.2). The coin-shaped entities are digitized single pad hits (*digis*), different sizes corresponding to different amplitudes. All digis of the same color are assigned to one cluster (depicted as bubble of the same color). Digis that are split between adjacent clusters are shown in mixed color.

The described cluster finding algorithm is robust and works reliably for simulated as well as real data from our beam tests. As an alternative, a cluster finding algorithm based on a Cellular Automaton (CA) algorithm has been developed. The advantage of this algorithm is its high parallelism, making it very well suited for a close-to-the-metal implementation in the final $\overline{P}$ANDA DAQ scheme.

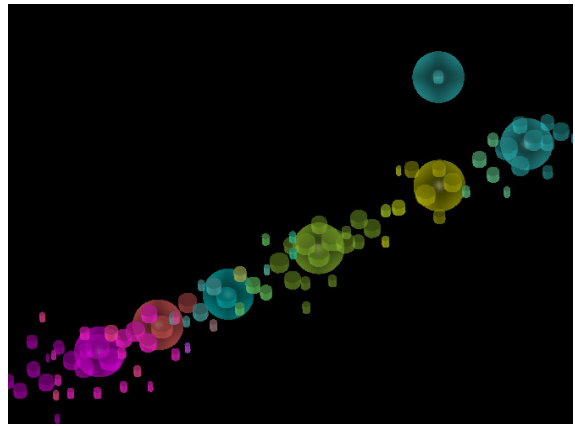

**Figure 10.6:** TPC cluster finding algorithm at work. The digitized single-pad hits (*digis* - output of the PSA algorithm, shown as coin-like shapes) are combined to larger *clusters*. Their amplitudes are summed up, the cluster position is calculated as the amplitude-weighted mean of the single digi positions.

## 10.7  Pattern Recognition

One of the biggest advantages of the TPC as central tracker in the purely time-based trigger and DAQ paradigm that we are facing in $\overline{P}$ANDA, is its independence on other detectors when it comes to pattern (track) recognition. Typical non-curling tracks from primary particles consist of $\sim$ 30-50 clusters forming consistent 3-dimensional helical arcs in the chamber. While the data rates remain challenging, the pattern recognition problem as such can be solved without relying on any information from other detectors.

We have developed two conceptually different methods for pattern recognition during development of the TPC reconstruction software. The standard pattern recognition algorithm is a track follower based on a conformal mapping method involving the so-called Riemann Sphere [57, 58], a well-known entity from complex analysis.

### 10.7.1 The Riemann Transformation

Common to both methods is the employment of the so-called Riemann Transformation. It is a *stereographic projection* of points from a plane - in this case $\mathbb{R}^2$ - onto the Riemann Sphere, a sphere of diameter one sitting *on top* of the origin of the complex plane. In our application for pattern recognition, the complex plane is replaced by $\mathbb{R}^2$.

The stereographic projection is defined on the entire sphere except one point (here: the north pole). It is smooth, bijective and conformal, but not isometric, thus preserving angles but not distances. The transformation rule for a point on the plane $\mathbf{x}_i = (x_i, y_i, z_i)$ reads:

$$
\begin{aligned}
x_i &= R_i \cdot \cos \phi_i \,/\, (1 + R_i^2), \\
y_i &= R_i \cdot \sin \phi_i \,/\, (1 + R_i^2), \\
z_i &= R_i^2 \,/\, (1 + R_i^2).
\end{aligned}
\tag{10.2}
$$

It is useful to note that the mapping of Eq. 10.2 does *not* increase the dimensionality of the problem, as a first glance might suggest. The most important feature of Eqs. (10.2) in the context of pattern recognition is that circles and lines *uniquely* map to circles on the sphere [59]. Since in turn a circle on a sphere uniquely defines a plane in space, there is a direct correspondence between a *circle* (on the plane) and a *plane* intersecting with the Riemann Sphere.

This fact has been exploited in the field of track fitting (e.g. [57, 58]), since it reduces the problem of fitting a set of points to a nonlinear mathematical entity (circle) to a linear one. In the scope of the helix pattern recognition problem it is used to gain additional criteria of hit proximity and track affiliation.

Figure 10.7 shows an example of such a transformation of a circular particle track (projections of helical tracks on the readout plane). The black dots in Fig. 10.7 correspond to reconstructed hits in the TPC (`Clusters`, Sec. 10.4) from simulated $\pi^+$-tracks. Thus they have been subject to effects like diffusion, readout clustering effects and electronics response (see Sec. 10.7.8). Figure 10.7 shows the transformed points of one of these tracks (marked red) as well as the associated best-fit plane.

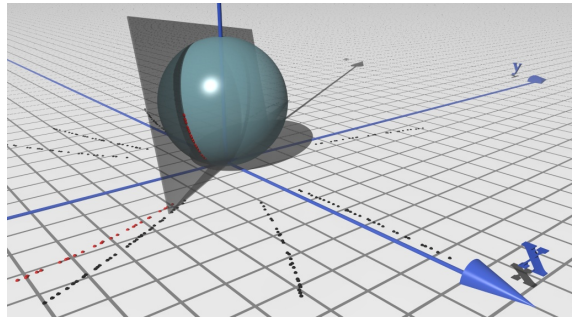

**Figure 10.7:** Example of the Riemann Transformation of a circular particle track ($\pi^+$). The tracks have been scaled by a constant factor to better match the scale of the unit Riemann Sphere. The plane defined by the transformed set of points is also shown.

### 10.7.2 Riemann Pattern Recognition - A Track Follower

The *Riemann Pattern* recognition as implemented for the $\overline{\text{P}}$ANDA TPC is a track following algorithm, that basically associates three-dimensional space points to track candidates by using different proximity criteria, both in detector coordinates $(x, y, z)$ as well as using the transformation described in Sec. 10.7.1.

Before *track building*, the clusters are presorted by z, radius or angle. The idea is that the pattern recognition goes from areas of low track density, where tracks can easily be separated by their proximity in space, to areas of high track density. The very first *track* is built and contains only one hit at this time, then the algorithm loops through the presorted clusters. Each hit is checked against each existing track. If one or several matching criteria (*"hit-track correlators"*) are fulfilled, the hit may then be assigned to the best matching track.

A hit-track correlator can be applicable or not, and if it is, it delivers a *matching quality*. Two correlators are applied:

- The *Proximity Correlator* checks proximity in space, by finding the nearest cluster in the track. It is always applicable, and the matching quality is the distance of the two cluster positions.

- The *Helix Correlator* checks the distance of the cluster to the prefitted *helix* that defines the track. If the track has not been fitted, the correlator is not applicable. The matching quality is the distance to the helix.

If the matching quality is smaller than a user-definable cut (*proximity-* and *helix-cut*), the track *survives* the correlator. These cuts are dynamically scaled, depending on a quality estimation of the so far existing tracklets. With better track quality, the *helix-cut* is narrowed, which gives better track separation power in areas of high cluster density, whereas the *proximity-cut* can be opened, which makes the process less prone to track splitting.

We call the number of applicable *and* survived correlators the *correlation level*. The hit is added to the track that reaches the highest level. If there is more than one track, the best matching quality decides to which track the hit is assigned. If no correlator is survived, a new track is built from the hit.

To avoid following the wrong track in an area where two or more tracks are crossing, clusters which match well to more than one tracklet can be excluded.

### 10.7.3 Helix Prefit

When a track has more than a user definable minimum number of hits, a helix fit [58] is performed in two steps:

- A plane fit on the Riemann sphere (see Sec. 10.7.1).

- A dip fit via the hit angles and z-positions.

A covariance ellipsoid is built from the residua of the hits to the average hit position in the Riemann space (weighted with 1/cluster error to be noise and outlier tolerant). This ellipsoid has three major axes, and the smallest axis (which is the eigenvector to the smallest eigenvalue of the covariance ellipsoid) is the normal vector to the plane the hits lie on.

With this simple method, problems arise especially for short tracks, which do not span a significant distance on the Riemann sphere and are therefore not curved sufficiently, and tracks where the clusters are not well aligned and show a wide spread on the measurement plane. Reasons for this can be delta electrons that broaden the track, noise hits which were assigned to the track and effects of the clustering.

In these cases often a plane is found that is almost perpendicular to the surface of the sphere. This is due to the fact that the plane fit minimizes distances of the hits to the plane, and not distances of the hits to the intersection of the plane with the sphere (i.e. the projection of the track). To prevent that, a check is performed which calculates the RMS of the distances of the hits to the intersection of the plane with the Riemann sphere for the planes corresponding to the two smallest eigenvectors. Then the plane with the minimum RMS is selected.

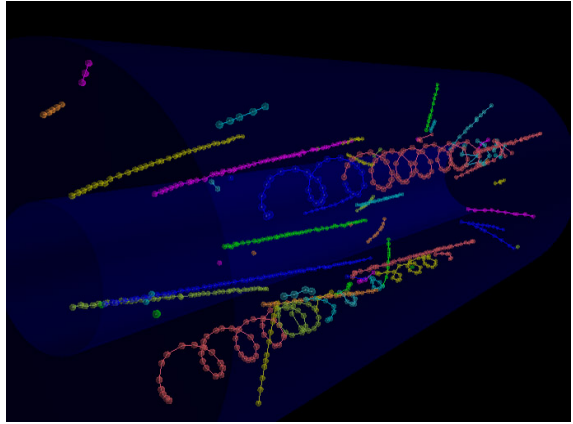

**Figure 10.8:** Performance of the Riemann pattern recognition for a real beam event taken with the prototype. Target tracks, curlers and steep tracks from interactions in the beam pipe are all recognized, merged and sorted properly.

This plane fit, projected back onto the pad-plane, delivers a circle, which has no constraints in radius and position of its center. It is the projection of the true helical track onto the readout plane.

In a second step, the winding sense and dip of the helix are fitted: The position of each cluster along the helix can be defined by its angle. The clusters are presorted by $z$. Iterating over all hits in the track, the angle of each hit is defined by the difference of its angle to the previous hit. This method grants consistent angles also for curling tracks.

A straight line fit of the hit angles versus the $z$ positions of the clusters is then performed, which delivers the dip $\vartheta$ of the track. After that, the hits are again sorted, now by their angle. For steep tracks ($\vartheta < 30°$ or $\vartheta > 140°$) the angles are not a good sorting criterion anymore, so these tracks are left sorted by $z$. This technique makes sure that the hits of the track are well sorted along the track, which is quite important for the Kalman filter. The winding ($\pm 1$) can simply be calculated by checking if the angle of the last hit in track is greater ($+1$) or smaller ($-1$) than the angle of the first hit.

### 10.7.4 Track merging

In the process of track building, the actual tracks might not be found as a whole. Especially steep tracks and curlers are prone to be split into several track pieces or *tracklets*: As the track-building goes from big to small radii, curlers cannot be found as a single track in one step. The same is true for steep tracks, where the sorting in radial direction does not reflect the track topology properly anymore. Besides, particles with low energy loss can lead to fragmentary tracks, or the track may exit and re-enter the chamber one or more times. In the real chamber, dead channels or chips can also cause gaps.

Therefore, a second level *tracklet merging* is performed. Similar to the track building process, the tracklets are presorted, and then compared to each other. Again, there are several *track-track correlators* which all (in this point the merging is different to the track building) have to be applicable and survived.

- The *Proximity Correlator* compares the position of the first and last hits of the two tracks. If the smallest distance is smaller than a definable *proximity-cut*, the correlator is survived.

- The *Dip Correlator* compares the dip angles of the two tracks. Therefore, both tracks have to be fitted. But not only has the absolute difference of the dip angles to be smaller than a definable *angle-cut*, also the relative z positions of the track have to match. Thus, the distance to the helix (defined by the track with more hits) of the nearest point of the smaller track is calculated. It has to be smaller than an adjustable *helix-cut*.

    If only one of the tracks is fitted, the helix-distances of all hits of the smaller track are calculated and compared with the helix-cut. If none of the tracks is fitted, the correlator is not applicable and the tracklets cannot be merged.

- Finally, the tracklets have to pass the *Helix Correlator*. For tracks with few hits, the helix fit might not be very accurate, and for straight tracks, some parameters of the helix (i.e. radius, center) are not well defined. Thus it is not reasonable to directly compare the helix parameters.

    Instead, a new track is created temporarily, containing the hits of both tracks. A helix fit is performed and a *helix-cut* on the RMS of the distance of the hits to the helix is applied.

    If the two tracks together do not have enough hits to be fitted, this correlator is not applicable.

The algorithm has been substantially improved over the last months. Track finding and merging works satisfactory for simulated as well as for real data, up to very high track counts. Fig. 10.8 shows the result of the pattern recognition in the large prototype (cf. Sec. 12) for a rather complicated target event including tracks from the primary vertex (end of the chamber in Fig. 10.8, low-momentum curlers as well as "spray" tracks almost parallel to $z$ from some event in the beam pipe.

### 10.7.5 Sectorization

The process of track building requires computational power in the order of $\mathcal{O}(n_{\text{clusters}} \cdot n_{\text{tracks}})$. In order to be performant also for large numbers of hits and tracks, the process is sectorized. The pad-plane is split into sectors. Track building and merging is done for each sector separately, and only then global merging is performed.

### 10.7.6 Multistep Approach

The pattern reconstruction efficiency for different track topologies depends strongly on the presorting of the clusters mentioned in 10.7.2. Performance is best for tracks in sorting direction (i.e. very steep tracks and very low momentum curlers for sorting along z, tracks with high transverse momentum from the interaction point for radial sorting, curling tracks for angular sorting). Thus, it is advantageous to run the pattern recognition more than once, and use a different presorting in each step. Tracks that reach certain quality criteria (i.e. a minimum number of hits and an RMS of distances of the hits to the helix smaller than a certain cut) are kept, the remaining clusters are sorted again, and the procedure is repeated.

This approach yields high efficiencies for all kinds of track topologies and a high track resolution power.

### 10.7.7 Track Seeding for the Fit

For track fitting, seed values for starting point, direction, momentum and charge have to be provided. These can easily be calculated from the helix prefit and the magnetic field. Moreover, tracks have to overcome a minimum number of hits in order to be passed to the Kalman filter, otherwise they are rejected.

### 10.7.8 Fast Hough Transform on the GPU

General purpose programming on Graphics Processing Units (GPUs) has been one of the hottest fields in parallel computing over the last years. Originating in the computer gaming industry, GPUs of today are versatile, conveniently programmable devices that offer massive *parallel* computing power.

Since the high level trigger of $\overline{P}ANDA$ is a performance critical system and the central tracker will play a decisive role as input, fast algorithms close to the hardware for pattern recognition are of great importance.

The Hough Transform [60] is a widely used, global method for pattern recognition. It is used for detection of a certain pattern in a given set of data points $X_i$. After choosing a fitting $N$-dimensional parameterization of a given pattern (e.g. a helix in the track finding problem at hand), every data point is transformed into the $N$-dimensional parameter space of that pattern parameterization. Consequently, every data point $X_i$ corresponds to an $N$-dimensional hyper-surface $p_i$ in the parameter space.

By construction, the hyper-surface representations of a set of data points lying *perfectly* on the chosen pattern *intersect in one single point* in the parameter space. In the more realistic case of a distribution of the $X_i$ around the pattern (think of the hits from a detector system with a finite resolution forming a track) one instead observes a region of elevated density of the $p_i$ in the parameter space. Consequently, the search for a pattern in the original data-set is transformed into a search for local maxima in an $N$-dimensional space.

A possible parameterization of a helix is

$$x = r \cdot \cos(t) + x_0$$
$$y = r \cdot \sin(t) + y_0 \qquad (10.3)$$
$$z = c \cdot t + z_0,$$

so in the case of helix detection $N = 5$. Here $t$ is the path *along* the helix curve and $c$ defines the pitch.

The search for maxima in a five-dimensional space is computationally non-trivial. A regular five-dimensional histogram in integer representation of moderate granularity would require memory allocation of the order of many terabytes. Alternatively, one can perform a tree-search in the parameter space instead, thus effectively trading memory load for computational cost. The recent developments in the field of massively parallel high-performance computing, however, make this kind of trade-off affordable.

The search implemented here works by iteratively dividing the parameter space into sub-spaces, so-called *nodes*, by bisection in each dimension [61]. In this way each node in the parameter space creates $2^5 = 32$ sub-nodes in every step of the algorithm. These nodes are then checked for intersection with the hyper-surfaces $p_i$. The number of hyper-surfaces $p_i$ crossing a node gives the node's *vote*. An (adjustable) threshold of minimal votes is required for every node after each iteration step, deciding if the node qualifies for further subdivision or is discarded. In its current implementation the algorithm terminates when a fixed iteration depth is reached. In the future this threshold should be determined for each track individually.

The algorithm has been implemented in `C++` and tested on simple simulated events. However, the calculations required lead to computation times $\mathcal{O}(10\,\mathrm{s})$ for an example of five tracks on recent hardware (Intel Core2$^{\mathrm{TM}}$ 2.66 GHz). This is due to the still very high number of nodes present (up to $\mathcal{O}(10^5)$) before converging on the maxima.

Since all of the nodes, the hyper-surfaces $p_i$, and hence intersection checks are completely *independent* of each other, the high number of nodes makes a massively parallel implementation attractive. Therefore the algorithm has been implemented in `C` and `C++` on a NVIDIA GPU using CUDA.

The implementation on the GPU has proven to be significantly faster than the identical code on the CPU (GPU hardware used for development and testing: NVIDIA GTX 285$^{\mathrm{TM}}$). Currently, in a very early version of the implementation, a Speedup factors of $\sim 20$ compared to single-thread performance on the CPU is observed. Without changes to the algorithm itself speed increases of another factor of $\sim 3$-5 should be reachable. The pure GPU calculation time only accounts to $\sim 5\,\%$ of the total computation time in the examples shown. Moving more tasks from the CPU to the GPU will lead to significant performance improvements.

### 10.7.9 Closing Comments

A strong feature of such an implementation is its scaling behavior. More tracks (e.g. more hits) will not lead to an increase of combinatorics and thus an increase of computational cost following some power law, as it would be the case for any algorithm performing hit-to-hit comparisons. Instead, increasing the number of hits just increases parallelism (linearly), resulting in a much more shallow increase of execution time, of course depending on the actual hardware setup.

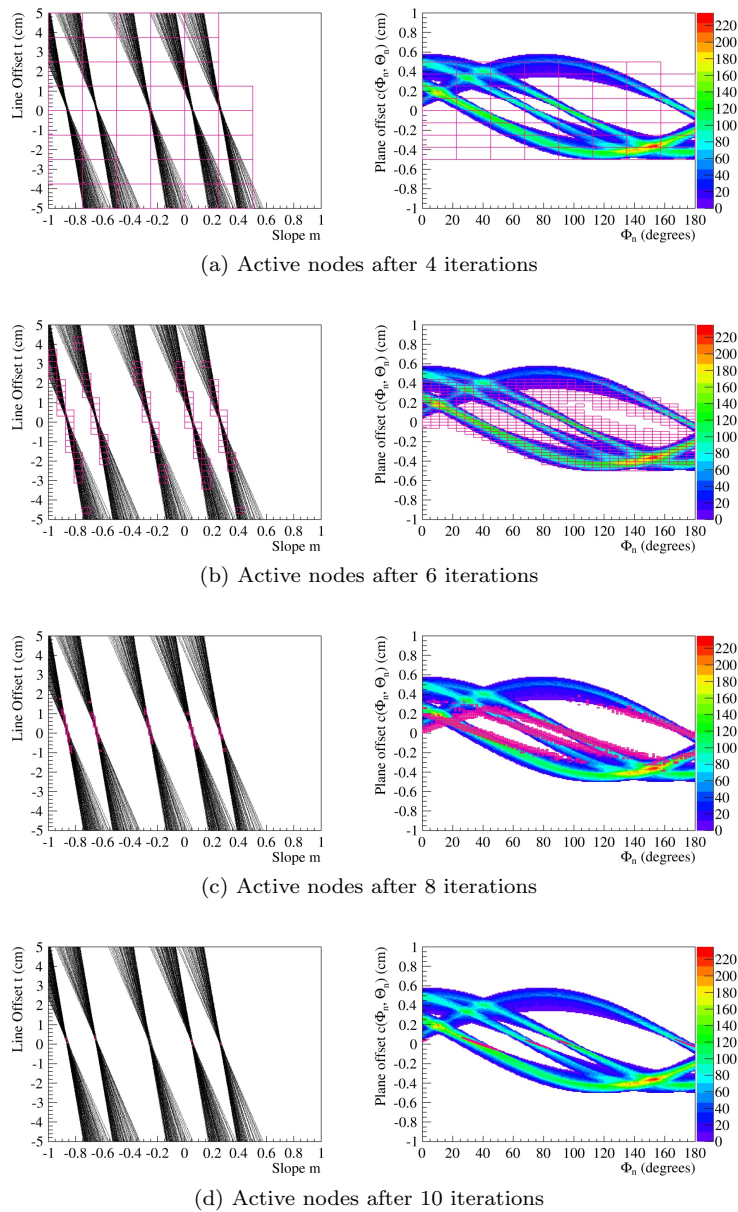

(a) Active nodes after 4 iterations

(b) Active nodes after 6 iterations

(c) Active nodes after 8 iterations

(d) Active nodes after 10 iterations

**Figure 10.9:** Tree search scheme of the FHT algorithm.

Exploiting the potential of massively parallel algorithms and hardware might prove to be crucial in order to keep computing times under control for the online trigger system of $\overline{\text{P}}$ANDA (cf. discussion of Sec. 8.5). The final pattern recognition scheme for the TPC could be a combination of a massively parallel algorithms and more conventional methods, like the very successful track following algorithm described in Sec. 10.7.2.

## 10.8 Track Fitting - GENFIT

After the pattern recognition tools have determined sets of detector hits which comprise particle trajectories, the best estimates for the track parameters, i.e. the particle positions and momenta with their covariances, have to be obtained. Space points measured by the TPC have to be fitted in combination with hits from other detectors with different geometries such as planar strip or pixel detectors in the Micro Vertex Detector of $\overline{\text{P}}$ANDA. This has motivated the development of a generic toolkit for track fitting in complex detector systems, called GENFIT [62], which is now the standard track fitting tool in

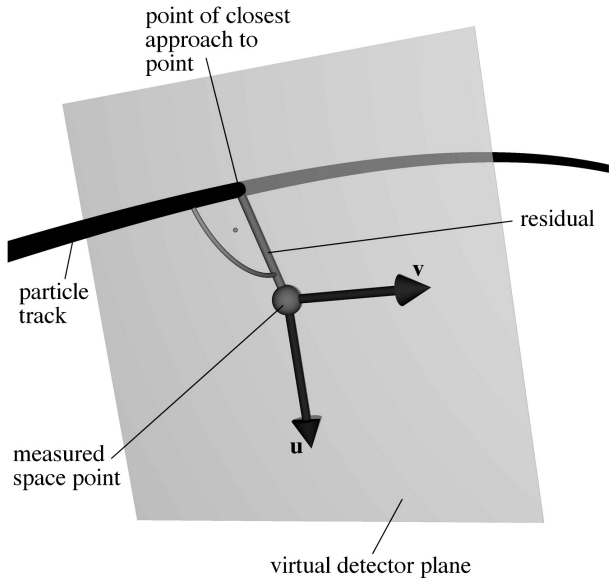

point of closest
approach to
point

residual

particle
track

v

u

measured
space point

virtual detector plane

**Figure 10.10:** Virtual detector plane (spanning vectors $u$ and $v$) for a space-point hit.

$\overline{\text{P}}$ANDA as well as in the Belle-II project. Applications in other experiments like ILD are currently under consideration.

The fact that GENFIT is applicable to a very wide range of experiments, independent of the specific event topology, detector setup, or magnetic field arrangement, is due to its completely modular design. Fitting algorithms are implemented as interchangeable modules. At present, the framework contains a validated Kalman filter [63]. The implementation of another algorithm called Deterministic Annealing Filter (DAF) [64] is currently ongoing. The DAF is an iterated Kalman filter which has the virtue of being able to dynamically assign reduced weights to noise hits in planar detectors or to outlier hits in the TPC. Other algorithms like Gaussian Sum Filters [65] can be implemented easily in the GENFIT framework.

Track parameterizations and the routines required to extrapolate the track parameters and their covariance matrices through the experiment are also implemented as interchangeable modules. This allows the use of well established track extrapolation tools (e.g. GEANE [66]) as well as the development and evaluation of new track extrapolation tools. Different track parameterizations and extrapolation routines can be used simultaneously for fitting of the same physical tracks, which allows a direct comparison in terms of execution time, resolution, and efficiency.

Representations of detector hits are the third modu-lar ingredient to the framework. The hit dimensionality and orientation of planar tracking detectors are not restricted in any way. Tracking information from detectors which do not measure the passage of particles in a fixed physical detector plane, e.g. drift chambers or TPCs, is used without any simplification. This goal is achieved via the concept of virtual detector planes, which are calculated dynamically each time a hit is to be used in a track fit. This allows to maintain complete modularity of GENFIT because the fitting-algorithm modules treat all hits in the same manner. In the case of space point hits in the TPC, the virtual detector plane is defined to be perpendicular to the track and to contain the point of closest approach of the track to the hit, as illustrated in Fig. 10.10. This allows the fitting algorithm to minimize the orthogonal distances of the track to the hits without projecting the hits onto predefined planes. The projection of hits onto planes defined by pad rows is common practice in TPC reconstruction.

GENFIT is implemented as a very light-weight C++ library, which is available as free software [67].

## 10.9 Tracking Performance

The following sections describe first results concerning the tracking performance of the $\overline{\text{P}}$ANDA TPC as evaluated with the present software framework:

- Reconstruction efficiency,
- Momentum resolution,
- Particle identification by $\mathrm{d}E/\mathrm{d}x$.

It is important to cross-check these results by experiment. For this purpose the large prototype has also been implemented in the simulation framework. A comparison of the resolution from experiment and simulation already shows very good agreement (cf. Sec. 12.2.1).

### 10.9.1 Single Track Reconstruction Efficiency

A detailed efficiency study of the Riemann pattern recognition has been performed for single-track events. A pad-plane with 16 segments was used. Tracking performance is excellent for tracks between $25°$ and $145°$, as can be seen in figures 10.11 and 10.12.

Scalability to large track multiplicities up to 3000 in terms of tracking performance and computational

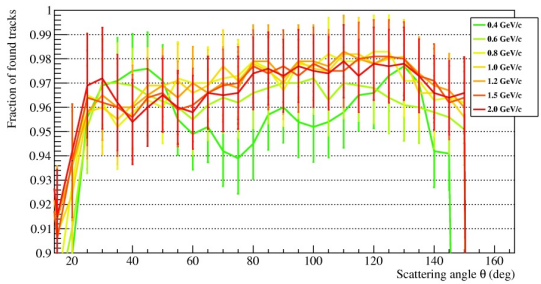

**Figure 10.11:** Fraction of found tracks (if a tracklet contains more than 50% of the clusters that belong to the track, it is found) vs. scattering angle $\theta$ for single track events. Regardless of the momentum, more than 94% of the tracks are found. Only for very steep tracks performance decreases significantly.

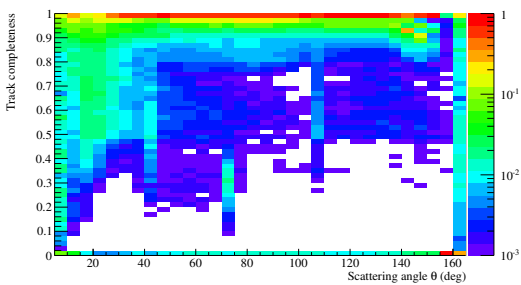

**Figure 10.12:** Track completeness (defined as the fraction of clusters of the track that are found in the biggest tracklet) vs. scattering angle $\theta$. This plot combines single track events with momenta from 0.2 to 2.0 GeV.

cost still has to be evaluated in detail, but first preliminary tests look very promising.

### 10.9.2 Momentum Resolution

The $\overline{\text{P}}$ANDA Central Tracker (CT) has to be able to reconstruct momenta of charged tracks with a resolution in the low percent regime [24]. To project the momentum resolution performance of the final $\overline{\text{P}}$ANDA TPC, an extensive simulation study has been performed. For a range of momenta especially relevant for the CT, $\mu$ tracks have been simulated leaving the target at a set of fixed angles $\theta$, undergoing the full simulation chain as described in Sec. 10.1.

Track extraction has been performed based on the available Riemann Pattern Recognition (c.f. Sec.10.7.2) without using any Monte Carlo information. The extracted tracks were fitted using GEN-FIT with Runge Kutta track representation (c.f. Sec. 10.8), immediately yielding the distribution

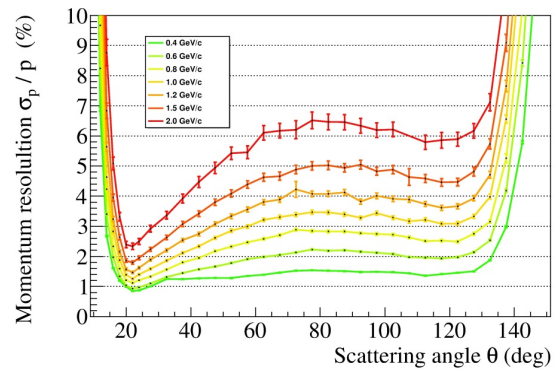

**Figure 10.13:** Momentum resolution $\sigma_p/p$ obtained from $\mu$-tracks reconstructed in the TPC alone. Each data point corresponds to a Gaussian fit to the momentum distribution as reconstructed from 5000 tracks. The error bars show the squared sum of the statistical uncertainties of sigma and mean of the final fit.

of reconstructed momenta.

From the Central Tracker point of view, for charged particles originating from the primary interaction point in the target, track reconstruction and hence determination of the particle momentum will greatly benefit from the high-precision space points measured in the Micro Vertex Detector (MVD) of $\overline{\text{P}}$ANDA. However, for charged tracks from secondary vertices outside the MVD acceptance or tracks with insufficient number of hits in the MVD, the CT alone will have to be able to reconstruct tracks and their momentum sufficiently well. It is thus important to assess the momentum reconstruction capabilities of the TPC both with and without MVD (or GEM) hit data.

Figure 10.13 shows the momentum resolution obtained with reconstructing tracks in the TPC only without relying on any external detector information. In this study only primary $\mu$ tracks were considered. Secondary tracks, as well as all tracks not meeting the a required *completeness* of 50 % (c.f. Sec. 10.9.1).

It is important to note that Fig. 10.13 shows the reconstruction performance as a function of the *total momentum* $p$, not the transverse momentum $p_t$. The total momentum is reconstructed during fitting taking into account the measured curvature as well as the helix dip. This is possible because the TPC measures consistent three-dimensional tracks also without relying on any external references.

A second study has been conducted in order to project the performance of the TPC together with the MVD and the GEMs. The reconstruction scheme was the following:

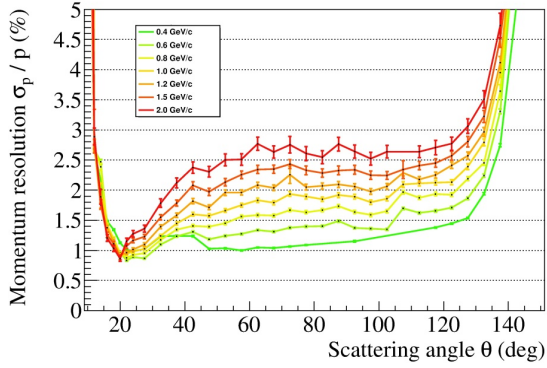

**Figure 10.14:** Momentum resolution $\sigma_p/p$ obtained from $\mu$-tracks reconstructed in the TPC and then merged with hits from the MVD and GEM detectors, using the same data sample as in Fig. 10.13. In each MVD (GEM) detector plane the closest hit inside a road width of 2 mm (5 mm) was added to the TPC track. No Monte Carlo information has been used for the matching.

- find tracks in the TPC only using real Riemann pattern recognition,

- fit the found tracks,

- extrapolate the tracks through the MVD,

- pick up MVD and GEM hits inside a certain road width around the track,

- re-fit the combined track.

The result of this method is shown in Fig. 10.14. The momentum resolution improves by a factor of $\sim 2$ when adding MVD hits to the TPC hits, as can be seen by comparing to Fig. 10.13.

## 10.10 Reconstruction of $\Lambda$ Decays

The Central Tracker of $\overline{P}$ANDA will have to be able to reconstruct events with complex topologies without the help of other detectors, e.g.

- decays of neutral particles, e.g. $\Lambda$ hyperons,

- kinks in charged particle tracks, e.g. from $\Xi$ decays,

- $\pi \to \mu$ decays (important for background rejection in the muon range system of $\overline{P}$ANDA).

As an example for the reconstruction performance of the TPC for such topologies, e.g. the invariant mass resolution, the vertex resolution and the reconstruction efficiency of secondary vertices, we investigate here the reaction

$$p\overline{p} \to \Lambda\overline{\Lambda} \to p\pi^- + \overline{p}\pi^+ \quad . \quad (10.4)$$

To this end, 10000 events of this reaction were simulated at a $\overline{p}$-beam energy of 4 GeV. This was done using the EvtGen direct event generator with the $\Lambda\overline{\Lambda}$ generated over the whole phase space (PHSP flag). The present simulation does not take into account the real angular distribution of $\Lambda$ and $\overline{\Lambda}$ from this reaction.

The kinematics of the reaction products of the simulation can be seen in Fig. 10.15. A strong forward boost is seen especially for the protons and antiprotons.

In Fig. 10.16 the number of primary particles leaving at least one MC point in the TPC is shown. This gives an upper limit for the geometrical acceptance of 52 %.

The number of reconstructed charged candidates can be found in Fig. 10.17; 30.5 % of the events have exactly two positive and two negative charged candidates.

For the analysis of this channel MC-based particle identification (PID) is used (e.g. the particle hypothesis is set to the one from the associated MC track). For each reconstructed event, lists are built of the track candidates for protons, $\pi^-$, antiprotons and $\pi^+$.

For further analysis, only events are considered which have at least one $\Lambda$ and one $\overline{\Lambda}$ candidate. For all combinations of protons and $\pi^-$ (antiprotons and $\pi^+$) the point of closest approach between the two tracks is calculated using a Newtonian method on only the available fits of the particle candidates. The center point of the line connecting the two points of closest approach on the tracks is taken as the vertex position. It should be stressed that no assumption on the vertex position is made for the reconstruction. In Fig. 10.18 one can see the distance from the calculated decay vertex to the MC vertex for all the different combinations where the MC vertex is taken as the creation vertex of the proton (antiproton). Resolutions of 0.4 mm and 1 mm are achieved for the $x/y$ and $z$ coordinate, respectively.

The tracks are then extrapolated to this vertex and the 4-momenta at this point are used to create a $\Lambda$ candidate. The candidate where the distance between the two tracks is the smallest is chosen. One can see the invariant mass distribution of the $\Lambda$ ($\overline{\Lambda}$) in Fig. 10.19. Within 3 $\sigma$ of the mean fitted to the mass spectrum in total 3592 $\Lambda$ $(\overline{\Lambda})$ were reconstructed with an invariant mass resolution of about 2 MeV. It should be noted that no kinematic vertex fit has been performed yet, which is expected to further improve these values.

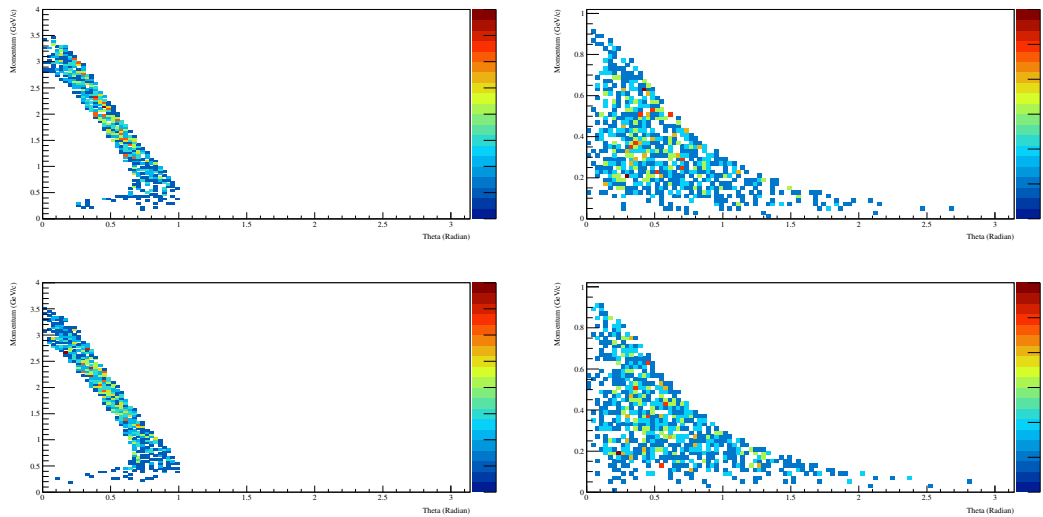

**Figure 10.15:** Kinematic distributions (momentum vs. laboratory polar angle) of the decay particles from $\Lambda/\overline{\Lambda}$ decays. Top left: protons, top right: $\pi^-$, bottom left: antiprotons, bottom right: $\pi^+$.

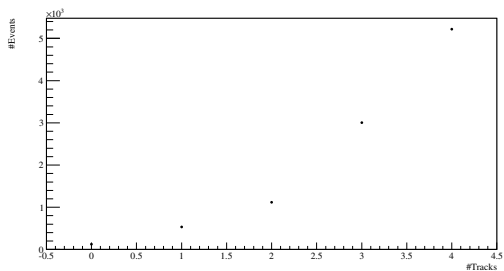

**Figure 10.16:** Number of primary tracks with at least one MC point in the TPC.

The global reconstruction efficiency of this process is 18 %. Taking into account a detector acceptance of 52 % the final TPC reconstruction efficiency is given by 34 %.

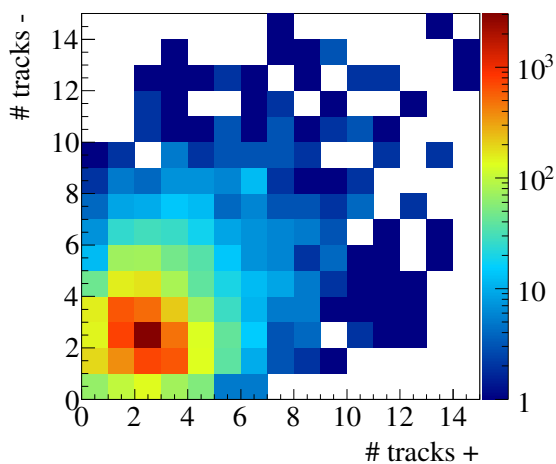

**Figure 10.17:** Number of reconstructed charged particle tracks.

## 10.11 Particle Identification by dE/dx

Charged particles can be identified with the help of their momenta and their specific energy loss. A TPC can measure both parameters and is therefore able to perform a complete particle identification (PID). The average specific energy loss of a particle crossing matter represented by the primary ionization is described by the Bethe-Bloch formula. It is based on three assumptions: First, the transfer of energy does not change the direction of flight of the ionizing particle, second, the gas molecules are at rest and third, the ionizing particle is much heavier than an electron. A modified version of the Bethe-Bloch formula, which makes the dE/dx behavior directly readable, can be written as a function of

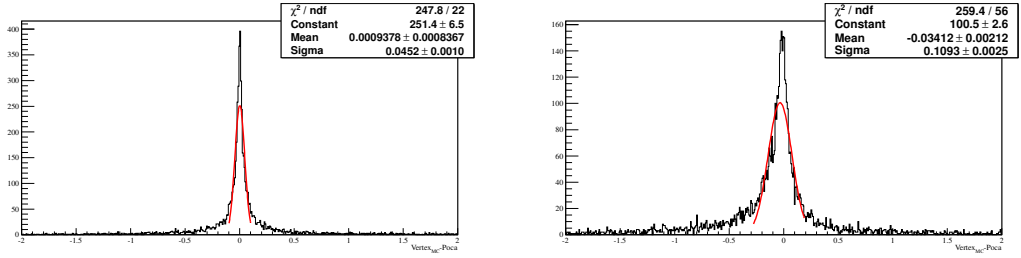

**Figure 10.18:** Vertex resolution for $\Lambda/\overline{\Lambda}$ decays in the $\overline{\text{P}}$ANDA TPC.

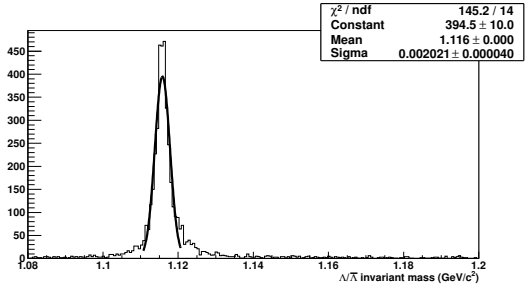

**Figure 10.19:** Reconstructed $\Lambda/\overline{\Lambda}$ mass.

the particle velocity $\beta$ and the charge number Q:

$$\frac{dE}{dx} = \xi \cdot \frac{1}{\beta^2} \cdot Q^2 [K + \ln(Q^2) + \ln(\gamma^2) - \beta^2 - \delta(\beta\gamma)] \quad (10.5)$$

In this equation $K$ is representing a constant and $\xi$ the electron density of the gas, while $\delta(\beta\gamma)$ denotes the density function, first introduced by Fermi. This correction accounts for polarization effects due to the electrical field of the relativistic particle. The specific energy loss as function of the momentum can be subdivided in three regions. At low momenta and therefore non-relativistic velocities a decrease with $1/p^2$ is visible, which causes large ionization. For momenta corresponding to three to four times the mass of the particle, a minimum is reached, followed by the relativistic rise proportional to $\ln(p^2)$. The relativistic rise is an effect of the deformed electrical field of the ionizing particle, leading to an increase of the transverse component of the field. Saturation is reached for very large values of $\ln(\beta\gamma)$ at the Fermi plateau. There the relativistic rise ends and the energy loss becomes independent from $\beta\gamma$. The most interesting part in $\overline{\text{P}}$ANDA would be in the transition of the region from low momenta to the the relativistic rise. Therefore the resolution of dE/dx measurements has to be at the level of a few percent in order to give a handle for particle identification in this momentum regime. By measuring hit charges, using gain correction from the krypton calibration and applying the truncated mean method, the energy loss of a track can be determined. With the additional information of the momentum calculated from the track curvature the particle can be clearly identified.

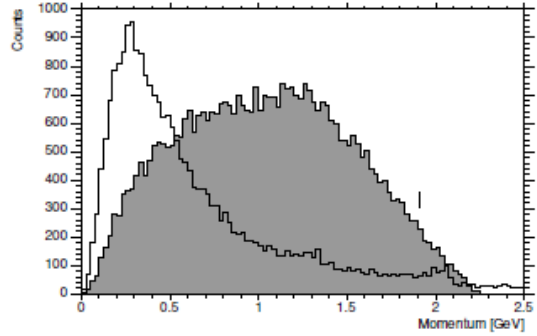

**Figure 10.20:** Momentum distribution of kaons from $\eta_c \Rightarrow 4$K decay (dark plot) and background pions (transparent) [24].

Figure 10.20 shows the benefit of particle identification below 1 GeV/$c$ for the $\overline{\text{P}}$ANDA central tracker. It shows the momentum distribution of kaons produced by the $\eta_c$ decay. Within the same figure the momentum distribution of background pions is shown, generated at the $\eta_c$ resonance. About half of the kaons have momenta below 1 GeV/$c$, where also most of the background pions are produced. Particle identification with with Cherenkov detectors can not be used at such low momentum values. The TPC would therefore add significant information for the PID.

A simulation study has been performed to study the motivated momentum region below 1 GeV/c in regard of the separability of kaons and pions [55]. Therefore three large samples ($\pi^+$, $K^+$, $p$) of 100.000 tracks have been produced each using the full $\overline{\text{P}}$ANDA root framework. A particle gun generator was used to distribute both scattering an-gle and momentum ($p \in [0.2, 1.2]$ GeV/$c$) homogeneously. The full digitization and reconstruction chain as outlined in the sections 10 has been used.

During reconstruction a track fit is performed using a Kalman filter (embedded in GENFIT), giving the reconstructed momenta. Energy loss is then extracted by a dEdx task. The track fit is used to "walk" along the particle's reconstructed trajectory in steps of fixed length $\Delta x$. At each step a plane is constructed perpendicular to the track and all TPC hits lying between this and the last plane are collected. These can either be the clusters obtained at the very end of the reconstruction or MC points.

In the first case the deposited energy $\Delta E$ over the length $\Delta x$ has to be inferred from the `Cluster` signal amplitudes. When working on MC points, the energy loss can be directly obtained. Whatever the method, the single values will be distributed around the mean value given by the Bethe-Bloch Formula.

In the results presented here the hits are collected over track pieces of a fixed length $\Delta x = 3$ mm, resembling the pad size on the TPC readout plane. The single energy deposits associated to the hits of one such step will be distributed according to straggling functions, which have been optimized on ALICE TPC data. However, the number of points will be small. Remembering the long tail of the straggling functions it is clear that this potentially leads to large fluctuations of the mean value.

A commonly used and robust approach to reduce the dependencies of these fluctuations is the truncated mean method: A certain fraction of the highest and/or lowest values is simply discarded. In this way one reduces statistics but on the other hand keeps the fluctuations of the mean values to a minimum, as the outliers from the underlying distribution's tail are likely to be completely omitted. Of course one has to take care of the systematic shift of the mean value that is introduced.

Figure 10.21 shows the results of the MC-based study. In the scatter plot three distinct bands $\frac{dE}{dx}(p)$ are visible, completely overlapping for $p \gtrsim 1200$ MeV/$c$. For lower momenta an efficient identification of either pion, kaon or proton seems possible.

Energy resolutions have found to be fairly good for low momenta. Example distributions and resolutions are given in Fig. 10.22 for a fixed momentum of $p = 400$ GeV/$c$, where contribution to PID of the TPC will be possible. When going to higher momenta (towards minimum ionization), the resolution decreases to $\sim$ 9-11 %.

An MC-based energy loss study was chosen in or-der to acquire a theoretical limit for the TPC subdetector. A similar study on reconstructed hits can in principle be done without modifications of the dE/dx-code itself.

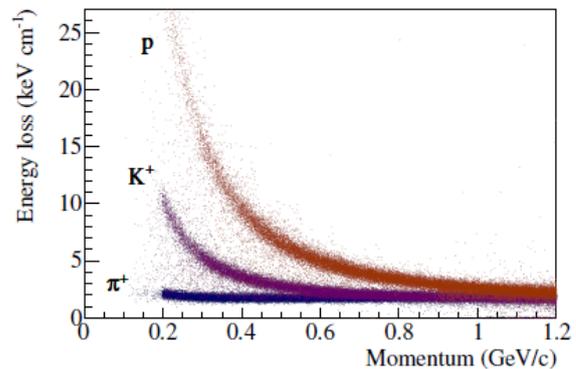

**Figure 10.21:** Scatter plot of dE/dx samples from Monte Carlo simulations of $\pi^+$, $K^+$ and protons at scattering angle $\Theta = 90°$. Each point consists of the reconstructed momentum from the Kalman tracker and the mean of the dE/dx distribution from 3 mm samples of the track. The truncated mean method has been used to suppress very high energy loss values (the highest 40 % are discarded). For each particle type $\sim 10^5$ tracks were included in the study [55].

However, the current implementation of the signal processing algorithms lacks an adequate pulse shape analysis: Only a local-minimum approach is currently available to disentangle pile-up signals on the readout pads. This is sufficient for tracking, where only the time and space information of a reconstructed hit is required. For dE/dx-analysis, however, one needs the signal height, which is proportional to the energy loss over the given pad. Therefore it is not only required to separate signal peaks in time, but also to subtract the signal's tail from the following/previous signal. Once these tail cancellation algorithms are in place, the same study can be repeated for fully reconstructed simulation data.

## 10.12 Effect of Space Charge

Accumulations of space charge in the active volume of the TPC lead to distortions of the drift field and hence the drift paths of the electrons to be detected on the readout plane. In 1 we explained how the application of GEM foils as gas amplification devices can help to minimize the feedback of slow ions into the drift volume, and how this suppression can be characterized.

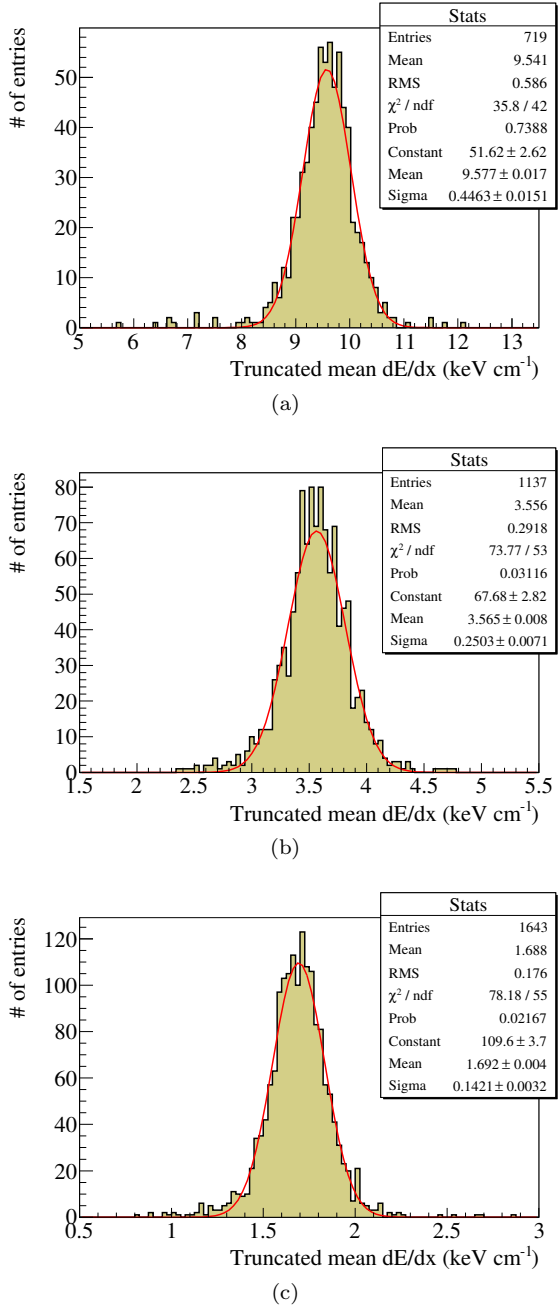

(a)

(b)

(c)

**Figure 10.22:** Distribution of mean dE/dx values obtained from analyzing 2000 tracks (truncated means, upper 40 % of values discarded) for pions (a), kaons (b) and protons (c) at $p = 400\,\text{GeV}/c$ each. Scattering angle of the tracks has been restricted to $\Theta = 30°$, resulting in a close-to-optimal track length. Gaussian fits have been applied. The corresponding energy loss resolutions are 8.3 % (a), 7.0 % (b) and 4.7 % (c) [55].

For simulating the effect of the remaining space charge we conveniently chose the *ion back-flow factor* $\epsilon$ as single input parameter, giving the number of back-drifting amplification ions per incoming primary electron. In the course of the studies presented here $\epsilon = 4$, which is equivalent to a *suppression factor* $\eta = 0.25\,\%$.

### 10.12.1 Simulation of Space Charge Buildup

For a realistic assertion of the projected space charge accumulation in the $\overline{\text{P}}$ANDA TPC we start with a sufficiently large set of events $S$ with sample size $s$ simulated using the Dual Parton Model (DPM) background generator for $\bar{p}p$-reactions. After passing this input data through GEANT (c.f. Sec. 10.1), we register the combined energy deposit of the full set of background events in the TPC active volume $\rho_i(r, z, \phi)$. Assuming instantaneous electron drift, we can regard this as the distribution of primary ion space charge.

In general, at time $t_1$ the charge density $\rho(t_1, r, z, \phi)$ in a volume element $dV$ centered around a point $(r, z, \phi)$ is given by

$$
\begin{aligned}
\rho(t_1, r, z, \phi) = \frac{e}{dV} \cdot \int_0^{t_1} & N_p(t, r, z, \phi) \\
& + N_{in}(t, r, z, \phi) \\
& - N_{out}(t, r, z, \phi)dt
\end{aligned}
\tag{10.6}
$$

where $N_p(t, r, z, \phi)$ is the rate of *primary* ions created in $dV$ and $N_{in}(t, r, z, \phi)$ and $N_{out}(t, r, z, \phi)$ are the rate of ions drifting into and out of the volume element respectively.

In addition, the back-drifting ions from the amplification stage at the readout side of the chamber are modeled by adding a contribution proportional to the projection of the total amount of newly added primary charge onto the readout plane (assuming instantaneous electron drift):

$$
N_{\text{in}}(t, r_i, z_0) = \epsilon \cdot \sum_j N_{\text{p}}(t, r_i, z_j).
\tag{10.7}
$$

Such a "template space charge distribution" from 10000 DPM background events is shown in Fig. 10.23a.

It has to be noted at this point that the shown simulations have been done using an old geometry model with insufficient detector wall thickness. In this light the following results should be regarded as an upper limit, as the correct wall material description can be expected to absorb some part of the incoming particles in the low momentum region. This might have a visible effect especially on the slow, elastically scattered protons which in turn dominate the energy deposition in Fig. 10.23a.

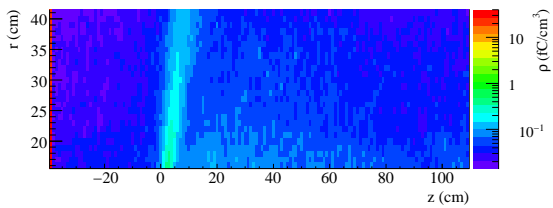

(a) Template distribution of ion space charge from 10000 DPM background events. The strong band leaving the interaction point (0,0,0) under $\sim 90°$ is due to slow, elastically scattered protons. The red band at the gas amplification at $z \sim -40$ models the amplification ions feeding back into the drift volume, assuming instantaneous electron drift.

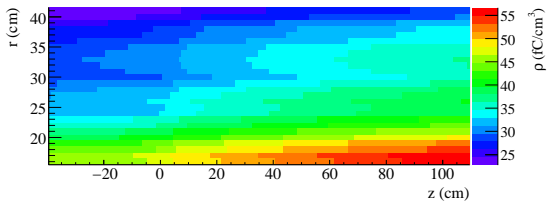

(b) Equilibrium space charge in the TPC chamber assuming azimuthal symmetry of the deposited primary charge.

**Figure 10.23:** Simulated initial space charge distribution from one time frame $t_1$ (a) and final equilibrium space charge after time integration (b).

We choose the sample size $s$ such that the associated accumulation time $t_1$ (given by $s$ divided by the interaction rate $2 \cdot 10^7 \, \mathrm{s}^{-1}$) is small compared to the ion drift time inside the TPC ($T_{ion} \sim 85 \, \mathrm{ms}$). Then we can regard the obtained initial space charge distribution $\rho_i(r, z, \phi)$ to be independent of the ion drift. In addition, the following assumptions are made to simplify the simulation of the space charge distribution in the TPC:

- Constant luminosity: We assume that $N_p$ is constant on time scales of interest to us.

- Azimuthal symmetry: Owing to the typical cylindrical geometry of a TPC chamber we will treat the problem in cylindrical symmetry so that the resulting charge density map can be represented in the $(r, z)$-plane and we drop the dependency on $\phi$: $\rho(r, z, \phi) = \rho(r, z)$.

- For simplicity we neglect the electrostatic forces between the ions themselves. In this model ion drift proceeds along straight lines with constant velocity $\mathbf{u}_{drift}^{I^+}$.

### 10.12.2 Modeling the Ion Drift

To represent the volume elements of 10.6 we choose a fixed binning in both coordinates $(r, z)$. Each bin thus describes a ring-shaped volume in the TPC. The bin width $\Delta z$ is chosen to correspond the accumulation time $t_1$ of the sample $S$. The actual time integration is then performed by repeated bin-wise shifting of the total charge distribution while adding the template charge distribution of Fig. 10.23a at the same time, until equilibrium is reached. The final equilibrium space charge map is shown in Fig. 10.23b.

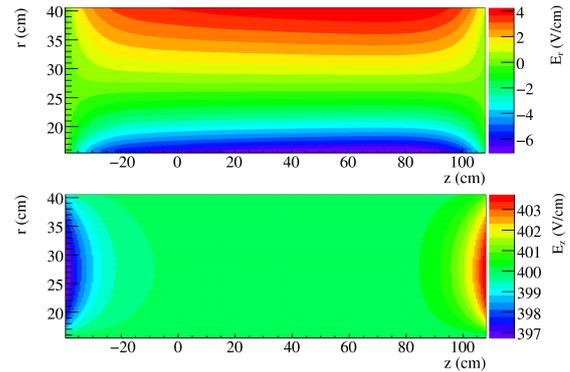

**Figure 10.24:** Distorted TPC drift field as calculated with a finite element software from the equilibrium space charge as shown in Fig. 10.23b. While effects in drift direction are small and localized at the drift cathode and readout plane respectively, there is a persistent field gradient along the radial direction.

### 10.12.3 Effect on the Drift Field

From the equilibrium space charge map in $(r, z)$ (Fig. 10.23b) one can obtain the corresponding electrostatic distortion field. Figure 10.24 shows the result of a finite element calculation (DOLFIN), where the resulting field has already been superimposed with the ideal drift field of $400 \, \mathrm{V \, cm^{-1}}$.

### 10.12.4 Resulting Electron Drift Distortions

The important question to answer is how these projected field distortions will affect the drift of electrons towards the readout plane. It can be answered by solving the equation of motion of electrons in the full $\mathbf{E}$ and $\mathbf{B}$ field configuration (c.f. [10]):

$$m\frac{d\mathbf{u}}{dt} = e\mathbf{E} + e[\mathbf{u} \times \mathbf{B}] - K\mathbf{u} \quad (10.8)$$

Equation 10.8 is integrated with a fourth order Runge Kutta algorithm. The obtained drift paths for electrons of different starting positions are then compared to ideal straight line drifts as expected

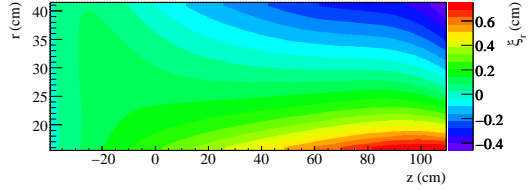

(a) Total drift distortions in radial direction (cm)

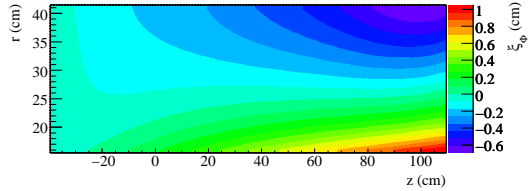

(b) Total drift distortions in azimuthal direction (cm)

**Figure 10.25:** Drift distortions of electrons in inhomogeneous $\mathbf{E} \times \mathbf{B}$ field. The graphs show the deviations from a straight line drift experienced by an electron which start their drift at points $(r, z)$ in the TPC.

for an ideal drift field. Figure 10.25 shows a map of these deviations as a function of the starting position of the drift. Absolute offsets of up to 1 mm can be observed in certain regions of the chamber. Deviations along the drift axis are not shown, as they turn out to be negligible.

It should be noted that although both fields are assumed to exhibit cylindrical symmetry there are nevertheless drift distortions perpendicular to the $(r, z)$-plane. This is due to the well known $\mathbf{E} \times \mathbf{B}$ term in the solution to Eq. 10.8.

### 10.12.5 Recovery of Drift Distortions

In order to recover the distortions discussed in Sec. 10.12.4, they have to be measured directly during the operation of the TPC. A possible solution would be to artificially create a well defined pattern of tracks in the TPC and analyze its image as measured by the detector. A comparison with the expected image directly yields the distortions.

Straight line ionization tracks from UV-lasers can provide such a pattern and have already been used for calibration in other drift chambers (e.g. in STAR [31]). To assess the potential of this method we modeled a laser system in the simulation framework. The laser beams are parametrized as straight line ionization tracks with constant ionization density and a Gaussian beam profile. Figure 10.26 shows an example of primary ionization along the laser rays in the simulation. This example laser grid geometry has been chosen under the requirements of complete chamber illumination and minimal number of track crossings, but is has not been optimized otherwise.

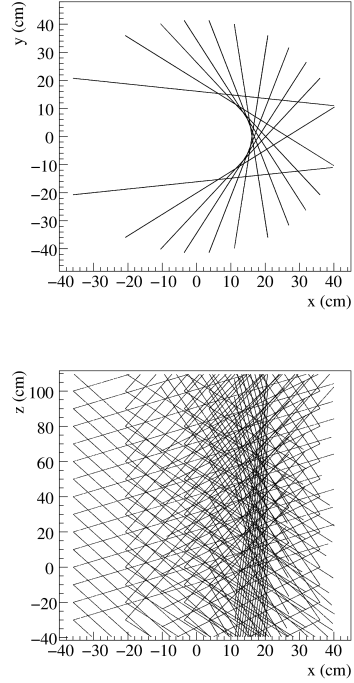

**Figure 10.26:** Primary ionization caused by laser tracks in the TPC volume as seen from the top of the chamber (a) and from the side (b).

The reconstruction of the laser tracks is greatly simplified by the knowledge of the geometry. For each laser-hit two residuals are obtained in the $x$-$y$-plane (assuming the point of closest approach to be exactly the true origin of the measured cluster): In radial direction and, perpendicular, in $\phi$ direction. Residuals in $z$ are ignored. The introduced error is small due to the small variation of the distortion maps along the $z$-axis (c.f. Fig. 10.25). Always the nearest available track is chosen for residual calculation of a given detector hit.

### 10.12.6 Quality of Recovery

The obtained 2-dimensional raw data is then fitted and smoothed by a bi-cubic spline fit. Figure 10.27 shows an example of a reconstructed drift distortion map obtained from the laser-fitting algorithm in the spline representation. The grid of Fig. 10.26 has been used, with a Gaussian laser beam profile ($\sigma = 400\,\mu m$) and an ionization density of 40 e$^-$ cm$^{-1}$, which is a conservative setting.

To be able to judge the quality of the result shown in Fig. 10.27, we compared the reconstruction results to our original drift distortion input (Fig. 10.25), as shown in Fig. 10.28. This distribution does not directly give the principle reconstruction accuracy of the described method, but it also contains intrinsic offsets and uncertainties due to the different representations of the linearly interpolated input deviation make and the reconstructed spline fit. It therefore should be regarded as an upper limit.

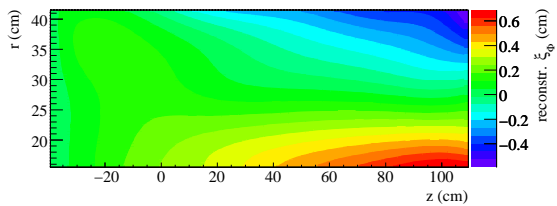

**Figure 10.27: Reconstructed** drift distortions in azimuthal direction based on one laser event (c.f. Fig. 10.25b).

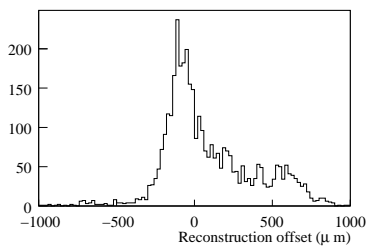

**Figure 10.28:** Distribution of total reconstruction offset compared to original distortion. **Mean:** 86.4 μm; **RMS:** 292.1.

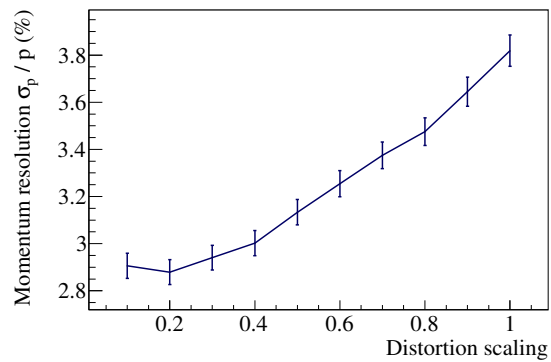

**Figure 10.29:** Effect of distortions with different scaling settings on momentum resolution.

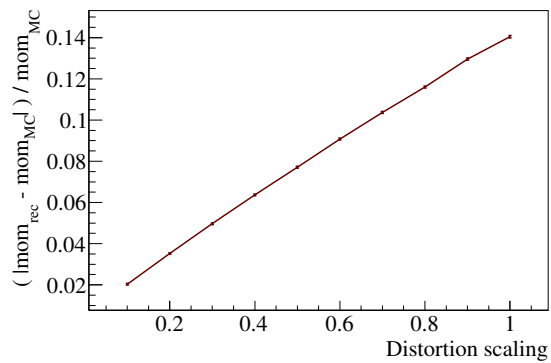

**Figure 10.30:** Relative error of reconstructed momentum compared to Monte Carlo value as a function of distortion scaling.

Applying a Gaussian fit to the peak of the distribution (including the shoulder but neglecting the "background") lets us estimate the possible accuracy of the recovery method. It yields $\sigma_{\text{Gauss}} \sim$ 180 μm and a mean of $\sim$ 80μm, showing that the total systematic error of the laser correction method is small and the overall precision of the presented distortion reconstruction and fitting method is slightly better than the aspired spatial resolution of the P̄ANDA TPC.

Such fit quality studies have been conducted for several scenarios (spline parameters, laser geometry), all yielding similar results. It is thus safe to say that we are able to reconstruct drift distortions with a precision better than $\mathcal{O}(300\,\mu\text{m})$, not taking into account mechanical distortions of the calibration system.

To estimate the momentum reconstruction and resolution as a function of the magnitude of present distortions, a sample of 1.0 GeV/$c$ $\mu$-tracks originating from the interaction point was simulated. The "scattering angle" $\theta$ was evenly distributed over the full CT acceptance. From the reconstructed tracks the momentum resolution as well as the resulting momentum offset due to the distortions can be studied (c.f. Fig. 10.29 and Fig. 10.30).

## 10.12.7 Application in the Simulation Framework to Correct Distortions

The results of the previous sections enable us to study both the impact of present space charge effect as well as their correction on track reconstruction within our simulation framework. The correction takes place directly inside the reconstruction part of the simulation chain, by applying shifts of the cluster coordinates in accordance with the obtained spline fit from the laser track data just before handing the tracks over to the fitting algorithms.

To study the effect of drift distortions on physics events, we simulated a test sample of 1000 pion tracks ($\pi^+$) at a momentum of 0.5 GeV/$c$ and uniformly distributed scattering angle $\theta$. Figure 10.31 visualizes the effect of uncorrected (red) and corrected (green) drift distortions compared to the ideal situation of completely homogeneous electrical field (black).

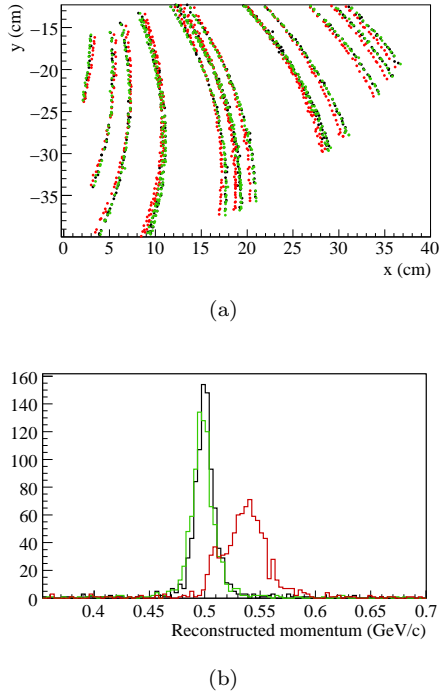

(a)

(b)

**Figure 10.31:** Effect of drift distortions and their correction for $\pi^+$- tracks at 0.5 GeV/$c$ momentum. Black represents data from an ideal simulation with perfectly homogeneous drift field. The track data subjected to uncorrected space charge effects is shown in red, the corrected case is displayed in green. In (a) the effect on clusters as reconstructed on the readout plane is shown. Displacements of several mm are visible with the naked eye, as well as a tilting/stretching of tracks. The effect on momentum reconstruction is imaged in (b): Space charge effects distort and shift the spectrum of reconstructed momenta, the correction method is able to restore a close-to-ideal distribution.

In Fig. 10.31a clusters in a small area of the TPC readout-plane for the three different scenarios are shown for a few selected events. Displacements of the order of several mm are clearly visible in the uncorrected case (as expected), while the correction moves them back on the ideal tracks nicely.

Figure 10.31b illustrates the large impact of drift distortions on momentum reconstruction. The initial distribution (black line) of reconstructed momenta is significantly deformed and shifted. Both effects can be understood within the scope of the preceding sections: Depending on the scattering angle of the individual track, it is more or less affected by drift distortions (compare to Fig. 10.25), leading to the broadening of the distribution. The asymmetric nature of the drift distortions with respect to the radial coordinate causes the measured curvature of the track to appear smaller than it actually was, leading to the shift to higher momenta. However, the correction method described in this section is able to fully recover the original position and shape of the distribution. A Gaussian fit of the ideal and corrected case reveals the difference of ideal and corrected case to be smaller than 1 % for both sigma and mean, respectively.

### 10.12.8 Final Remarks

We can conclude that we successfully showed the general principle of drift distortion correction using a grid of laser beams traversing the active gas volume of the TPC on simulation basis. Mechanical feasibility has to be studied, as well as the impact of structural uncertainties on the reconstruction quality and accuracy.

Alternatively, a simpler concept would be to just measure the *integrated* drift distortions over the full drift length and recover the $z$-dependence based on simulation models. As the distortions show a rather smooth structure along the drift direction (c.f. Fig. 10.25), this might prove to be a sufficient and mechanically much less challenging alternative.

As closing comment, a more detailed discussion of the presented simulation can be found in [55] and [68].

## 10.13  Event Deconvolution - Monte Carlo Studies

The time an electron needs to cover the complete drift distance in the TPC (a so called *drift frame*) is approximately 50 µs, depending on the gas mixture and the applied drift field. At an $\bar{p}p$ annihilation rate of $2 \cdot 10^7$ events per second, with an average time spacing of 50 ns between two events, thus approximately 1000 annihilations happen during one drift frame.

In order to reconstruct exclusive channels, it is essential to be able to assign the recorded signals in the central tracker to the correct physical event. To investigate this issue, a Monte Carlo simulation has been done. The individual events have been generated using the standard DPM $\bar{p}p$ background Monte

Carlo event generator. In order to simulate the time distribution of the signals in the TPC these events have been mixed with an exponential distribution for the inter-event spacing, which corresponds to an annihilation rate of $2 \cdot 10^7 \, \mathrm{s}^{-1}$. For each physics event 1000 background events have been added. For the physics events (only) also the information of the MVD has been retained in the simulation. Here we study the two cases where tracks originate from the primary vertex or where neutral particles (such as $\Lambda$s) which decay into two charged tracks inside the central tracker are produced in the primary annihilation.

It should be noted that the criteria for event Deconvolution described below have deliberately been chosen as simple as possible, keeping in mind that all these operations will have to be done online. The strategy to disentangle the events is the following:

1. We assume stand-alone track reconstruction in the TPC, up to an unknown event-time-offset.

   - In the present study we assume this is possible with high efficiency and precision.
   - Note that the track reconstruction needs to be done online in order to implement a trigger logic.

2. Under the assumption of tracks originating from the primary vertex we extrapolate to the beam axis. The offset along the beam from the target directly gives the event time of the track. If an event time is defined by outside detectors (as assumed in this study) the offset can be used to for a cleaning cut.

3. Remaining tracks are extrapolated to the Micro Vertex Detector and correlated with the hits in this detector. This yields additional selection criteria.

4. In a second sweep a $V^0$ hypothesis is tested to find candidates of neutral decays, for example $\Lambda$s. If two tracks with opposite charge in the TPC appear close to each other a $V^0$ candidate is formed. Note that the reconstructed 3-momenta are sufficient to reconstruct the direction of flight of the neutral particle. With this direction we can extrapolate again to the beam axis and determine the offset with respect to the target region.

5. Detectors outside of the TPC, like the SciTil or the GEM detectors can also contribute to the timing information. This has not yet been taken into account in the simulations presented below.

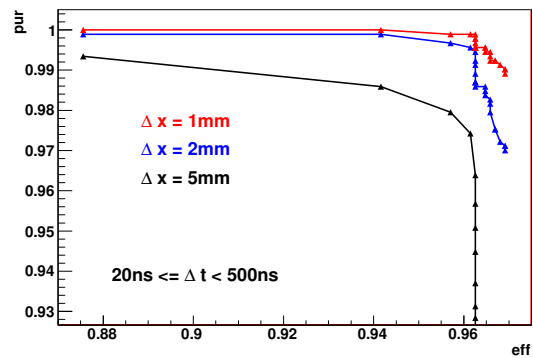

**Figure 10.32:** Event deconvolution performance for tracks from the primary target. Red: 1 mm cut on MVD residuals, blue 2 mm, black 5 mm. The points on the curve correspond to different event time cuts ranging from 20 ns to 500 ns.

### 10.13.1 $Y(4260) \rightarrow J/\psi \pi^+ \pi^-$ Channel

In order to evaluate the event deconvolution performance for events where all tracks come from the primary vertex the decay of the charmonium-like $Y(4260)$ into $J/\psi \pi^+ \pi^-$ where the $J/\psi$ decays into two muons. So there are 4 charged tracks per physics event. To each physics event a background of 1000 DPM events has been added as described above. The physics event was given the time $t_0 = 0$, while the background events were distributed around this time. The track parameters have been taken from the Monte Carlo truth at the inner boundary of the TPC, assuming perfect track reconstruction and an external event time definition.

In order to suppress pile-up events the following cuts have been made:

- Remove all tracks with $p_t < 100 \, \mathrm{MeV}/c$.
- Extrapolate track to the $z$-axis and cut around the target region in a time window ranging from 20 ns to 500 ns (see Fig. 10.32).
- Extrapolate the remaining tracks to the hits in the MVD and cut on the residuals. At least one hit surviving the cut is required in order to accept the track to the physics event.

Figure 10.32 shows the purity of this approach versus the achieved efficiency. Here we define the efficiency $\epsilon$ as the ratio of the number of recovered physics tracks $N_{\mathrm{reco}}$ divided by the number of total physics tracks simulated $N_{\mathrm{phys}}$. The purity $\Phi$ is given by the expression

$$\Phi = 1 - \frac{N_{\mathrm{bkg}}}{N_{\mathrm{phys}} + N_{\mathrm{bkg}}}$$

Where $N_{bkg}$ is the number of background tracks wrongly attributed to the physics event. A purity of 1 means that no pile-up track has been kept. If the number of physics tracks is equal to the number of falsely kept background tracks the purity would be $\frac{1}{2}$ and it goes to zero if there are much more background tracks kept than physics tracks were present in the actual event.

The red, blue and black lines correspond to MVD residual cuts of 1 mm, 2 mm or 5 mm, respectively. Note that even the 1 mm cut is comparably loose when compared to the expected resolution of the devices. The points on the curves correspond to different time cuts around the the target ranging from 20 to 500 ns.

With the tightest cut on the MVD residuals and a wide 500 ns cut on the event time still a single track event assignment efficiency of 97% with a purity of about 99% is achieved.

Of course in practice the stand-alone track reconstruction in the TPC will have some finite resolution. In Fig. 10.33 the corresponding efficiency plot is shown for a simulation where the true track values were subjected to Gaussian smearing by 1% in the momentum and by 0.5 mm in the track position at the inner boundary of the TPC. One sees the drop in efficiency down to 80% for the 1 mm residual cut. The magenta colored curve shows a scenario where at least two hits in the MVD are required but with a relatively relaxed cut of 3 mm on the residual. In summary, even with these more realistic smeared track parameters the event deconvolution through target pointing criteria and correlation with the MVD data seems feasible for tracks originating from the primary vertex.

## 10.13.2 $\Lambda \to p\pi^-$ Channel

In order to study the case when no hits in the MVD are present because a neutral particle has decayed inside the TPC volume into two opposite charge tracks a sample of $\Lambda \to p\pi^-$ decays has been simulated with added background. In this case only the target pointing of a reconstructed $V^0$ has been used. Additional information from detectors outside the TPC would further improve the performance. Figure 10.34 shows the performance of the algorithm for different event time cuts: red 100 ns, blue 150 ns, black 200 ns. The points on the curve correspond to different cuts for the distance of closest approach of the two charged tracks originating from the decay of the $\Lambda$.

The crucial point here is the vertexing algorithm.

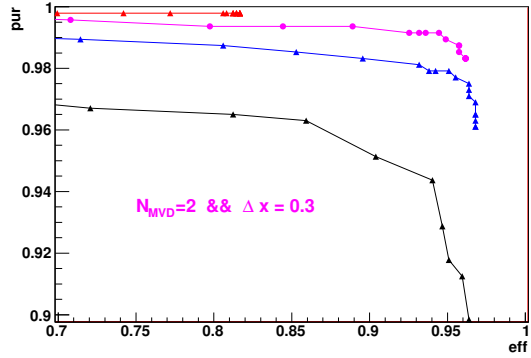

**Figure 10.33:** Event deconvolution performance for tracks from the primary target with smeared resolution: Red: 1 mm cut on MVD residuals, blue 2 mm, black 5 mm. Magenta: 3 mm residual cut but at least two hits found in the MVD. The points on the curve correspond to different event time cuts ranging from 20 to 500 ns.

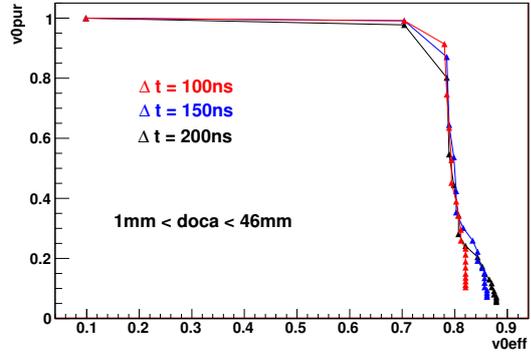

**Figure 10.34:** Event deconvolution performance for neutral $\Lambda$ tracks from the primary target. Red 100 ns event time cut, blue 150 ns, black 200 ns. The points on the curve correspond to different cuts for the distance of closest approach of the two charged tracks originating from the decay of the $\Lambda$.

As a crude approximation we have just cut on the distance between the endpoints of a track to define a $V^0$ vertex. It turns out that for $V^0$s decaying inside the TPC using a linear extrapolation to the beam axis and applying the event time cut all events were reconstructed background free. For $\Lambda$s decaying closer to the $z$-axis a true vertexing algorithm is needed for higher efficiencies.

# 11 Detector Tests

In order to verify the design choices for the $\overline{\text{P}}$ANDA TPC, several detectors with GEM amplification have been built and tested:

- a $10 \times 10\,\text{cm}^2$ triple GEM detector for ion back-flow studies,

- a small TPC with $10 \times 10\,\text{cm}^2$ active area and $7.7\,\text{cm}$ drift length for resolution studies and front-end tests,

- a large prototype TPC with $30\,\text{cm}$ outer diameter and $73\,\text{cm}$ drift length.

The following sections report key results obtained with the small detectors. The design of the large prototype and first results from beam tests are described in Sec. 12.

## 11.1 Gain Measurements

The effective gain of a GEM detector is determined by measuring the current at the readout anode $I_{\text{anode}}$ for a given rate $R$ of incident X-rays, each X-ray conversion producing $N_{\text{ion}}$ ionization electrons:

$$G_{\text{eff}} = \frac{I_{\text{anode}}}{e N_{\text{ion}} R} \quad . \qquad (11.1)$$

Defined in this way, Eq. 11.1 corresponds to the effective gain seen be the readout, and takes into account charge losses in the GEM structures. The effective gain curves for three different gas mixtures, Ar/CO$_2$ (70/30), Ar/CO$_2$ (90/10) and Ne/CO$_2$ (90/10), are shown as a function of the voltage applied to the drift cathode in Fig. 11.1.

During these measurements, the potentials of the GEM foils were defined by the cathode voltage via a resistor chain. Table 11.1 lists the voltage settings corresponding to a cathode voltage of 4000 V, which is also referred to as 100% standard settings.[1]

## 11.2 Ion Backflow Measurements

In order to prove the feasibility of the PANDA TPC, it is extremely important to study the accumulation of space charge in the drift volume and its effect on track reconstruction. In addition to the primary ionization, there is a backflow of slow ions ($v_{\text{Ion}} \sim 1.7\,\text{cm/ms}$) from the amplification stage, which has to be suppressed as much as possible in the GEM structures. To study this suppression, a $10 \times 10\,\text{cm}^2$ dedicated triple-GEM test detector has been built, which allows an easy exchange of the GEM foils and an independent setting of all voltages. The single projection readout plane consists of strips with a pitch of $200\,\mu\text{m}$. A copper X-ray tube ($8\,\text{keV}$ $K_\alpha$) is used for irradiation. Fig. 11.2 shows a picture of the whole setup.

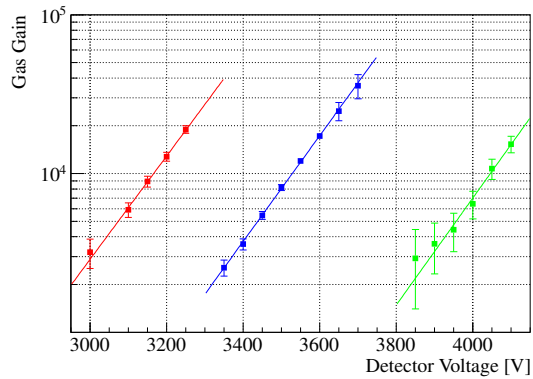

**Figure 11.1:** Effective gain of a triple GEM detector for three different gas mixtures as a function of the voltage applied to the drift cathode. Ne/CO$_2$ (90/10) in red, Ar/CO$_2$ (90/10) in blue and Ar/CO$_2$ (70/10) in green.

**Table 11.1:** Electric potentials and fields for the gain measurements, defined by the cathode voltage through a resistive voltage divider. Here, the conversion gap between the drift cathode and the first GEM is 3 mm, the transfer gaps between GEMs and the induction gap are 2 mm, respectively.

| Electrode | Potential (V) | Field (kV/cm) | $\Delta V$ (V) |
|---|---|---|---|
| Cathode | $-4000$ | 2.43 | |
| GEM1 top | $-3271$ | | 400 |
| GEM1 bot | $-2871$ | 3.65 | |
| GEM2 top | $-2142$ | | 364 |
| GEM2 bot | $-1778$ | 3.65 | |
| GEM3 top | $-1049$ | | 320 |
| GEM3 bot | $-729$ | 3.65 | |
| Anode | $0$ | | |

---

1. These are the average settings for the COMPASS GEM detectors.

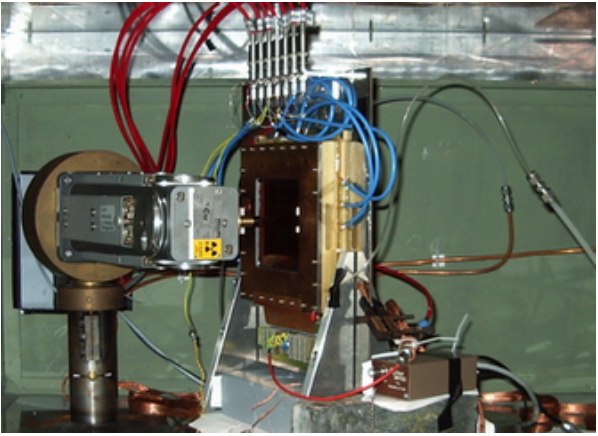

**Figure 11.2:** X-ray irradiation setup of a triple-GEM detector for ion backflow studies. The detector is mounted in vertical position.

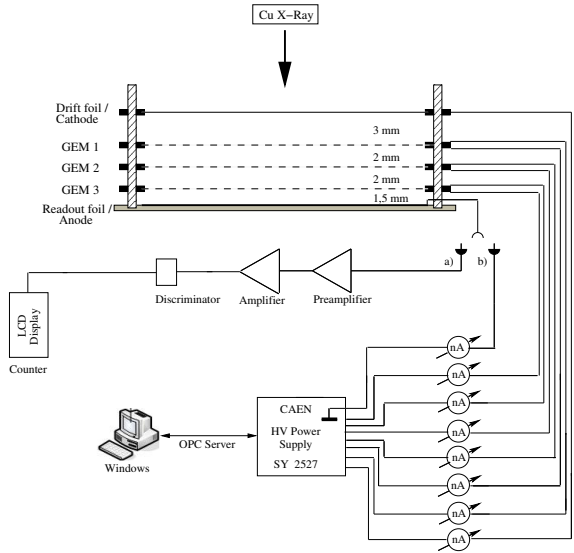

**Figure 11.3:** Schematic view of the experimental setup of the ion backflow studies: Via a), the rate of the Cu X-ray was determined and via b), the currents on all electrodes of the triple GEM were measured.

The currents on all electrodes, including the GEM foils, are measured with custom-made high-voltage current meters with a resolution of a few tens of pA. Figure 11.3 presents a schematic view of the experimental setup.

With this setup different field configurations and foil geometries within the GEM stack have been tested. All measurements were carried out without magnetic field and in an Ar/CO$_2$ (70/30) gas mixture.

Figure 11.4 shows the value of the ion backflow, defined by the ratio between ion current on the cathode and electron current on the anode, as a func-tion of the induction field, i.e. the field between the last GEM and the readout anode. A minimum ion backflow of 0.8% has been achieved. The corresponding settings comprise an induction field between the readout plane and the last GEM foil of 5 kV/cm and a transfer field between this and the middle foil of 0.16 kV/cm. This asymmetry is essential for the ion suppression. A rather low drift field of 250 V/cm was used and the transfer field between the topmost and the middle foil was set to 4-6 kV/cm, which still ensures a stable and safe operation. Table 11.2 summarizes the settings of fields and GEM voltages. The chamber was operated at a rather high effective gain of $\sim 10^4$ during these measurements. Tests with the large prototype TPC (cf. 12) have shown that the TPC can be operated at a gain of $\sim 2 \cdot 10^3$, thus reducing the number of back drifting ions per electron arriving at the GEM stack.

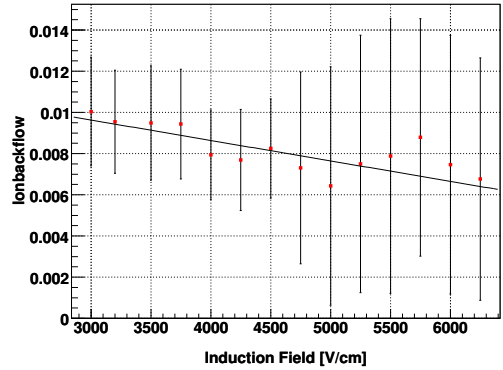

**Figure 11.4:** Ion backflow depending on the induction field value.

| Drift Field | 0.25 kV/cm |
|---|---|
| $\Delta$U$_{GEM1}$ | 330 V |
| Transfer Field 1 | 4.5 kV/cm |
| $\Delta$U$_{GEM2}$ | 375 kV |
| Transfer Field 2 | 0.16 kV/cm |
| $\Delta$U$_{GEM3}$ | 450 V |
| Collection/Induction Field | 5.0 kV/cm |

**Table 11.2:** Chamber settings for a minimal ion backflow at an effective gain of about 10$^4$.

The presence of a magnetic field helps to reduce the ion backflow even further [69]. An ion backflow value of 0.25% has been measured at a field of 4 T in a Ar/CH$_4$/CO$_2$ (93/5/2) gas mixture, as can be seen from Fig. 11.5. The reduction of ion backflow with increasing magnetic field can be explained by a reduced transverse diffusion in the GEM holes and PASA/ALTRO electronics [71] originally designed for the ALICE TPC. Figure 11.7 shows the residual distribution for the coordinate along the small side of the pads for tracks crossing the chamber approximately parallel to the readout plane. All clusters, which were selected by the pattern recognition algorithm to belong to the track, are considered. This introduces a small bias to the results, because these clusters are also used to define the track. This is in particular true for low multiplicity events; therefore a minimum of four clusters per track is required. The resolution is obtained by fitting the residual distribution with two gaussians and calculating the weighted mean of the standard deviations. Spatial resolutions of 170 μm and 200 μm, averaged over the full drift length, were achieved for tracks parallel and perpendicular to the readout plane, respectively.

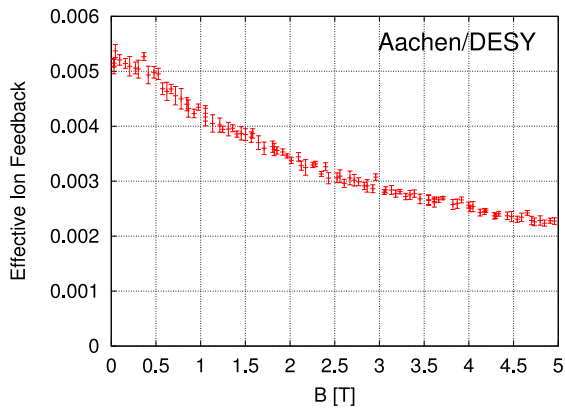

**Figure 11.5:** Ion backflow depending on the magnetic field.

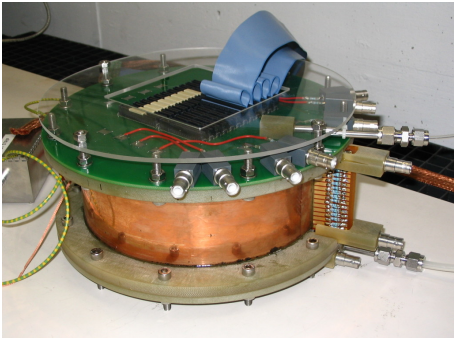

**Figure 11.6:** Picture of the TPC test chamber with the readout plane on top and the field-cage resistor chain on the right hand side. The GEM foils are mounted on the readout plane, the drift direction is from bottom to top.

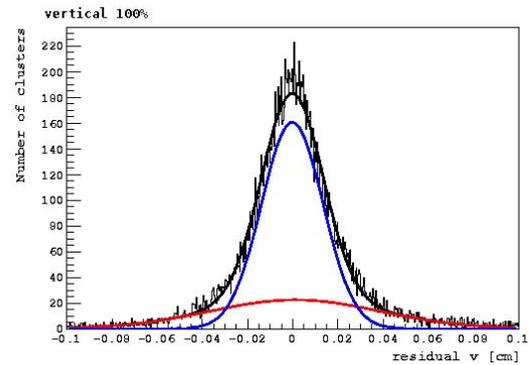

**Figure 11.7:** Residual distribution for tracks crossing the chamber parallel to the readout plane. The weighted mean of the standard deviations of the two Gaussians is 170 μm.

The spatial resolution along the drift direction was found to be 240 μm.

therefore an increased electron transparency. For the $\overline{P}$ANDA TPC with its low diffusion gas mixture, a similar ion backflow can be achieved at a field of 2 T. At a gain of $2 \cdot 10^3$, this corresponds to four back drifting ions per electron reaching the GEM stack.

## 11.3 Test TPC

A small-size GEM-TPC detector ($10 \times 10\,cm^2$ active area, 77 mm drift length) has been built and tested using cosmic muons [70]. Fig. 11.6 shows a picture of the fully assembled detector, which was operated with an $Ar/CO_2$ (70/30) gas mixture for the measurements presented here.

Several track topologies were studied by choosing different orientations of the TPC with respect to the incoming muons. For these tests, the readout plane consisted of rectangular pads with a pitch of $1.0 \times 6.2\,mm^2$; the signals were read out with the

# 12 Large Prototype

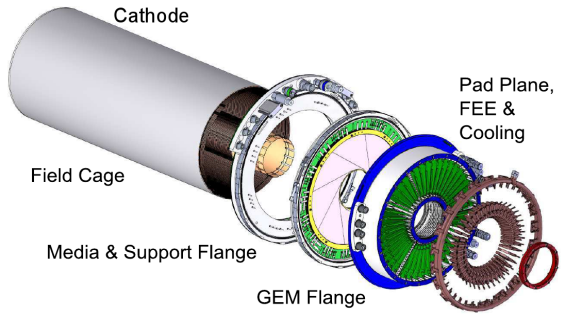

**Figure 12.1:** Explosion view of the GEM-TPC.

| Material | Thickness |
|---|---|
| Aluminized Mylar | 200 nm |
| Kapton | 25 μm |
| Kapton | 125 μm |
| Rohacell | 2 mm |
| Kapton | 125 μm |
| Rohacell | 2 mm |
| Kapton | 125 μm |
| Kapton | 25 μm |

**Table 12.1:** Materials and their thickness listed from the outside of the fieldcage to the inside.

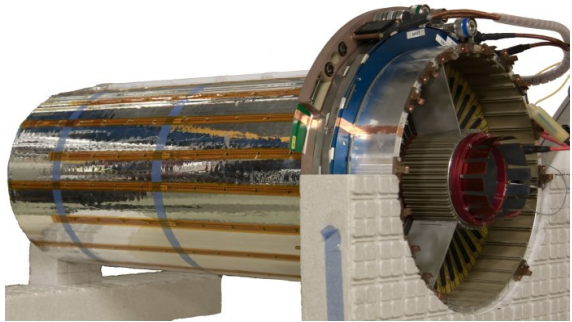

**Figure 12.2:** Photo of the large prototype.

## 12.1 Design

A large GEM-TPC prototype was built to be employed in a physics run in hadron collisions at intermediate energies within the existing FOPI spectrometer at GSI. The size was chosen such to fit inside the Central Drift Chamber (CDC) of FOPI and to have the possibility to host different target types within the GEM-TPC detector. In addition design requirements for the usage of the same chamber within the Crystal Barrel experiment at ELSA in Bonn were taken into account.

The large prototype has a total drift length of 727.8 mm, an inner diameter of 104 mm and an outer diameter of 308 mm. The prototype mainly consists of three parts: the field-cage, the readout part and the media flange. Figure 12.1 shows an explosion view of the GEM-TPC prototype with the different detector parts, while Figure 12.2 shows a photo of the large prototype that was built.

### 12.1.1 The Fieldcage

The lightweight field-cage structure consists of a self-supporting sandwich made of a Rohacell core, Kapton insulation layers and two skins of fiber glass material, arranged in two concentric cylinders. The materials of the fieldcage and their thickness is listed in table 12.1. The downstream end-cap is made of the same structure. Electrical shielding to the outside world is provided by an additional Kapton layer with aluminum coating. A homogeneous electric field in the drift direction is provided by the HV plane at the downstream end cap and precision concentric cylindrical field cage rings along the barrel that cover the inner and outer radius and are stepwise degrading the HV up to ground potential at the anode side. To improve the homogeneity, the cylindrical field cage consists of two sets of copper strips on both sides of the Kapton foil. The potential on each ring is defined by a resistor chain. The strip foil is made out of a 25 μm thick Kapton foil with over 700 copper strips with a pitch of 1.5 mm on each sides. The upstream side of the field cage is connected to the media flange made of fiber glass material, which provides mechanical stability and serves as the mounting structure of the GEM-TPC to the external support. The radiation length for a track perpendicular to the beam axis is 0.6 % and for a track in forward direction, crossing the drift cathode, it is 0.6 %.

### 12.1.2 The GEM Flange

The GEM-Flange is designed to hold up to four GEM foils for the gas amplification. The foils used for the prototype are standard foils produced at

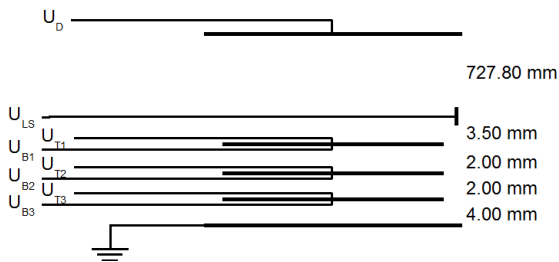

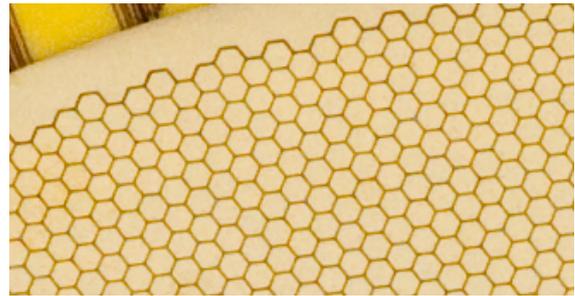

**Figure 12.3:** Scheme of the GEM-stack used for the large prototype where $U_D$ is the drift voltage, $U_{LS}$ is the voltage of the last strip before the GEM foils and $U_T$ and $U_B$ are the top and bottom voltages of the respective GEM foil.

**Figure 12.4:** Closeup view of the pad-plane.

CERN with a thickness of 50 µm and a hole pitch of 140 µm. To keep the charge released in a discharge as low as possible each foil is segmented into 8 sectors on one side, each with its own loading resistor. A schematic of the employed GEM-stack is visible in Figure 12.3.

### 12.1.3 The Readout

For a more uniform charge distribution across the pads a new pad-plane with hexagonal pad shapes was designed. The optimum outer pad radius was determined to be 1.5 mm by Monte Carlo simulations of the P̄ANDA TPC, taking into account a magnetic field of 2 T. These simulations show that the spatial resolution and hence the momentum resolution does not improve for outer radii of hexagonal pads below 1.5 mm, but is already dominated by diffusion in the chamber.

Since the pad-plane is used to close the gas volume it had to be designed in a gas tight way. This was done by using a four layer build-up with tracks on all layers and staggered connections only between neighboring layers. To avoid crosstalk between signals of different pads the connections from the pads to the connectors were designed to minimize crossing of tracks or narrow parallel tracks. In Fig. 12.4 a close-up of the pad-plane is shown. Field inhomogeneities on the outer part of the pad-plane were avoided by placing copper matching the shape of the pads.

For the readout the pad-plane is equipped with 42 front-end cards, each equipped with 4 T2K/AFTER chips [72]. The T2K/AFTER chip is an analog sampling chip with a 511-cell switched capacitor array per channel, a maximal sampling frequency of 50 MHz and a multiplexed output. The design value of the equivalent noise charge (ENC) is around 600 electrons at a 10 pF input capacitance. Figure 12.5 shows the 2-dimensional noise distribution on the pad-plane. One can see that the noise is rather uniformly distributed around a mean value of $1.6\,ADC$ channels which corresponds to an ENC of 625 electrons. The visible ring structure is due to single high noise channels on each FE-chip.

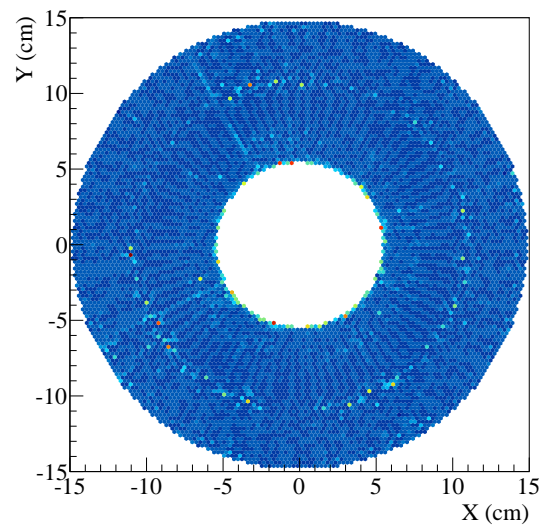

**Figure 12.5:** 2-Dimensional noise distribution on the pad-plane.

Since the drift velocity depends strongly on the temperature, a cooling system for the front-end cards was installed to avoid heating the pad-plane (see 12.1.5)

### 12.1.4 Material Budget

The total radiation length seen by charged particles traversing the large prototype has been evaluated in the same fashion described in 3.2, employing a detailed geometry of the prototype within the PAN-DAROOT framework. To calculate the radiation length the origin of the tracks was set to the tar-get position in the FOPI spectrometer (cf. Sec. 12), corresponding to a drift distance of about 65 cm in the large prototype. Figure 12.6 shows the obtained radiation length versus the polar angle of the track.

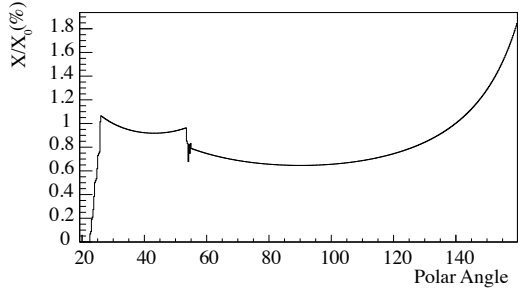

**Figure 12.6:** Radiation length plotted against the polar angle.

Starting from 20° to 55° the track goes through the drift cathode, therefore the radiation length is bigger than in the fieldcage in the area between 55° and 160°. After the fieldcage the radiation length increases significantly (not shown in Fig. 12) due to the material of the media and GEM flange. For experiments with FOPI this high material budget in backward direction is not an issue since no further detectors are located behind the prototype and in this region there is no overlap with the CDC.

### 12.1.5 The Cooling System

A water-driven cooling system was built to cool each front-end card, each ADC and the voltage regulators independently. A mixture of water and glysantin G48/BASF (Glycol & Ethandiol) is circulating in a closed pipe system and flowing through heat exchangers coupled to copper plates. The copper plates are put in contact with the elements that have to be cooled. The water temperature is kept to a constant value of 20° by a UC080T-H chiller system and distributed through flexible polyurethane pipes. Each front end card is sandwiched between 2 copper plates put in direct contact through heat conducting pads to the 4 FE chips. Figure 12.7 shows the copper plates and the heat exchangers attached to one FE card. Figure 12.8 shows the complete cooling ring connected to the readout electronics. The length of the pipes was chosen such to guarantee the same impedance for all the attached FE cards. The same system provides the cooling of the ADC and the voltage regulators through a serial connection of the pipes following the cooling ring for the FE cards.

Temperature sensors were applied on the pad plane and along the external surface of the field-cage in order to monitor the thermal fluctuations. The envisaged temperature stability, necessary in order to maintain a constant drift velocity, yields $0.1\,K$. 12 PT100 sensors were placed on the pad-plane along the outer radius, while 210 Dallas Semiconductor DS18B20U 1-wire temperature sensors were distributed along the field cage cylinder.

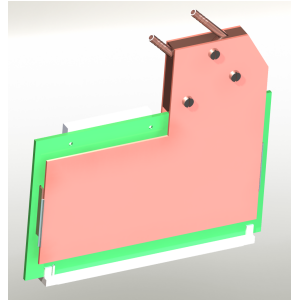

**Figure 12.7:** Copper plates utilized to cool the FE chips.

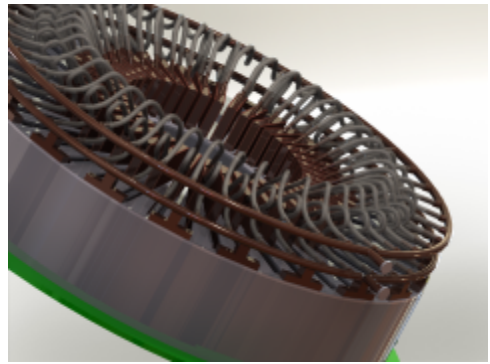

**Figure 12.8:** Complete cooling ring for connected to the the FE cards.

The region inside the FOPI spectrometer is characterized by rather large temperature fluctuations, mainly due to the heat dissipated by the RPC detectors. In order to stabilize the temperature along the field cage, compressed air was blown inside the 5 mm gap that separates the internal hole of the CDC from the external walls of the TPC barrel.

Figure 12.9 shows the temperature distribution as a function of the azimuthal angle $\phi$ and the z coordinate measured along the field cage. One can see that temperature fluctuations up to $1.5\,K$ appear along the field cage. The same fluctuations are measured on the pad-plane surface which shows that the temperature gradient surrounding the chamber is responsible for the measured variations.

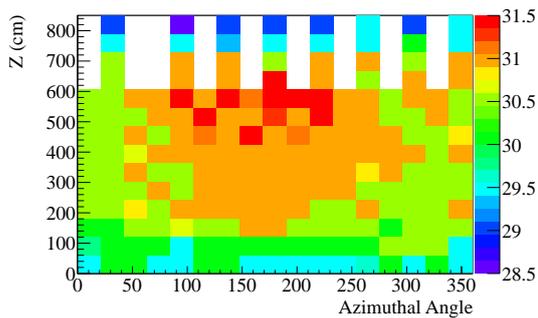

**Figure 12.9:** Measured temperature along the field cage as a function of the azimuthal angle $\phi$ and the z coordinate.

### 12.1.6 The Gas System

The prototype GEM-TPC has been operated both with an $Ar/CO_2$ (90/10) and a $Ne/CO_2$ (90/10) gas mixture. Figures 12.10 and 12.11 show the drift velocity and the transverse and longitudinal diffusion constants of $Ne/CO_2$ (90/10) and $Ar/CO_2$ (90/10), respectively, in a 0.6 T longitudinal magnetic field as a function of the electric drift field, calculated with MAGBOLTZ [28, 29].

An open gas circulation system was employed for the large prototype. The system is composed of a supply unit and analyzing devices that monitor the gas mixture and flow, the Oxygen content, the dew-point and the pressure. The total volume of the large prototype is about 45 l and a flow rate of 45 l/h was used for both $Ar/CO_2$ and $Ne/CO_2$ gas mixtures. A Brooks model 0254 was used to control the mass flow. This and the other parameters were monitored on a LabVIEW-based slow control system. A quadrupole mass spectrometer (QMS) was used to measure the $Ar/CO_2$ and $Ne/CO_2$ gas mixture during run time. A special measuring scheme was developed, where the TPC gas sample is measured in-between two calibration samples. This allowed a suppression of long-term drifts of the mass spectrometer and therefore the determination of the $CO_2$ concentration with an accuracy of less than 1 % was possible. Figure 12.12 shows an example of this measurement for a time interval of 2.5 days. The ratio Ar to $CO_2$ is shown for the analyzed gas sample and the calibration sample, the extracted $CO_2$ concentration is shown as well. The error band corresponds to 1 % fluctuations. A constant over-pressure of 18 mbar was kept inside the chamber and an Oxygen content between 6 and 8 ppm was reached.

## 12.2 Results

To test its operation, the large prototype was installed inside the FOPI [73] spectrometer at GSI (Darmstadt, Germany). FOPI is a large acceptance, fixed-target heavy ion experiment. It was designed to study the properties of compressed nuclear matter formed by collisions of heavy ions at energies from $0.1\,AGeV$ to $2.0\,AGeV$. The detector consists of sub-detector systems which nearly have a complete azimuthal symmetry. This nearly $4\pi$ coverage of the solid angle, allows for an almost complete event characterization. FOPI is able to identify light charged particles (pions, kaons, protons ...) and intermediate mass fragments. Hadron resonances and neutral hadrons can also be reconstructed from their decay products.

In Fig. 12.13 FOPI (grey color) as well as the GEM-TPC (violet color) are shown. The FOPI spectrometer consists of a central drift chamber (CDC) which is surrounded by a scintillator barrel (Barrel) and a RPC barrel (RPC) with an intrinsic time of flight resolution of 200 ps and 80 ps respectively. In forward direction another drift chamber (Helitron) and a scintillator time of flight wall (PLAWA) with a time of flight resolution of 120 ps - 150 ps are installed. The CDC, Helitron and the two barrels are surrounded by a 0.6 T solenoid magnet (not shown in Figure 12.13).

The FOPI spectrometer delivers a vertex resolution of few millimeters in the $x-y$ plane and a resolution along the beam axis of around 5 cm. The reached momentum resolution for particles traversing the CDC is $4-10\,\%$. Vertex and especially secondary vertex resolutions can be significantly improved by using the GEM-TPC prototype as an additional tracker.

The TPC was mounted in the inner hole of the central drift chamber of FOPI. With this configuration it is possible to use the tracking and time-of-flight detectors of FOPI as a reference. All detectors of FOPI have a $2\pi$ acceptance in the azimuthal angle. The acceptance in the polar angles are listed in table 12.2.

The data acquisition (DAQ) of FOPI is based on the Multi Branch System (MBS) and designed for a trigger-rate of around 600 Hz. The DAQ of the GEM-TPC is based on a different system since the VME readout like in FOPI would be too slow to handle the large amount of data coming from the TPC ($\sim 40\,MB/s$). This system is a slightly modified version of the COMPASS readout [74]. In order to merge the two systems the following strategy was adopted. The Trigger Control System module

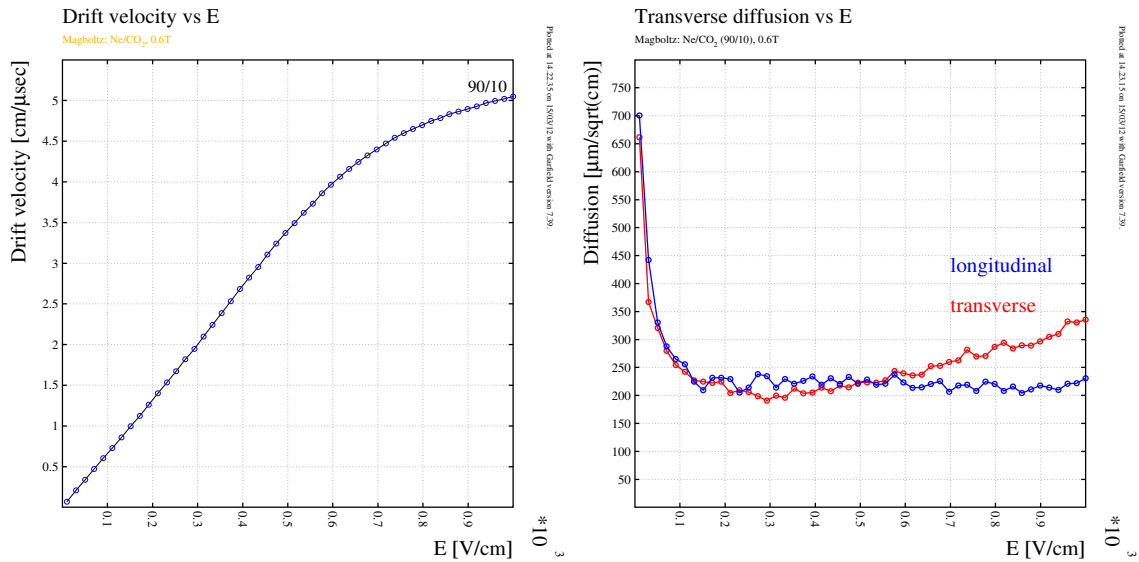

**Figure 12.10:** Properties of a Ne/CO₂ (90/10) mixture in a 0.6 T magnetic field. (Left) Drift velocity as a function of the electric drift field. (Right) Transverse (red) and longitudinal (blue) diffusion as a function of the electric drift field.

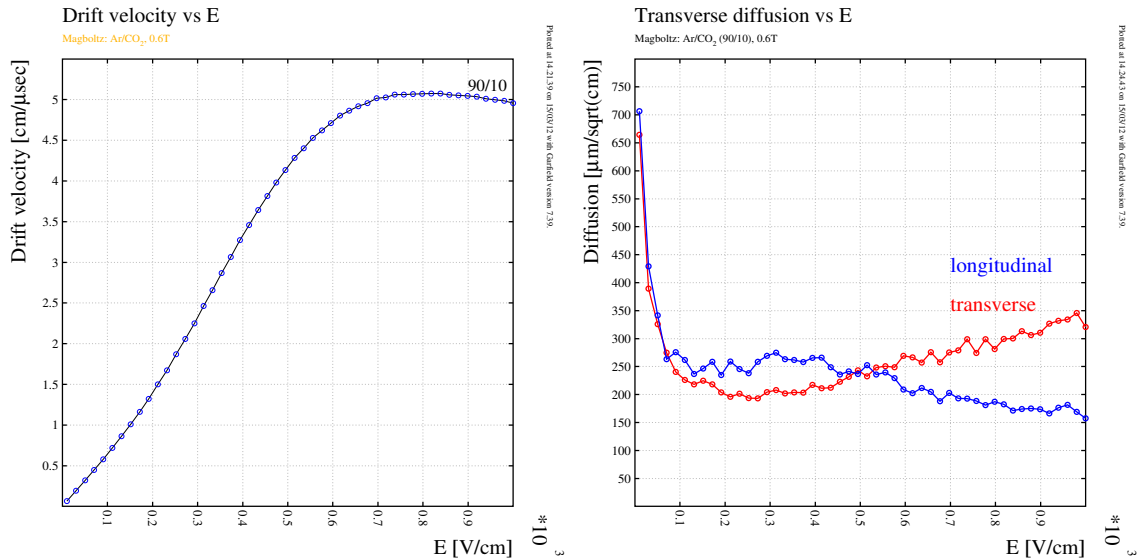

**Figure 12.11:** Properties of an Ar/CO₂ (90/10) mixture in a 0.6 T magnetic field. (Left) Drift velocity as a function of the electric drift field. (Right) Transverse (red) and longitudinal (blue) diffusion as a function of the electric drift field.

(TCS) of the GEM-TPC gets control signals from the FOPI main trigger and sends out triggers to the GEM-TPC readout modules (GeSiCA) accordingly. TCS also sends out to the FOPI system its dead-time signal, which amounts to 2.5 ms. Additionally the time-stamps from the MBS event header are communicated to the GeSiCa, in order to achieve the synchronization of the data word of the different sub-detectors. Optical fibers are used to send the data of the prototype via the S-Link protocol

[74] to the event builder were it is merged with the FOPI data stream.

Furthermore a lot of different programs to monitor and control the high voltage, low voltage, temperatures, gas mixture, DAQ and data were developed and improved.

Several different settings of the large prototype were tested during the measurement within FOPI.

- Several drift fields: 150, 200, 300, 360 V/cm

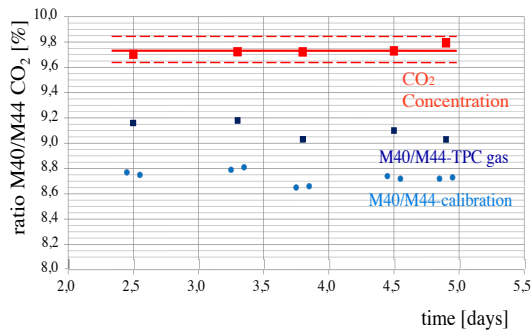

**Figure 12.12:** QMS analysis of the $CO_2$ content in a $ArCO_2$ 90/10 gas mixture along 2.5 days of data taking.

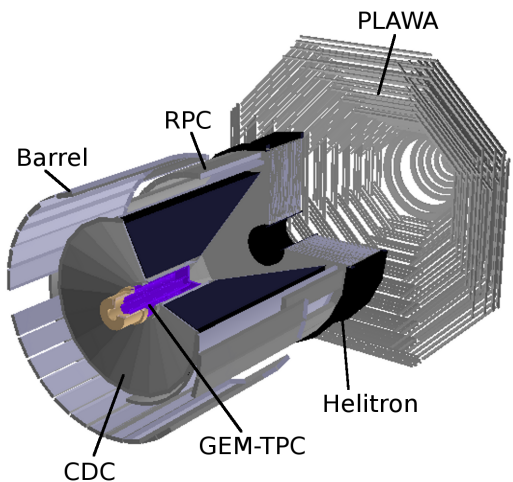

**Figure 12.13:** The FOPI spectrometer (gray color) with the GEM-TPC (violet color) prototype mounted in the inner bore of the central drift chamber (CDC).

| Detector | Acceptance |
|----------|------------|
| CDC | 30° - 140° |
| RPC | 36° - 67° |
| Helitron | 10° - 30° |
| Barrel | 60° - 110° |
| PLAWA | 10° - 35° |
| GEM-TPC | 5° - 175° |

**Table 12.2:** Acceptances of the FOPI detectors and the GEM-TPC.

with the corresponding drift velocities: 0.9, 1.4, 2.2, 2.9 cm/μs [75].

- Different settings for the gain for the GEM-stack. The nominal setting of the high voltage, referred to as 100 %, are given in Table 11.1.

- Two gas mixtures, namely $Ar/CO_2$ and

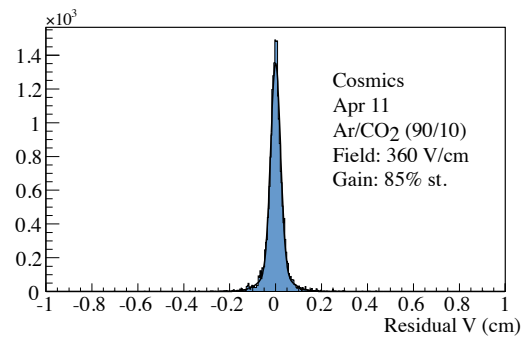

(a) Residual measured with cosmic tracks. The experimental distribution is fitted with two Gaussian functions. The weighted mean of the sigmas of the two Gaussian curves is 230 μm.

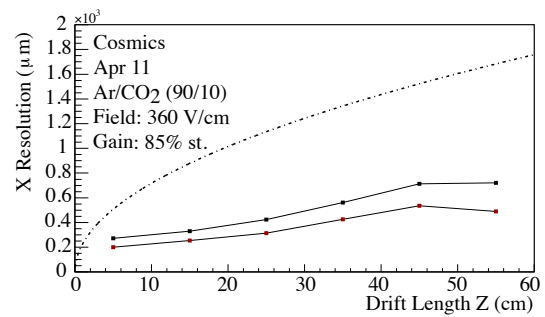

(b) Chamber resolution (along $x$) as a function of drift length. The dashed line is the transverse diffusion for single electron drift as calculated with GARFIELD. The dots show the results from residual distributions for tracks in 10 cm bins along the drift direction (cf. (a)). The red dots are the sigmas of the narrow Gaussian, while the black dots above correspond to the weighted mean of both Gaussian curves. Statistical error bars are included, but smaller than the data points.

**Figure 12.14:** Figure (a) shows the residuals of one slice in $z$. Figure (b) shows the $z$-dependence of the residuals. The results were obtained from cosmic data.

$Ne/CO_2$ both in a 90/10 mixture by weight.

The behaviour of the chamber under the different experimental conditions is discussed in the following paragraphs.

## 12.2.1 Results with Cosmics

As a first test, tracks from cosmic rays were measured, using the FOPI-Barrel as a trigger. The TPC was operated at a drift field of 360 V/cm with an $Ar/CO_2$ (90/10) mixture, which translates into a drift velocity of $\sim 2.9$ cm/μs. The high voltage settings for the GEM corresponded to 85 % of the reference values.

The track reconstruction in the GEM-TPC was done by first finding the track points with the stan-dard three-dimensional clustering algorithm (cf. Sec. 10.6). The obtained clusters were then used in a four-dimensional Hough transformation for linear straight line pattern recognition, a modification of the algorithm described in Sec. 10.7.8. The obtained residuals are shown in Fig. 12.14. According to the definition of $U$ and $V$ given in 10.8 and considering that most of the analyzed cosmic tracks are vertical, the $V$ residuals represent in good approximation the resolution in the $x$-coordinate. A spatial resolution of $230\,\mu m$ for the first slice along the $z$-axis was obtained. It should be noted at this point that the transverse diffusion for an electron at the end of the large prototype ($z = 70\,cm$, cf. Fig. 12.14) in $Ar/CO_2$ (90/10) and no magnetic field is roughly 1.5 times *larger* than the maximum transverse diffusion in the final TPC with full PANDA field configuration. The same procedure was tested with simulation. The full-scale simulation (cf. Sec. 10.1) of the large prototype was used with events containing mono-energetic muons with a kinetic energy between 10 and 100 GeV. A total statistics of $1 \cdot 10^6$ vertical muon tracks was simulated. After the Monte Carlo and digitization procedures the full analysis chain was utilized and the residuals extracted in the same way used for the real data. Figure 12.15 shows the chamber resolution calculated for simulated tracks for the first 10 cm drift (panel a)) and as a function of the drift length (panel b)). The same comparison with the Garfield calculation shown in Fig. 12.14 is shown also in Fig. 12.15 together with the results of the Gaussian fits. The red dots represent the sigmas of the narrow Gaussian, while the black dots above correspond to the weighted mean of both Gaussian curves. The agreement between simulations and experimental data is satisfactory but small differences remain for large drift length slices that should be studied systematically.

The simulated cosmic tracks were also analyzed in terms of completeness and splitting. The completeness of the track is defined as the percentage of clusters that belong to a certain track and are also associated to this track. An average completeness of 91.4% was obtained for the simulated cosmic tracks. The splitting defines the number of segments one track is split into by the pattern recognition algorithm. About 5% of all the cosmic tracks are split and mainly (84%) into two segments.

Cosmic events were collected also utilizing a $NeCO_2$ mixture (90/10) as drift gas and the drift field was set to 360 V/cm. Several settings of the GEM high voltage were tested, ranging from 70% to 74% in steps of 1%. This data sample was employed to

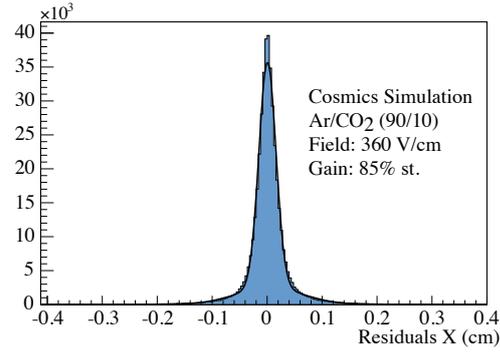

(a) Residuals in $V$ direction (approx. parallel to the $x$−axis for cosmics) for the first 10 cm of drift. The resolution obtained from the weighted mean of two fitted gaussian curves is $259\,\mu m$

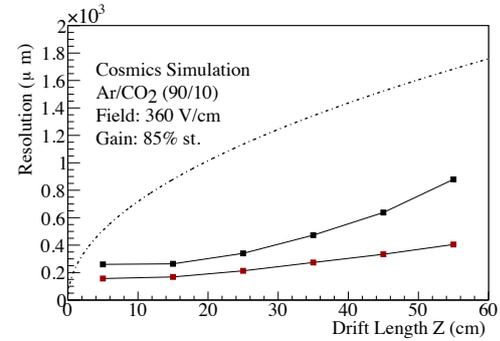

(b) Chamber resolution (along $x$) as a function of drift length. The dashed line is the transverse diffusion for single electron drift as calculated with GARFIELD. The dots show the results from residual distributions for tracks in 10 cm bins along the drift direction. The red dots are the sigmas of the narrow Gaussian, while the black dots above are from the weighted mean of both Gaussian curves. Statistical error bars are included, but are smaller than the data points.

**Figure 12.15:** Figure (a) shows the residuals of one slice in $z$. Figure (b) shows the $z$-dependence of the residuals. These results are extracted from simulations (see text for details).

check the efficiency of the GEM-TPC chamber and the total efficiency of the track reconstruction algorithms. The cosmic tracks reconstructed in the CDC detector have been used as an external reference to evaluate the efficiency. Events containing a single track were selected and the curvature radius of the identified particles in the CDC was determined. Applying an additional straight line fit and evaluating the respective reduced $\chi^2$ for each track, its degree of straightness was evaluated. Figure 12.16 shows the residual (defined as $\frac{\sqrt{\chi^2}}{Ndf-2}$) for the straight line fit as a function of the track radius. Straight tracks are selected applying the following cuts on these variables: residual $< 0.1$ and

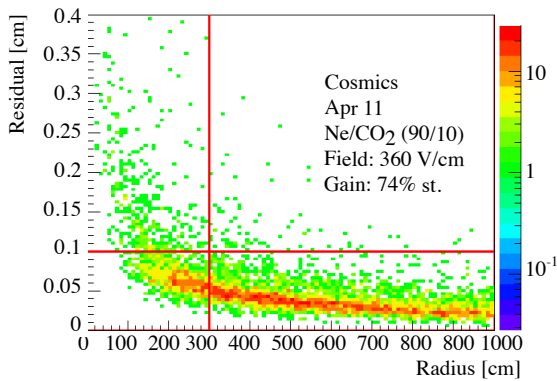

**Figure 12.16:** $\chi^2$ for the straight line fit as a function of the track radius reconstructed in the CDC. The full lines identify the cuts applied to select straight tracks.

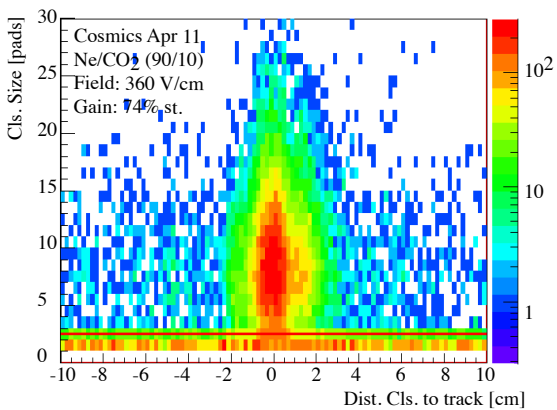

**Figure 12.17:** Cluster size as a function of the cluster distance from the CDC reference track.

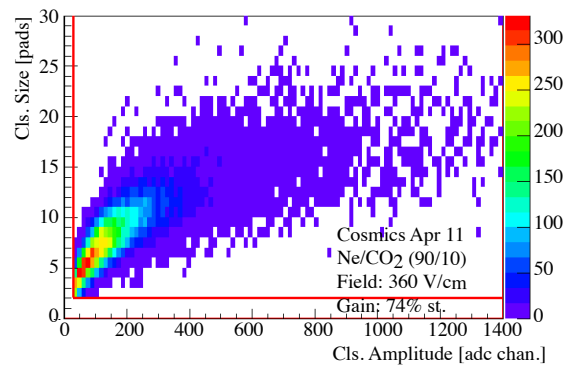

**Figure 12.18:** Cluster size versus cluster amplitude.

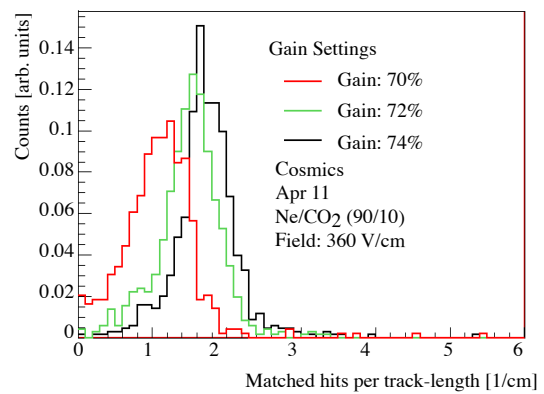

**Figure 12.19:** Distribution of the number of matched clusters normalized to the track length in the GEM-TPC (see text for details). The different curves refer to different settings of the GEM gain.

Radius $> 300$ cm. A fiducial volume is defined in the GEM-TPC, to exclude the detector edges and account for the scarce position resolution of the CDC along the beam axis (about 5 cm). The first and last 5 cm of the chamber (Z direction) are excluded and a maximal radius of 13 cm is defined. The selected straight tracks in the CDC are extrapolated to the GEM-TPC within the fiducial volume defined above and used as a reference to evaluate the efficiencies.

First, the efficiency of the GEM-TPC chamber is extracted counting the number of extrapolated segments in the fiducial volume that are correlated with identified clusters in the GEM-TPC. Clusters were selected which lay within a maximal distance of 2 cm in both the X and Y coordinates respect to the extrapolated CDC track in the GEM-TPC fiducial volume. No matching in the Z coordinate is required due to the insufficient spatial resolution of the CDC. In order to reduce the contribution of the electronic noise arising in the GEM-TPC a minimal cluster amplitude ($\geq 30$ ADC channels) and size ($\geq 2$ pads) were requested. Figure 12.17 shows the cluster size as a function of the distance between the clusters and the reference CDC track. Figure 12.18 shows the size of all the clusters as a function of the cluster amplitude after the selections mentioned above were applied. These two histograms demonstrate that after the applied selection most of the clusters do not stem from electronic noise and hence are suited for the calculation of the efficiencies.

Figure 12.19 shows the number of clusters matched to the extrapolated CDC track divided by the length of the extrapolated track segment in the GEM-TPC. This quantity is referred hereafter as the parameter $P$. The different curves refer to 3 different gain settings for the GEM stack. A minimal length of 12 cm is required for the extrapolated segment in the GEM-TPC volume. One can see how decreasing the GEM gain the average number of clusters normalized to the track length shifts as expected to lower values. The efficiency of the chamber for the different GEM gain settings can be extracted varying the threshold value of the parameter $P$. Table 12.3 shows the obtained efficiency for three different gain settings and applying either no cut on $P$ or selecting two threshold values: $P = 0.3 - 0.5$. These extracted efficiencies include the pulse shape analysis and cluster finding procedure described in 10.5 and 10.6.

**Table 12.3:** Chamber efficiency for different GEM gains and different cuts on the parameter $P$ (see text for details).

| | Efficiency | | |
|---|---|---|---|
| Gain | All | $P \geq 0.3$ | $P \geq 0.5$ |
| 74 % | 96.2 % | 95.9 % | 95.5 % |
| 72 % | 91.6 % | 90.9 % | 89.2 % |
| 70 % | 79.5 % | 75.2 % | 71.5 % |

Given the identified clusters matched to the external reference track in the CDC, the full reconstruction chain based on the PandaRoot framework [23] was applied to the data. This analysis chain includes additionally to the clustering algorithm the pattern recognition (cf. Sec 10.7.2) and the track fitting algorithms. The efficiency is defined as the probability of obtaining a fitted track candidate in the GEM-TPC if a matching between the clusters in the GEM-TPC and the reconstructed track in the CDC is available.

When applying the full reconstruction chain to this sample, the combined efficiency of the pattern recognition and track fitting is found to be 100 % for all the gain settings and selections of the cut on the $P$ variable.

### 12.2.2 Results with Beam Data

Several beam species were utilized to collect data for the joint GEM-TPC/FOPI system at GSI. Two test experiments took place in November 2010 and April 2011 and $^{84}$Kr, $^{197}$Au and $^{22}$Ne beams at 1.2 AGeV, 1.0 AGeV and 1.7 AGeV kinetic energies respectively were colliding on an Al target of 2 % interaction length inside the GEM-TPC. The beam parameters were set to an average particle rate of $5 \cdot 10^6$ particles/spill with a total spill length of about 10 sec and a duty cycle factor of 50 %. One physics run employing a $\pi^-$ beam at 1.7 GeV/$c$ colliding on different solid targets took place in June 2011. For this secondary beam a lower rate of 25.000 $\pi^-$/spill with a total spill length of 3.5 sec and a duty cycle factor of 42 % was available.

The major aim of this experiment was to measure charged and neutral kaons produced almost at rest in a nuclear medium and to employ the GEM-TPC together with the other FOPI sub-detectors to increase sensibly the momentum resolution and the secondary vertex identification.

Figure 12.20 shows two typical events in the GEM-TPC when particles are produced from reactions happening upstream (panel a) ) and on the Al target (panel b)) employing a $^{22}$Ne beam. On the left panel the three-dimensional view of the hit points is shown where the vertical axis corresponds to the beam axis. On the left panel the projection of the event on the pad-plane is depicted, the color-code is proportional to the cluster amplitude. The chamber settings correspond to 360 V/cm drift and 86 % of the standard GEM gain. The ArCO$_2$ mixture was used as drift gas. The noisy pads were rejected on-line by applying a threshold of 4.5 times the $\sigma$ value of the amplitude distribution stemming from electronic noise only. This threshold values are the same for all the data presented hereafter. The curled tracks shown in panel a) refer to low energy electrons strongly bent in the 0.6 T magnetic field. Even if the 3D representation on the right hand side of the panel shows the raw hits, the excellent tomographic capability of the device is evident. Panel b) shows the particle tracks detected in the TPC after a collision of the $^{22}$Ne beam with the Al target. An average length of 14 cm characterizes the tracks reconstructed in the GEM-TPC and due to the fact that the average particle momentum exceeds 1 GeV/$c$, the tracks are rather straight. The momentum of each track candidate can be reconstructed in the CDC, where the path length of each particle yield in average 80 cm. The additional space points in the GEM-TPC can be combined to the CDC points to improve the accuracy of the determination of the bending and hence of the particle momentum.

The track reconstruction in the GEM-TPC was done according to Sec. 10.7.2 with the Riemann track follower. The procedure is exactly the same described for the PANDA TPC except for the fact that the pad-plane of the large prototype is not sectorized. The distribution of the number of reconstructed tracks for $^{22}$Ne+Al collisions at 1.7 AGeV is shown in Fig.12.21. A maximal multiplicity of 100 particles per event was reached, on average 19 tracks per $^{22}$Ne collision, with an average number of 23 hits per track have been reconstructed in the GEM-TPC.

The resulting residuals expressed along the two coordinates $U$ and $V$ (see Figure 10.10 in Sec.10.8)

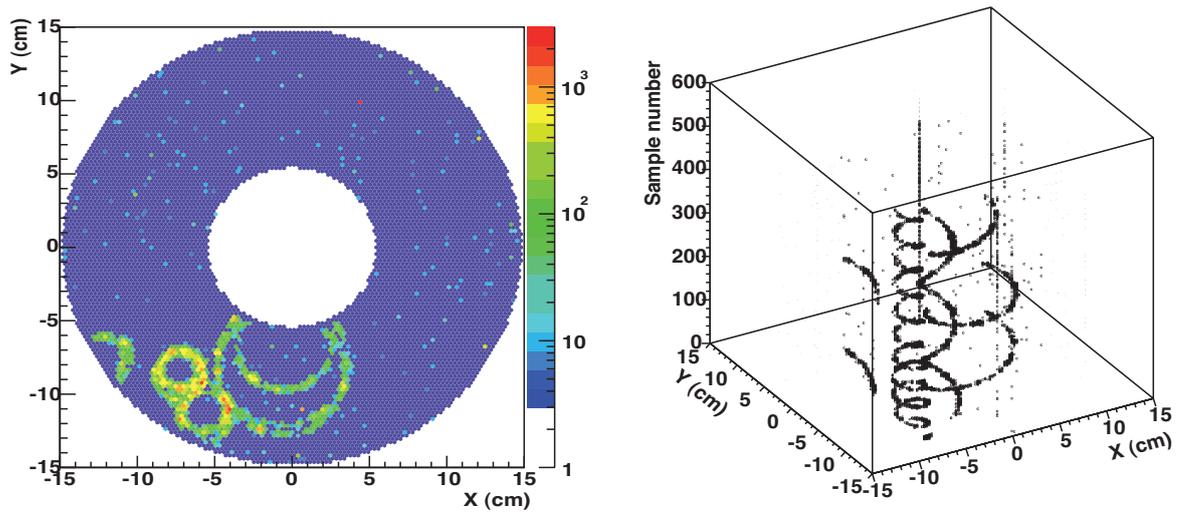

(a) Typical event characterized by an interaction of the $^{22}$Ne beam upstream the target region. The curled tracks represent the tracked electrons in the chamber.

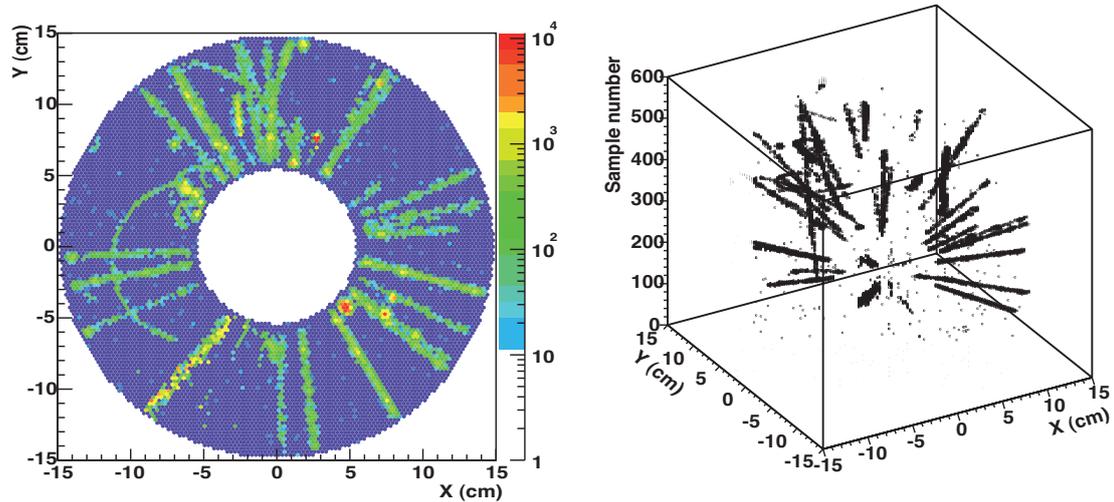

(b) Typical event characterized by an interaction of the $^{22}$Ne beam with the Al target.

**Figure 12.20:** Pictures (a) and b) show two typical events in the GEM-TPC. The left panel shows the samples from the ADC in both pad-plane position and sample time while the right picture shows the projection onto the pad-plane. The color code corresponds to the amplitude of a signal.

that define a plane perpendicular to the track and that contains the hit are shown in Figure12.22. The residual width for the U and V coordinates is extracted fitting the residual distributions obtained for the different slices in the z-coordinate with a double Gaussian, like it was done for the cosmic data. In Fig.12.22 the residual distribution for the first 10 cm drift is shown in panel a) together with the result from a fit with three Gaussian function. Panel b) in Fig. 12.22 shows the sigma of the narrow Gaussian fits as a function of the drift length. It should be pointed out that in the case of particle tracks stemming from beam reactions, U and V do not correspond univocally to cartesian coordinates. This preliminary result agrees rather well with the resolution obtained for cosmic tracks. The residual distribution obtained for the beam data displays a larger background in respect to the cosmic residuals. This is probably due to the fact that the rather high GEM gain leads to overflow and this effect is more present for beam data where the particle energy loss is about 5 times larger than for cosmic tracks. The overflow spoils the determination of the cluster barycenter and hence the resolution. Further studies will exclude such tracks.

The homogeneity of the tracks distribution was checked displaying the chamber occupancy for reactions with the $\pi^-$ beam. Figure 12.23 shows the track occupancy as a function of the radial and z coordinates of the GEM-TPC chamber. The scale is linear in z. This distribution was obtained with a reduced drift field of 235 V/cm drift, 80% of the GEM gain and a sampling frequency of 16 MHz, corresponding to a drift window of 31.93 µs. The resulting active length with these settings is 55 cm, shifted to the end of the chamber. The location of the target set in the middle of the GEM-TPC chamber, can be recognized in Figure 12.23 and the resulting acceptance is well suited to measure low momentum kaons produced in the $\pi^-$-induced reactions. One can see that a good homogeneity is guaranteed for a minimum distance of 0.5 cm from the field-cage walls. Further distortion effects of the electric field are currently under study.

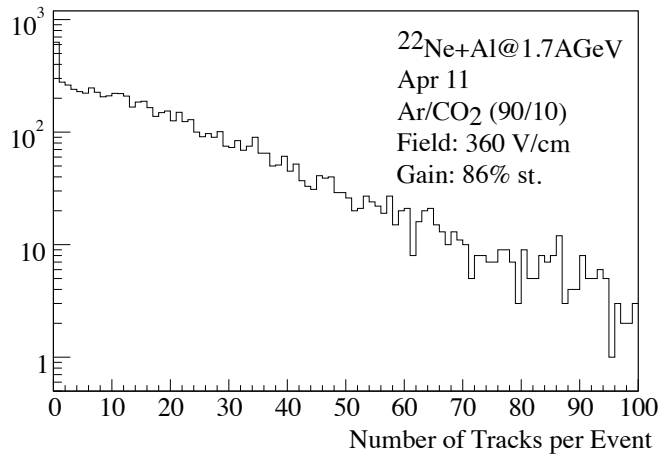

**Figure 12.21:** Multiplicity distribution of the reconstructed tracks in the GEM-TPC detector for $^{22}$Ne+Al collisions at 1.7 AGeV.

A similar matching procedure like the one used to correlate the cosmic tracks identified in the CDC with the hits recorded in the GEM-TPC has been used for the heavy ion collisions data. Figure 12.24 shows the projection on the X-Y plane of the recorded tracks in one $^{22}$Ne +Al collision. The pink lines shows the circle fits applied to the tracks in the CDC, in the insert the GEM-TPC fiducial area is magnified and the blue points represent the reconstructed clusters. The correspondence between the tracks measured in the two detectors is evident but a relative mis-alignment between the detector is visible as well. Currently the alignment of the GEM-TPC chamber is being studied using cosmic and beam data. Figure 12.25 shows an event which might be identified with the decay of a $\Lambda$ hyperon ($\Lambda \rightarrow p + \pi^-$) This event was also selected from the

$^{22}$Ne data sample and this qualitative picture shows clearly how the particle identification and momentum determination capability of the CDC detector can be combined with the secondary vertex information delivered by the GEM-TPC to improve the reconstruction of hadron decays. As already mentioned, the simultaneous fit of the CDC and GEM-TPC hits is currently being implemented to achieve a global track representation in the combined detection system.

## 12.2.3 Calibration Results with Krypton

During the beam time in Apr 2011 the source containment could be integrated into the gas system and first data from Krypton decays could be taken with Ar/CO$_2$ (90/10), Ne/CO$_2$ (90/10) for different GEM gain settings during the April beam time and in the beginning of June. A typical event with clusters from $^{83m}$Kr decays is shown in Fig.12.26. The size and overall charge of the clusters is much higher as for cosmic tracks and thus can be clearly separated from background events or noise. As a first approach for a relative gain-calibration of the readout-pads a modification of the so-called "leader-pad method" was chosen. In its original form it was developed and successfully tested for the HARP TPC and is described in [76]. Clusters associated with the highest energetic decay of 41.6 keV are selected by cuts on total ADC charge and pads per cluster. The accumulated charge of each pad in a selected cluster is calculated and the energy entries of the highest 3 pads (so-called "leader-pads") that carry the main charge are filled into a his-

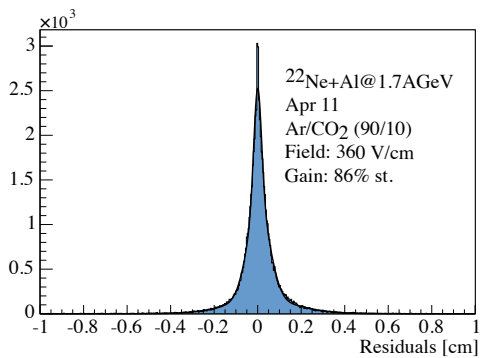

(a) Residuals in V and U direction for the first 10 cm of drift. The data are fitted with 3 Gaussian function. The line shows the sum of the resulting fitted curves.

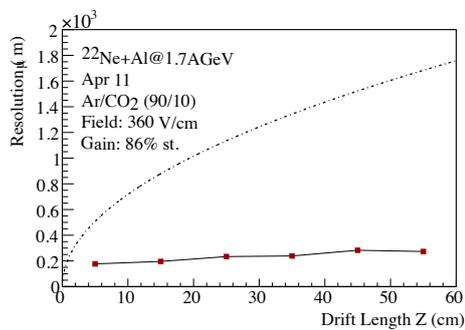

(b) Residual distributions for tracks reconstructed from $^{22}$Ne+Al at 1.7 AGeV collisions as a function of the drift length. The dashed line shows the transverse diffusion for single electron drift calculated with GARFIELD, the red dots represent the sigmas of the narrow Gaussian fits.

**Figure 12.22:** Picture (a) shows the residuals of one slice in z. Picture (b) shows the z-dependency of the residuals. These results are extracted from experimental $^{22}$Ne+Al at 1.7 AGeV collisions (see text for details).

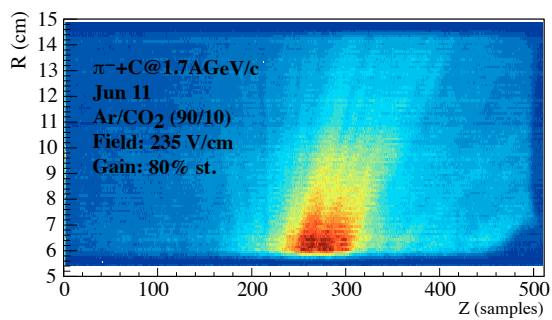

**Figure 12.23:** Occupancy plot of the GEM-TPC detector obtained with tracks produced in $\pi^-$+C at 1.7 GeV/c reactions (see text for details). The z-axis is displayed with a linear scale, units are arbitrary.

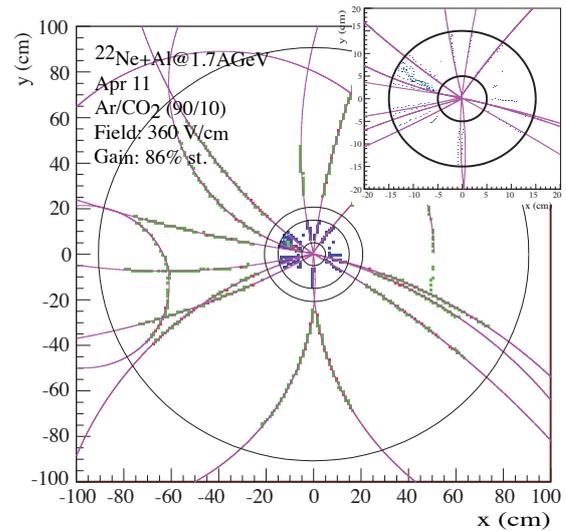

**Figure 12.24:** A matched event in TPC (two inner circles) and CDC (two outer circles) in a x-y projection. The pink lines indicate the reconstructed tracks of the CDC while the green point are CDC hit points and the blue points are hits in the TPC

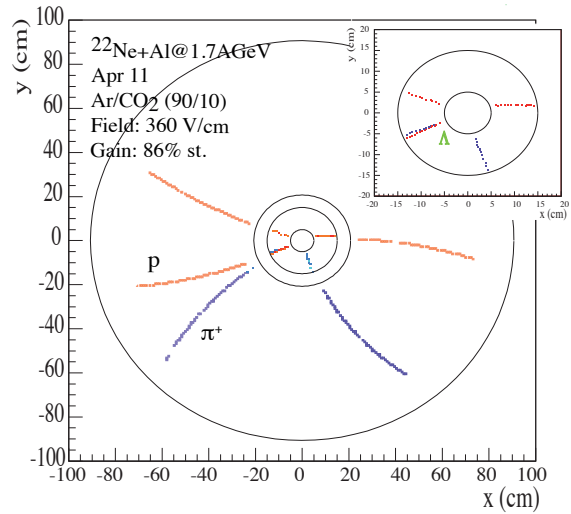

**Figure 12.25:** $\Lambda$ candidate.

togram for each pad. The parameters and pad size of the HARP TPC are different, i.e. the main charge was deposited on only one single pad. Regarding the pad size of the prototype TPC and the width of the observed clusters the number of 3 pads for this method was found to be optimal. Less pads introduce a bigger error on the estimation of the central peak position in each bin while more pads introduce a higher background by lower energetic hits. Normalized gain equalisation factors for the whole pad plane are calculated from the peak positions and applied as a correction factor to the raw data hits. A plot of the equalisation factors for each pad can be seen in Fig. 12.27. The sector borders of the GEM detector responsible for the higher gain can be clearly seen as these lines are almost blind and subjacent pads have to register a lower overall charge. The same effect is true for pads close to the field cage as they do not register the full cluster charge. Thus the position of the leader-pad cannot be determined clearly and the gain of such pads would be over-corrected. Such pads were left out in this map and are shown with a correction factor of 0. Different methods have to be applied for such cases as it was already shown in [76] where a truncated-means method had been shown to be suitable. For the GEM-TPC prototype a suitable method to treat such pads still had to be tested or developed.

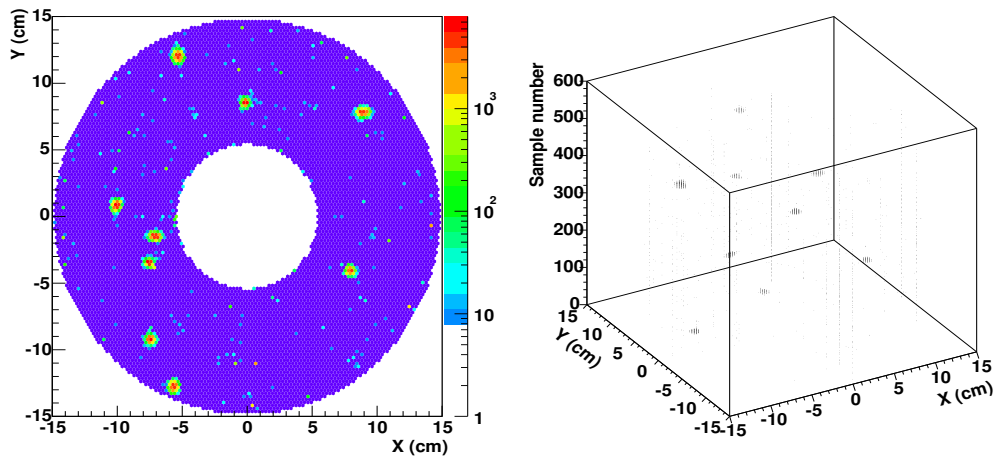

**Figure 12.26:** A typical krypton calibration event with several decays inside the TPC volume

One damaged GEM sector in the upper region can be clearly seen in the gain map. This sector showed a very noisy spectrum during the krypton runs and no usable clusters were observed. As a consequence the gain equalisation constants could not be determined and were set to 0 in this plot to show this effect.

Several iterations of the leader-pad method running the full cluster reconstruction were used to correct for possibly wrongly reconstructed clusters in a previous iteration. After 3 iterations the relative change in the equalisation constants is about 1-2%, the resulting leader-pad peaks before and after correction are shown in 12.28. Gain variations show a periodic structure that has a dependence on the FE chip channels as well as on geometrical properties and the GEM foils as discussed above.

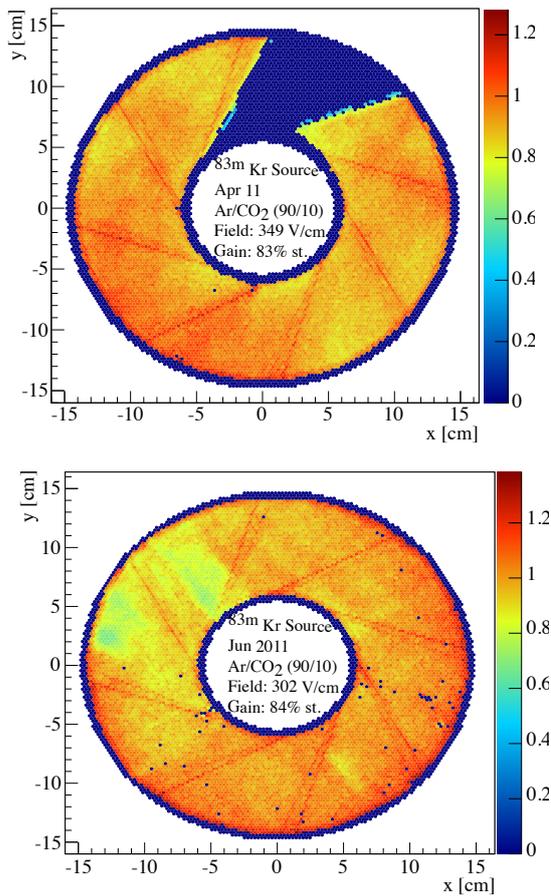

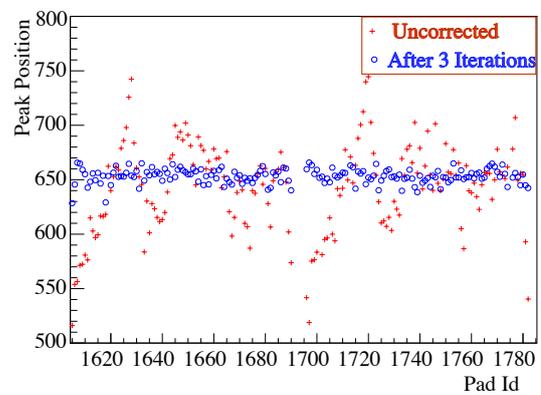

**Figure 12.28:** Peak positions of leader-pads uncorrected (red) and after 3 iterations of corrections. The corrected distribution is compatible with a flat distribution within the error bounds (not shown here for better visibility of the data points). The periodic structure can be seen on all pads and it repeats for each new FE chip while additional effects due to pads at sector borders or next to the field cage can be observed.

**Figure 12.27:** Gain equalization constants plotted for each channel over the pad plane. The overall gain shows small spatial fluctuations while the pads below a gem sector boundaries need to be corrected as the deposited charge on the central pad below such borders is lower. At the inner and outer field cage no central energy deposition of krypton clusters was observed, different methods have to be applied here. One damaged GEM sector was found during the krypton calibration. Even though this sector cannot be calibrated a quick check for damaged sector is possible with the krypton method.

# 13 Quality Control and Risk Assessment

## 13.1 Field Cage

Building gaseous detectors, cleanliness, especially the absence of dust and lints and low humidity is a prerequisite for good quality especially if gluing with epoxy glues is involved. Dealing with high voltages of several 10 kV as involved in the development of the $\overline{P}$ANDA TPC gives even harder constraints to a proper surrounding. Thus building and handling of the field cage will be done in a clean-room surrounding of class ISO 4. All materials especially the ones involved in the production of the sandwich walls of the field-cage vessels will be stored in zero-humidity compartments and processed solely in the clean room.

All base materials will be quality checked before their application in production. Especially their electrical properties concerning resistivity and HV-stability will be subject to tests. Their properties in terms of detector aging will be investigated in a dedicated setup (cf. Sec. 13.4).

All parts of the field defining system, especially the foils carrying the strip-line electrodes will be checked optically and electrically for any deviations of resistivity, quality of the surface and deviation from their ideal shape on small and large-area scale. The field homogeneity, which has to be expected from the field-cage foils will be checked by measuring individual resistivity values for every adjacent strip. The final specimen produced will be quality checked with respect to the major characteristics, e.g. the high-voltage stability for representative time period.

The test procedures, tools, process descriptions which will be applied for the production of the final detectors have already been developed and successfully applied during the development and construction of the Large-Prototype TPC.

## 13.2 GEM Detectors

COMPASS was the first large-scale experiment to use GEM-based detectors. A lot of experience has been gained during the construction of these detectors. Quality assurance procedures for all detector components have been established in order to guarantee a uniform and stable operation of the chambers [13]. These have successfully been applied to the large prototype, and will be adopted to a large extent for the $\overline{P}$ANDA TPC.

Since the active detector surface required for the $\overline{P}$ANDA TPC is relatively small compared to other experiments which use the GEM technology (it is e.g. only about a quarter of the total active area of GEM detectors in COMPASS), the production of the required number of GEM foils at CERN-TS/DEM does not require any additional resources in this laboratory. Nevertheless, first contacts with external companies have been established in order to explore alternative providers. Test samples of GEM foils have been provided by one company and are currently being tested.

The testing and handling of GEM foils and the construction of GEM detectors will be done under clean-room conditions (class 10 000 or better). The foils and all mechanical parts in contact with high voltage (HV) will be tested for HV stability and leakage currents before being mounted, and after each mounting step. For use in a TPC, a good uniformity of gain over all GEM foils is essential in order to achieve the desired $dE/dx$ resolution. This will be tested using optical methods to verify uniform hole shapes and thus transparency, and will be measured on control samples in a dedicated test setup. The final detectors will be subjected to extensive tests using X-rays, radioactive sources and cosmics, before they are mounted to the TPC.

## 13.3 Pad plane

The proper reconstruction of tracks inside the drift volume of the TPC relies on the constant drift velocity without distortions as well as on the knowledge of the shape and layout of the pad-electrodes on the pad-plane where these charges are projected and collected.

The design of the hexagonal pads and their relative positioning among each other and with respect to the optical axis of the detector has to be known within an accuracy of below 0.1 mm. To check the quality of these configurations, a system for optical inspection has been used for quality control. It consists of a large-area planar scanning table with 3D axis motion and a digital optical microscope.

The measurements of the geometries, e.g. of the mounting holes of the pad-plane reveal small devia-tions from its perfect circular shape which are well within the design values.

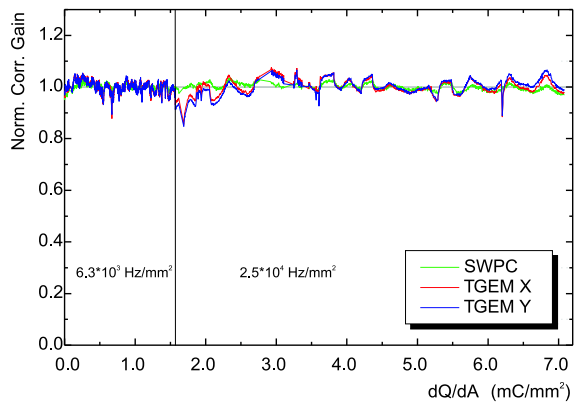

**Figure 13.1:** Normalized gain as a function of the integrated charge for a COMPASS triple GEM detector in a mixture of Ar/CO$_2$ (70/30).

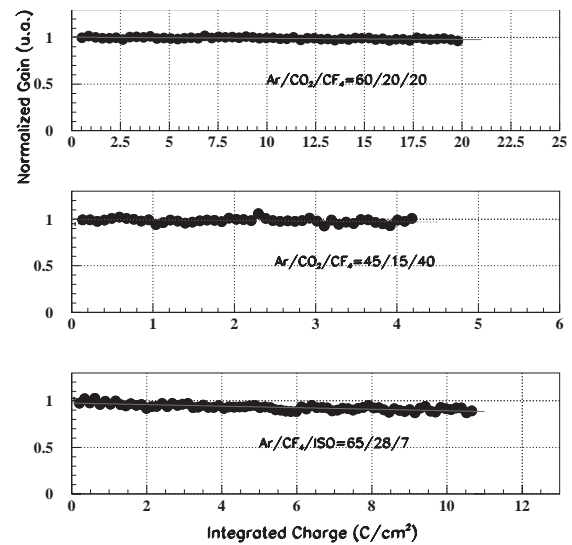

**Figure 13.2:** Normalized gain as a function of the integrated charge for a triple GEM detector of the LHCb experiment in various gas mixture containing CF$_4$.

## 13.4  Aging

In COMPASS, the GEM detectors, operated at a gain of $\sim$ 8000 in an Ar/CO$_2$ (70/30) gas mixture, are exposed to extremely high particle fluxes of charged particles ($> 1 \cdot 10^5$ mm$^{-2}$s$^{-1}$ close to the beam axis), imposing considerable requirements on the radiation tolerance of their performance. During aging tests, more than 7 mC/mm$^2$ were collected on the readout strips using an intensive X-ray beam without loss of gain or energy resolution [77]. This corresponds to the total charge collected in more than 7 years of operation in COMPASS. Figure 13.1 shows the variation of the effective gain of a triple GEM detector with 2-D readout under irradiation, compared to a single-wire proportional counter irradiated in parallel.

No degradation of performance was observed in nine years of operation, both with $\mu$ and hadron beams, with local charge accumulation exceeding 10 mC/mm$^2$ in some areas.

In aging studies performed for the LHCb triple GEM tracker, operating with an Ar/CO$_2$/CF$_4$ (60/20/20) gas mixture, the total integrated charge was as high as 200 mC/mm$^2$ without degradation of the performance [78], as shown in Fig. 13.2.

Owing to the facts that the anti-proton beam does not cross the GEM detectors in P̄ANDA, and that the detectors are located in the backward hemisphere of the target, the operating conditions for the GEM detectors in P̄ANDA are expected to be less harsh. At the foreseen gain of 2000, and assuming half a year of running at 80% duty cycle, the total integrated charge is estimated to be about 0.1 mC/mm$^2$, much below the numbers quoted for other experiments. The fact that the P̄ANDA TPC will operate with the noble gas Ne instead of Ar is not expected to make any difference concerning aging, as it is either the quench gas or impurities introduced into the gas volume which cause aging. All materials in contact with the detector gas will be subject to aging tests using a dedicated setup consisting of a reference detector, a heatable sample box, which contains the material under investigation and is flushed with detector gas, and a second detector downstream of the sample box.

## 13.5  Ion Backflow

As mentioned already in chapter 11 the ion backflow (IB) is one of the main parameters, that has to be controlled for a continuously operating TPC. Two contradicting requirements have to be kept in mind. On the one hand large gas avalanches, especially in the first GEM foil, are favored for optimal resolution. Ionization charges lost in the very beginning of the charge avalanche process contribute to a deterioration of point and energy resolution. However in regard of ion backflow a very small amplification in the first GEM stage is favored to minimize the backflow of ions where the ions can directly escape into the drift volume. The triple GEM amplification with the electric field settings indicated in Sec. 4.3, which is the current baseline solution for the $\overline{\text{P}}$ANDA TPC, provides an ion backflow of 0.25%, corresponding to four back drifting ions per electron reaching the GEM stack at a gain of $2 \cdot 10^3$. We have shown that corrections based on a conventional laser system restore the required point resolution of the detector even at the maximum luminosity of $2 \cdot 10^{32} \, \text{cm}^{-2}\text{s}^{-1}$.

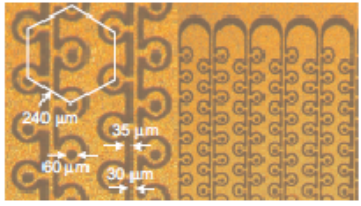

**Figure 13.3:** Microscope photograph of a Cobra micro hole electrode with the dimensions given in the figure [79].

Currently, there are two different strategies of further reducing the ion backflow, which will be pursued in the future, and which potentially eliminate the problem completely. The first strategy will be the further optimization of the MPGD devices. There are promising new results using GEM foils with additional patterning on one side, which are already available and will be investigated in the near future. The second strategy will be to investigate an alternative amplification concept of the primary electrons, that does not produce ions in the first place. However, as the alternative concept has so far never been applied to the field of tracking detectors some R&D effort have to carried out. Therefore this alternative concept certainly requires a longer time perspective.

### 13.5.1 Strategy 1: Reduction of IB by Optimization of Amplification System

The next logical attempt to improve the performance after monitoring and calibrating the TPC will be to minimize the IB by optimizing the TPC by modifications of the gas amplification methods. One promising technique, which has to be evaluated for the specific need of the $\overline{\text{P}}$ANDA TPC is the application of a Cobra/2GEM device as it has been proposed by [79]. The new patterned micro-hole electrode named Cobra (see Fig. 13.3) was devel-oped with a geometry that is expected to improve the ion divergence away from the holes. It has thin anode electrodes surrounding the top side of the holes and creating strong electric field inside the holes (required for charge amplification). The more negatively biased cathode electrodes cover a large fraction of the area for a dramatic improvement of the ion-collection efficiency on the top side of the Cobra device. It was found, that when introduced as a first element (with the patterned area pointing towards the drift volume), preceding two GEMs in the cascade (Fig. 13.4), it drastically improves the ion trapping capability. The ion backflow (IB) as a function of the total gain for drift fields of 0.2 and 0.5 kV/cm, respectively, is shown in Fig. 13.5. With a drift field of 0.5 kV/cm in an Ar/CH$_4$ gas mixture the IB was $3 \cdot 10^{-6}$ and with a drift field of 0.2 kV an IB of $2.7 \cdot 10^{-5}$ was reached. These values are so far the best values ever achieved for IB measurements. It has to be investigated if this setup is applicable for the PANDA TPC without deteriorating other key parameters such as the momentum resolution. Especially the large scale application for tracking purposes has to be demonstrated to establish this as an alternative technique to an conventional amplification system with GEMs alone.

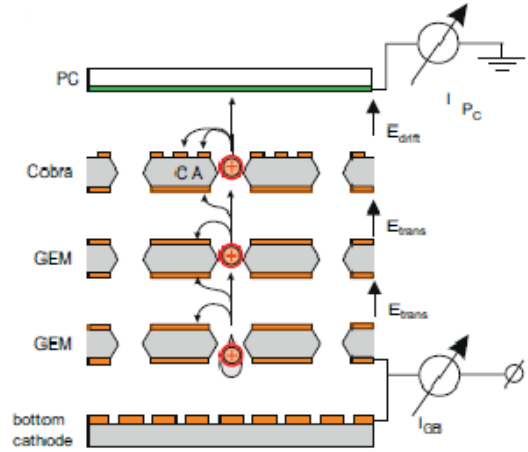

**Figure 13.4:** Scheme of a COBRA/2GEM amplification. Possible avalanche ion paths are shown [79].

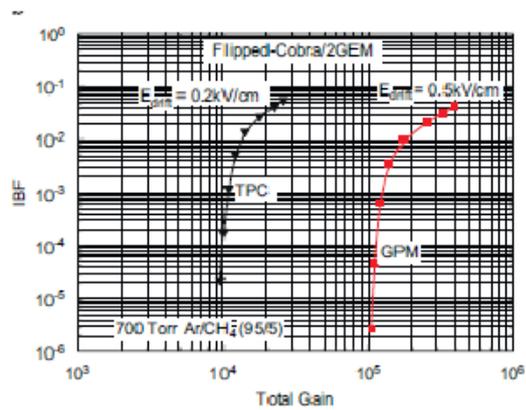

**Figure 13.5:** Ion backflow as a function of the total gain for drift fields of 0.2 and 0.5 kV/cm, respectively [79].

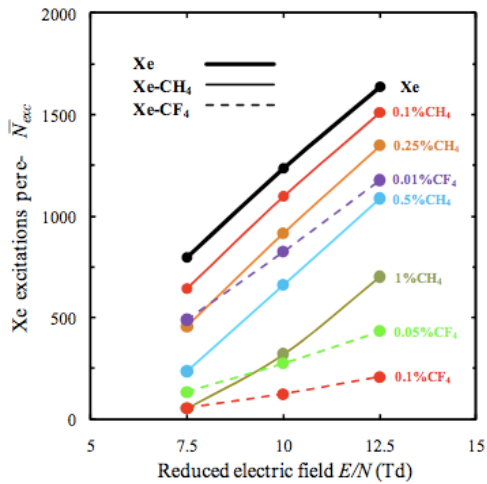

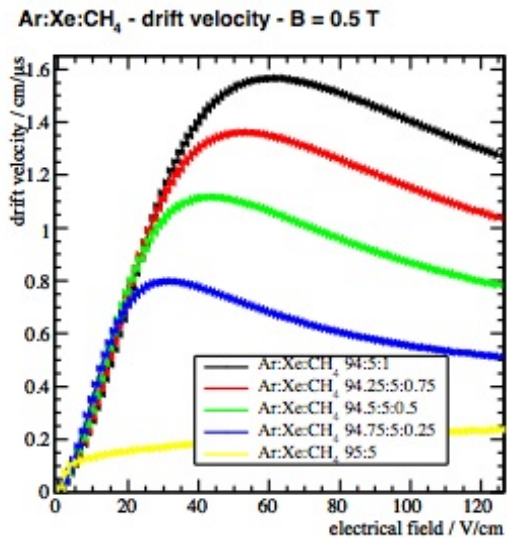

**Figure 13.6:** Light yield as a function of the electric field for pure Xenon and with a small percentage of CH$_4$ and CF$_4$ quencher, respectively [80].

**Figure 13.7:** Drift velocity for an Ar/Xe gas mixture as a function of the electric field for small contributions of CH$_4$.

### 13.5.2 Strategy 2: Prevention of IB by an Alternative Concept

An alternative approach would be to employ a completely new concept, which reduces the ion backflow to zero. Instead of amplifying the primary charges by charge avalanche of the gas in an amplification stage such as a GEM, Micromegas, Cobra or MHSP it is possible to only excite a noble gas by electro-luminescence (EL). This process produces VUV photons. The wavelength of the photons depends on the composition of dimers of the specific gas. In xenon the VUV photons of about 178 nm are produced. The number of photons depends on the number of collisions of the primary charges with the gas atoms. This process is linear to the length of the scintillation gap, where a specific electric field is applied. The following empirical

formula [81] is given to describe the EL gain:

$$\eta_{Xe} = 140(E/p - 0.83)p\Delta x \text{ (UV photons /e cm}^{-1}), \quad (13.1)$$

where $E$ denotes the electric field, $p$ the pressure and $\Delta x$ the length of the scintillation gap. As the photons leave the amplification region without further interacting with the gas atoms, no secondary interaction will take place, so that also the gain fluctuations are linear. Therefore EL also improves significantly the energy resolution of a TPC detector.

The EL process will reduce the IB to zero. The drawback of a detector operating in EL mode is that it works best in pure noble gases like Argon or Xenon. Pure noble gases have a very low drift velocity, which limits the readout speed of the detector. The key optimization, which has to be performed is to find the right gas mixture of a noble gas plus a minimal amount of quencher, which speeds up the drift velocity (and therefore the readout speed) and at the same time allows still an sufficient light yield. In conventional TPCs a quencher is added exactly for the reason to reduce the VUV photon to zero and to increase the drift velocity. In [80] a study to find a suitable quencher has been performed. In particular, the quenchers CH$_4$ and CF$_4$ have been investigated. Figure 13.6 shows that if the CH$_4$ contribution is below 1%, still a sufficient light yield can be reached. In the case of CF$_4$ even a minimal contribution of 0.1% will practically kill the whole EL light production. Figure 13.7 shows the increase of drift velocity as a function of the applied electric field for gases with $CH_4$ contributions from 0.25 to 1%. A 1% contribution of $CH_4$ increases the drift velocity significantly, so that the drift velocity becomes comparable to the ones used in current TPC detectors.

Also the readout plane would have to follow the different concept of light readout. Instead of a passive readout by pads, which is non amplifying, photosensors like Avalanche Photo Diodes (APD) or Silicon Photomultiplier (SiPM) have to be used, which would convert the photons back into an linear or exponentially amplified electric signal. The size of these devices would have to be in the order of the current pad size, which is about $3 \times 3 \, mm^2$. APDs as well as SiPMs can be built in these sizes (see [82]). With APDs or SiPM one has readout pixels, which are amplifying devices so that an additional pre-amplifier might not be needed.

In the scope of neutrinoless double beta gaseous TPCs ([82]) first preliminary results for such a readout are shown. However the proof of principle still has to be given that such a technique can be applied for a tracking TPC of the PANDA type. Therefore strategy 1 will be the short to mid term solution, while strategy 2 is a more fundamental study, which requires a long term perspective.